\documentclass{article}

\usepackage{graphicx}
\usepackage{natbib}
\usepackage{amsmath}
\usepackage{authblk}
\usepackage{color}
\usepackage{amssymb,latexsym}

\newcommand\Rey{\mbox{\textit{Re}}}  

\definecolor{verde}{rgb}{0,0.5,0}

\newcommand{\mylab}[3]{\raisebox{#2}[0mm][0mm]{%
\makebox[0mm][l]{\hspace*{#1}\textbf{#3}}}}
\def\spacce#1{\hskip #1pt}
\def\drawline#1#2{\raise 2.5pt\vbox{\hrule width #1pt height #2pt}}
\def\solid{\drawline{24}{.5}\nobreak}

\def\bdash{\hbox{\drawline{7}{.5}\spacce{2}}}
\def\bdashort{\hbox{\drawline{3.5}{.5}\spacce{1}}}

\def\dashed{\bdash\bdash\bdash\nobreak}

\def\bdot{\hbox{\drawline{1}{.5}\spacce{2}}}

\def\dotted{\hbox{\leaders\bdot\hskip 24pt}\nobreak}

\def\chndot{\hbox%
{\drawline{9.5}{.5}\spacce{2}\drawline{1}{.5}\spacce{2}\drawline{9.5}{.5}}\nobreak }
\def\circle{$\circ$\nobreak }

\def\trian{\raise 1.25pt\hbox{$\scriptstyle\triangle$}\nobreak}
\def\leftrian{\raise 0.75pt\hbox{$\displaystyle\triangleleft$}\nobreak}
\def\rightrian{\raise 0.75pt\hbox{$\displaystyle\triangleright$}\nobreak}

\def\dtrian{\raise 1.25pt\hbox%
{$\scriptscriptstyle\bigtriangledown$}\nobreak}

\def\squar{\raise 1.25pt\hbox{$\scriptstyle\Box$}\nobreak}

\def\diamon{\raise 1.25pt\hbox{$\scriptstyle\diamond$}\nobreak}

\def\solidtrian{$\blacktriangle$\nobreak}

\def\solidcircle{$\bullet$\nobreak}

\def\bra{\langle}
\def\ket{\rangle}

\def\beq{\begin{equation}}
\def\eeq{\end{equation}}

\def\aaa{{\it a}}
\def\bbb{{\it b}}
\def\ccc{{\it c}}
\def\ddd{{\it d}}
\def\eee{{\it e}}
\def\fff{{\it f}}
\def\ggg{{\it g}}

\def\citalajim03{Del \'Alamo \& Jim\'enez (2003)}
\newcommand\ie{{\it i.e.}\;}

\newcommand{\etal}{\textit{et al.}}
\newcommand{\eg}{e.g.}

\newcommand{\unit}[1]{\ensuremath{\, \mathrm{#1}}}
\title{Flow of a viscous nematic fluid around a sphere}
\author[1]{Manuel G\'omez-Gonz\'alez \thanks{manu@ucsd.edu}}
\author[1, 2]{Juan C. del \'Alamo}
\affil[1]{Mechanical \& Aerospace Engineering Department, University of
California San Diego.}
\affil[2]{Institute for Engineering in Medicine, University of California San 
Diego.}
\begin{document}
%
\maketitle
%
\begin{abstract}
%
We analyze the creeping flow generated by a spherical particle moving through a
viscous fluid with nematic directional order, in which momentum diffusivity is
anisotropic and which opposes resistance to bending.
\textcolor{black}{Specifically, we provide closed-form analytical expressions 
for the response function, \ie the equivalent to Stokes's drag formula for 
nematic fluids.}
Particular attention is given to the rotationally pseudo-isotropic condition
defined by zero resistance to bending, and to the strain pseudo-isotropic
condition defined by isotropic momentum diffusivity.
We find the former to be consistent with the rheology of biopolymer networks
and the latter to be closer to the rheology of nematic liquid crystals.
These ``pure'' anisotropic conditions are used to benchmark existing particle
tracking microrheology methods that provide effective directional viscosities
by applying Stokes's drag law separately in different directions.
We find that the effective viscosity approach is phenomenologically justified
in rotationally isotropic fluids, although it leads to significant errors in
the estimated viscosity coefficients.
On the other hand, the mere concept of directional effective viscosities is
found to be misleading in fluids that oppose an appreciable resistance to
bending. 
Finally, we observe that anisotropic momentum diffusivity leads to asymmetric
streamline patterns displaying enhanced (reduced) streamline deflection in the
directions of lower (higher) diffusivity.
The bending resistance of the fluid is found to modulate the asymmetry of
streamline deflection.  In some cases, the combined effects of both anisotropy
mechanisms leads to streamline patterns that converge towards the sphere.
%
\end{abstract}

\section{Introduction} \label{sec:Introduction}
%
Nematic fluids exhibit molecular or supramolecular alignment along a director
vector field, leading to anisotropic rheology. 
These fluids appear often in industrial applications and in biology. 
For instance, nematic liquid crystals are a key component in a vast variety of
displays for their high resolution and energy efficiency.  \textcolor{black}{The
interaction between flow, topological defects and nematic orientation in liquid
crystals are exploited in particle self-assembly 
\citep{Poulin.Stark.Lubensky.Weitz.1997, Loudet.Barois.Poulin.2000} and sensing
applications \citep{Brake.Daschner.Luk.Abbott.2003}.}
Anisotropic rheology is also observed in reconstituted gels when the polymer
phase is aligned \citep{Hasnain.2006, He.Mak.Liu.Tang.2008} and, perhaps more
importantly, the cytoplasm of animal cells is far from being isotropic.  The
filaments that make up the cytoskeleton can experience spatial order and
alignment both at the level of the mesh size ($\sim 10 nm$) and of the whole
cell ($\sim 10 \mu m$), leading to short- and long-range directionality
\citep{lubyphelps:00}.
Intracellular rheology has been proposed to modulate important cell functions
such as mechanotransduction, \citep{chien2007,Wang05,Chen04}, cell migration
\citep{kole:05,pollard:03,yanai:04} and intracellular organelle transport
\citep{lammer:05,lee:05,minin:06}.
Recent studies suggest that intracellular anisotropy may play an important role
in controlling the directionality of these cellular processes
\citep{sch:has:07,Rogers.Waigh.Lu.2008,DelAlamo.2008}.
%
%
In short, anisotropy is purpose.

Particle Tracking Microrheology (PTM) determines the viscoelastic shear modulus
of a medium from the measured motion of embedded microparticles
\citep{Mason.1995,Mason.1997,Mason.2000}.  This technique is particularly
suitable for probing minute quantities of biological materials, including live
cells \citep{Tseng.Kole.Wirtz.2002, crocker:07, DelAlamo.2008}.  
In active PTM, the motion of the probing particle is forced with magnetic or
laser tweezers, whereas in passive PTM this motion is caused by the thermal
excitation of the medium.  
These two approaches estimate the complex response function $\tilde{ \zeta}(s)$
that relates the drag force experienced by the probing particle and the
particle velocity as
$ \tilde {\mathbf f}(s) = \tilde {\mathbf \zeta}(s) \tilde {\mathbf v} (s), $
where $\, \tilde \cdot \,$ indicates Laplace transform and $s$ is the Laplace
frequency.
A crucial step in both active and passive PTM is to connect the measured
response function to the underlying rheological properties of the medium, which
is usually accomplished by assuming that the particle experiences a generalized
Stokes's resistance.
For a sphere of radius $a$ moving in a single-phase no-slip isotropic
continuum, the response function is modeled as
$ \tilde \zeta(s) = 6 \pi a \tilde \eta(s)$
\citep{Mason.1995,squires:10},
where $\tilde \eta(s)$ is the frequency dependent viscosity of the medium. 

The application of the generalized Stokes's law in complex biomaterials (e.g.
gels) is complicated by their multiphase nature as well as by interactions
between the material and the probing particle.
For instance, the Stokes's flow assumption can break down in live cells when
tracking endogenous particles that are anchored to the cytoskeleton
\citep{lin:90}.  Furthermore, the mobility of injected microspheres has been
reported to vary dramatically depending on their surface charge
\citep{Tseng.Kole.Wirtz.2002}.  Chemical interactions, polymer depletion near the microsphere
and network compressibility can also introduce substantial deviations from
Stokes's flow \citep{squires:08,mcgrath:00,valentine:04,gittes:97}.

Theoretical studies have analyzed the effect of both network compressibility
and slip between the probing particle and the material in two-fluid gels
composed by a polymer network viscously coupled to a solvent
\citep{Levine.Lubensky.2000,Fu:She:Pow:08}.  These studies have provided
frequency-dependent corrections to the generalized Stokes's law that can be used
to better interpret PTM experiments.  
The introduction of two-point PTM \citep{crocker:00,Levine.Lubensky.2001} has
allowed investigators to account for most non-Stokesian effects in the vicinity
of the probing particle by cross-correlating the motion between pairs of
distant probing particles.

Despite these advances, little is known about the effect of anisotropy in the
response function of a microrheology probe.  This lack of knowledge affects
both the active and passive modalities of PTM, as well as both single-particle
and two-particle microrheology.
Previous studies have provided directional diffusivity coefficients (effective
viscosities) of microrheological probes in nematic fluids,
$\eta_{\parallel}^{\mbox{\scriptsize{eff}}}$ and
$\eta_{\perp}^{\mbox{\scriptsize{eff}}}$, where $\parallel$ and $\perp$ denote,
respectively, the directions parallel and perpendicular to the nematic director.
Although these measurements provide useful information about anisotropic
diffusion processes, it is not clear how to relate $\eta_{\parallel \, /\,
\perp}^{\mbox{\scriptsize{eff}}} $ with the directional viscosities of
anisotropic fluids because the Stokes's formula is not applicable in this case.  

\begin{figure}
	\centering \includegraphics*[width=0.5
\columnwidth,keepaspectratio]{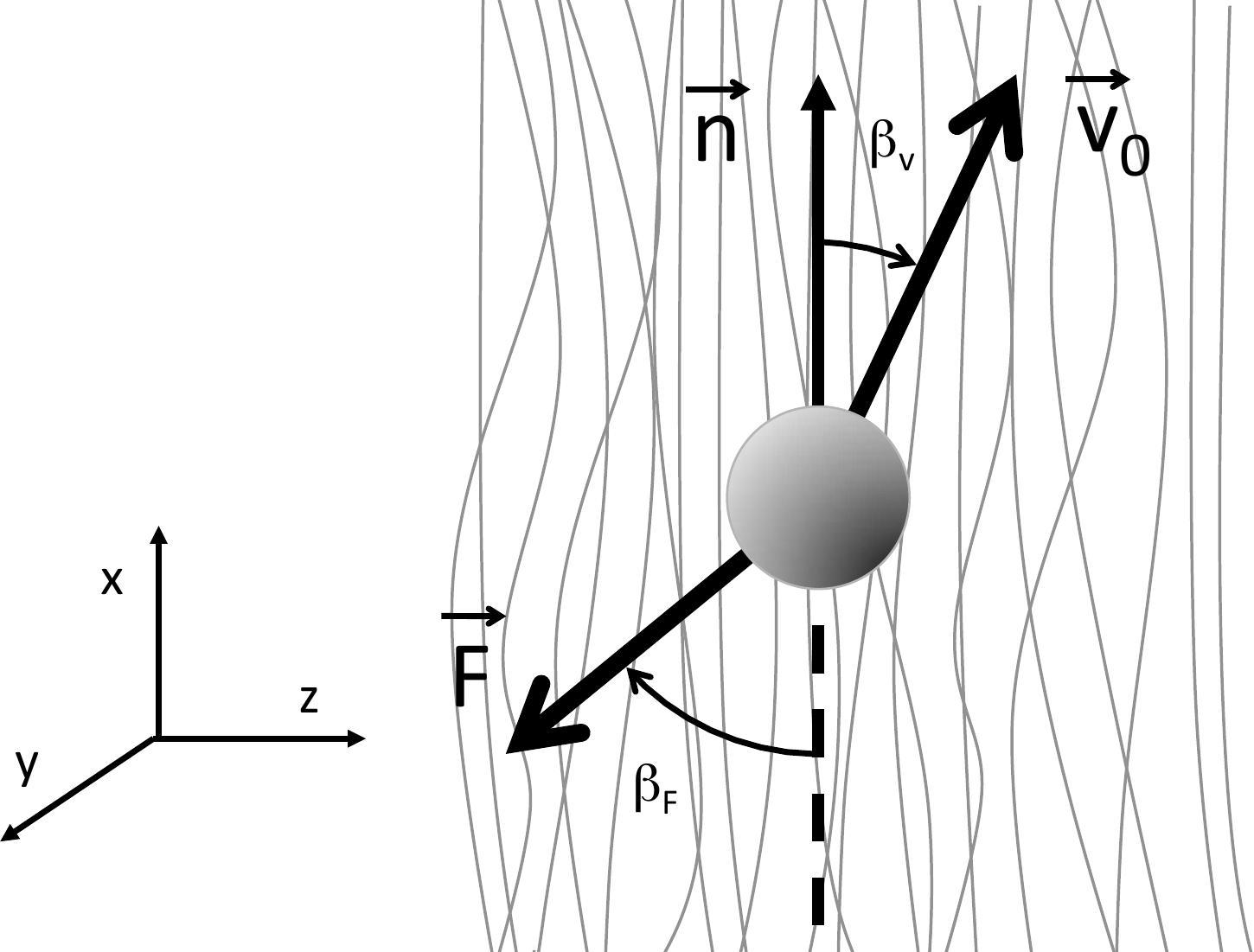}
	\caption{We consider an orthotropic gel with constant and homogeneous
nematic. The motion of the sphere \textcolor{black}{is assumed to not} alter the nematic.}
\label{fig:Sphere_Orthotropic}
\end{figure}

In this paper, we address the limitations of PTM in anisotropic fluids by
calculating an analytical expression for the response function of a microsphere
moving through an orthotropic two-fluid gel.
In \S  \ref{sec:General_Equations}, we formulate the mathematical problem under
the assumptions that the director remains constant and is not altered by the
motion of the sphere (see figure \ref{fig:Sphere_Orthotropic}) and that
there is  strong viscous coupling between the network phase and the liquid
phase. The anisotropy of the medium is modeled through the Leslie-Ericksen
constitutive equations \citep{Ericksen.1960, Leslie.1966}. 
Interestingly, the resulting flow 
equations are also valid for the case of a nematic fluid with high bending 
elastance \textcolor{black}{and free of defects}, thereby conferring a broader 
reach to the results of this theoretical study. 

The Green's function of the problem is obtained in \S
\ref{sec:Greens_Leslie_Ericksen} and is used in \S
\ref{sec:Response_Leslie_Ericksen} to determine an explicit analytical
expression for the anisotropic response function that depends on up to four
viscosity coefficients.
This solution is compared with numerical solutions that have previously been 
obtained for particular combinations of the viscosity coefficients
\citep{Heuer.Kneppe.Schneider.1992}, obtaining excellent agreement.
We find that the anisotropy of the Leslie-Ericksen constitutive relation
affects the response function through two independent mechanisms.  One is the
anisotropy in the stress-strain relationship and the other is the bending of
the fluid with respect to the nematic director.  
These mechanisms are studied in isolation by considering rotationally
pseudo-isotropic fluids in which the bending stresses are zero, and strain 
pseudo-isotropic fluids in which the stress-strain relation is isotropic. 

Analysis of our results and experimental data from nematic biopolymer networks
in \S \ref{sec:DPTM} reveals that the application of Stokes's drag law can lead 
to errors of orders of magnitude in PTM measurements even for moderate levels 
of anisotropy.
Section \ref{sec:velos} illustrates the effect of anisotropy on the flow
velocity field showing that, in some cases, it leads to streamline patterns
that challenge our intuition about creeping flow.

\section{Problem Formulation} \label{sec:General_Equations}
%
This section introduces the equations to be solved in subsequent sections.  
We start by considering a two-fluid gel formed by a dilute polymer network that
is permeated by a liquid.  We illustrate the possible sources of anisotropy in
this model before performing a number of simplifying assumptions that hold
under the conditions of a live-cell PTM experiment. 
The simplified equations are still of general scope, as they also describe the
flow in other anisotropic fluids such as nematic liquid crystals. 
We conclude the section by providing a physical interpretation of the different
viscosity terms appearing in the flow equations.

As mentioned above, we consider a two-fluid gel composed of a polymer network
viscously coupled to a solvent similar to previous PTM studies
\citep{Levine.Lubensky.2000,Levine.Lubensky.2001,Fu:She:Pow:08}.
The viscoelastic network is characterized by its displacement field $\vec{u}$,
and the viscous solvent by its velocity field $\vec{v}$.  
We assume that the network is dilute as the polymer network fraction $\phi$  in
the eukaryotic cytoskeleton is very small \citep[$\phi \sim 0.01 - 0.02$,
see][]{su:etal:07}.
For a dilute network, the two-fluid equations take the form
%
\begin{align}
	\label{eq:network} \rho \ddot{\vec{u}} - \nabla \cdot
\overline{\overline{\tau}}_{N} &= -\overline{\overline{\Gamma}} \cdot
(\dot{\vec{u}} - \vec{v}) + \vec{f}_u,
\\
	\label{eq:cytosol} \rho_F \dot{\vec{v}} - \eta \nabla^2 \vec{v} +
\nabla p &= \overline{\overline{\Gamma}} \cdot (\dot{\vec{u}} - \vec{v}) +
\vec{f}_v, 
\\ \label{eq:continuity} \nabla \cdot \vec{v} &= 0,
\end{align} 
%
where $\overline{\overline{\tau}}_{N}$ is the Cauchy stress tensor in the
network, $\overline{\overline{\Gamma}}$ is a tensor that represents the viscous
coupling between the network and the solvent and $\vec f_u$ and $\vec f_v$ 
model the stresses caused by the sphere in the gel.  

The directional order in the network may cause anisotropic dynamics not only
through constitutive equations of the polymer but also through the viscous
coupling between the polymer and the solvent, even if the solvent is isotropic. 
For instance, if we idealise the polymer phase in a nematic network as a square
array of circular cylinders, then the coupling tensor is defined by the two
coefficients
$$
\Gamma_{||} = \frac{\pi \eta} {\xi^2 \log(A/\phi)} \;\;\;
\mbox{and} \;\;\;
\Gamma_{\perp} = \frac{2 \pi \eta} {\xi^2 \log(A/\phi)},
$$
where $\xi$ is the distance between neighbouring filaments and A is a
non-dimensional shape factor that takes the value $A \approx 0.23$ under dilute
conditions \citep[$\phi < 0.02$, see][]{wan:03}.  Note that, in this case, the
viscous coupling is weaker in the nematic direction than in the perpendicular
directions by a factor of two.

An order-of-magnitude analysis of equation \eqref{eq:cytosol} after eliminating
the pressure indicates that viscous coupling is dominant for $(k\xi)^2 \ll 1 $
and $s \xi^2 \rho_F / \eta \ll 1 $, where $k$ and $s$ are respectively a
characteristic wavenumber and frequency.
These conditions are often met in live cell cytoplasmic microrheology, where
the network (cytoskeleton) mesh size is $\xi \sim 0.02 \, \mu m$, and the
density and viscosity of the cytosol (the solvent) are approximately equal to
those of water \citep{lub:etal:86,lubyphelps:00}.
Using the inverse radius of the probing particle $1/a$ as an upper bound to
$k$, the strong coupling limit is found for $a \gg \xi \sim 0.02 \,\mu m$ and
$s \ll 10^4 \, s^{-1}$.  In the forthcoming analysis, we assume that these
conditions hold so that $\dot{\vec{u}} = \vec{v}$, and equations
\eqref{eq:network} and \eqref{eq:cytosol} combine into
%
\begin{equation} \label{eq:Cauchy}
	\rho \dot{\vec{v}} - \nabla \cdot \overline{\overline{\tau}} = \vec{f},
\end{equation} 
%
where $\overline{\overline{\tau}}$ is prescribed, as a function of $\vec{v}$,
by the constitutive relations for the strongly coupled two-fluid gel.
Interestingly, this simplification makes the flow equations applicable to
single-phase fluids such as nematic liquid crystals as well. 
\textcolor{black}{Note that, due to the strong coupling and the incompressibility
of the solvent \eqref{eq:continuity}, the network motion in this model is also
divergence-free, even if the network is compressible.}

Equation \ref{eq:Cauchy} is completed with the Leslie-Ericksen constitutive
relation, which is generally used for anisotropic fluids with nematic order
\citep{Ericksen.1960, Leslie.1966,De_Gennes.1993},
%
\begin{eqnarray} \label{eq:Leslie_Eriksen}
	\tau_{ij} = & - & p \delta_{ij} + \alpha_1 n_k n_p \varepsilon_{kp} n_i
n_j + \alpha_2 n_i N_j \nonumber \\ & + & \alpha_3 n_j N_i +\alpha_4
\varepsilon_{ij} + \alpha_5 n_i n_k \varepsilon_{kj} + \alpha_6 n_j n_k
\varepsilon_{ki}.
\end{eqnarray}
%
This constitutive relation depends on the six Leslie viscosity coefficients,
$\alpha_i$.
It relates the stress tensor $\tau_{ij}$ with the strain rate tensor
$\epsilon_{ij}=(v_{i,j} + v_{j,i})/2$, the director $n_i$ and 
the rate of change of the director with respect to the background fluid.  The
latter is written as 
%
\begin{equation*}
	\vec{N} = \partial_t \vec{n} + (\vec{v} \cdot \nabla) \vec{n} - (\nabla
\wedge \vec{v}) \wedge \vec{n} /2,
\end{equation*}
%
where the first two terms represent the substantial derivative of $\vec{n}$ and
the third one represents the rotation of the fluid with respect to $\vec{n}$.

In general, the director field $\vec{n}$ can be obtained by imposing the
elastic and viscous torques to be in equilibrium.
For simplicity, however, we assume that the viscous torque is much smaller than
the elastic one, so that the director remains uniform and constant throughout
the fluid and $\vec N = - (\nabla \wedge \vec{v}) \wedge \vec{n} /2 $. 
This assumption is known as the small Ericksen number limit
\citep{Stark.Ventzki.2001} and is likely reasonable
to model passive PTM experiments, where the thermal energy driving the motion,
$k_B T$, is extremely low and not expected to alter the nematic configuration.
We further assume that the Reynolds number, $\Rey = \rho U a / \alpha_4$, is
very small.
Typical parameter values in PTM are $\rho \sim 10^3 \unit{kg} \unit{m^{-3}}$, 
$U \sim 10^{-6} \unit{m} \unit{s^{-1}}$, $a \sim 10^{-6} \unit{m}$, 
$\alpha_4 \sim 10^{-3} \unit{Pa} \unit{s^{-1}}$ and $Re \sim 10^{-6}$,
so the low-Reynolds-number simplification is reasonable in this problem.
These simplifications make the resulting equations analogous to those describing
the dynamics of nematic visco-elastomers \citep{Stenull.Lubensky.2004} when
only the viscous terms are retained in the latter.

It is possible to reduce the number of independent viscosity coefficients from
the six that appear in the Leslie-Ericksen constitutive equation
(\ref{eq:Leslie_Eriksen}) to four by introducing the Miesowicz coefficients
\citep{Miesowicz.1936},
%
\begin{align*}
	\eta_a &= \alpha_4/2,\\
	\eta_b &= (\alpha_3 + \alpha_4 + \alpha_6)/2, \\
	\eta_c &= (-\alpha_2 + \alpha_4 + \alpha_5)/2,
\end{align*}
%
and by using Parodi's relation \citep{Parodi.1970},
%
\begin{equation*}
	\alpha_6 = \alpha_2 + \alpha_3 + \alpha_5.
\end{equation*}
%
Under these simplifications and, using a Cartesian coordinate system in which
$\vec{n} = (1, 0, 0)$, the momentum balance equation \eqref{eq:Cauchy} can be
written as
%
\begin{align}
	\label{eq:fluid_equations_x} 
\partial_x p' &= \eta_b \nabla^2 u + \alpha_1 \partial_{xx} u + f_x, \\
	\label{eq:fluid_equations_y} 
\partial_y p' &= \eta_a \nabla^2 v + (\eta_c - \eta_a) \partial_x \omega_z +
f_y, \\
	\label{eq:fluid_equations_z}
\partial_z p' &= \eta_a \nabla^2 w - (\eta_c - \eta_a) \partial_x \omega_y +
f_z,
\end{align}
%
where $\vec{r} = (x, y, z)$ is the position vector, $\vec{v} = (u, v, w)$ is
the velocity vector, $\vec{\omega} = (\omega_x, \omega_y, \omega_z)$ is the
vorticity vector and $p' = p - \alpha_5 \partial_x u$ is a modified pressure.  
This form of the momentum balance equations is not the most suitable one to
solve for the flow around a sphere, but it is included here to facilitate the
physical interpretation of the Miesowicz viscosity coefficients.
The equations suggest that the two first Miesowicz coefficients, $\eta_a$ and
$\eta_b$, govern the diffusion of the momentum components perpendicular
($\perp$) and parallel ($||$) to $\vec n$, respectively.
In particular, $\eta_a$ is equivalent to the viscosity in an isotropic fluid
and the difference $\eta_b - \eta_a$ indicates the level of anisotropy in the
stress-strain relation due to the nematic order of the fluid.
Apart from the momentum diffusion Laplacians, equations
\eqref{eq:fluid_equations_x}-\eqref{eq:fluid_equations_z} have contributions
from $||$-derivatives of the vorticity.  These terms represent bending of the
fluid with respect to the nematic and are proportional to $\eta_c
- \eta_a$.  This second source of anisotropy is perhaps less intuitive than the
  anisotropy of the stress-strain relation but it can modify significantly the
response function of a sphere, as shown in \S
\ref{sec:Response_Leslie_Ericksen}.

Taking the curl of equations
\eqref{eq:fluid_equations_x}-\eqref{eq:fluid_equations_z} yields the vorticity
equation,
%
\begin{equation}
	\left[ \eta_a \nabla^2 + (\eta_c - \eta_a) \partial_{xx} \right]
\vec{\omega} + \left[ (\eta_b - \eta_a) \nabla ^2 + \alpha_1 \partial_{xx}
\right] \left\{ \nabla \wedge \left[ (\vec{u} \cdot \vec{n}) \vec{n} \right]
\right\} = \vec{0},
\label{eq:vorti}
\end{equation}
%
which is analogous to Laplace's vorticity equation for an isotropic, low-{\Rey}
flow, and where terms in addition to $\nabla^2 \vec{\omega}$ clearly show the
two anisotropy mechanisms mentioned above: the anisotropy in the stress-strain
relationship, which is proportional to $\eta_b - \eta_a$, and the bending
resistance of the fluid, which is proportional to $\eta_c-\eta_a$.
Unfortunately, the physical interpretation of $\alpha_1$ appears to be less
clear. This parameter bears relation to the compressibility of the fluid in the
direction of the nematic, as it can be deduced by inspecting the constitutive
equation \eqref{eq:Leslie_Eriksen}.  Nevertheless, our results suggest that
$\alpha_1$ affects the response function of a sphere less than the other
viscosity coefficients (see \S  \ref{sec:Response_Leslie_Ericksen}).

\section{ Green's function for the flow in an orthotropic fluid }
\label{sec:Greens_Leslie_Ericksen}
%
The response function of a particle in the linear regime can be calculated by
using a multipole expansion \citep{Microhydrodynamics, Liron.Barta.1992,
Ladyzhenskaya.1969}.  The first step is to calculate the Green's function for
the flow.  This task is simplified by defining a second modified pressure,
%
\begin{equation*}
	\overline{p} = p' + (\eta_c - \eta_a) \partial_x u = p + (\eta_c -
\eta_a - \alpha_5)\partial_x u,
\end{equation*}
%
so that equations \eqref{eq:fluid_equations_x} to \eqref{eq:fluid_equations_z}
become separated in $u$, $v$ and $w$, 
%
\begin{align}
	\label{eq:Motion_1} \partial_x \overline{p} &= (\eta_c + \eta_b - 
\eta_a + \alpha_1) \partial_{xx} u + \eta_b (\partial_{yy} + \partial_{zz}) u 
+ f_x, \\
	\label{eq:Motion_2} \partial_y \overline{p} &= ~~~~~~~~~~~~~~~~~~~~~~ 
\eta_c \partial_{xx} v + \eta_a (\partial_{yy} + \partial_{zz}) v + f_y, \\
	\label{eq:Motion_3} \partial_z \overline{p} &= ~~~~~~~~~~~~~~~~~~~~~~ 
\eta_c \partial_{xx} w + \eta_a ( \partial_{yy} + \partial_{zz}) w + f_z.
\end{align}

In these equations, the boundary conditions introduced by the sphere are
replaced by a point force at the origin before shifting to Fourier space.
Because the problem is linear, we assume a solution of the form
%
\begin{align}
	\label{eq:velocity_greens} \vec{v} &= \frac{\overline{
\overline{\mathcal{G}}} \cdot \vec{F} } {8 \pi} = \frac{1}{8 \pi} \left\{
\begin{array}{c}	 \mathcal{G}_{1j}F_j 	\\	\mathcal{G}_{2j}F_j \\
\mathcal{G}_{3j}F_j	 \end{array} \right\}, \\
	\label{eq:pressure_greens} \overline{p} &= \frac{\vec{\mathcal{P}}
\cdot \vec{F}}{8 \pi} = \frac{1} {8 \pi} \mathcal{P}_j F_j,
\end{align}
%
where $\mathcal{G}_{ij}$ is the Green's function for the velocity (also called
the Oseen tensor) and $\mathcal{P}_j$ is the Green's function for the pressure.
Solving for the Green's function in Fourier space renders
\textcolor{black}{(see Appendix \ref{sec:green_ap} for more details)} 
%
\begin{eqnarray}
	\label{eq:P_j} \frac{\widehat{\mathcal{P}}_{j}}{\sqrt{8/\pi}} &=& i k_j
\left[ \frac{1 - \delta_{1j}}{k^2} + \frac{\frac{1 - \delta_{1j}}{k^2}(\alpha_1
k_1^4 - \eta_a k^4) + \eta_a k^2 + (\eta_b - \eta_a) k_1^2 + (\eta_c - \eta_b)
k_1^2 \delta_{1j}}{\alpha_1 k_1^2 (k_2^2 + k_3^2) + \eta_b k^4 + (\eta_c -
\eta_b) k_1^2 k^2} \right], \nonumber
\end{eqnarray}
and
%
\begin{eqnarray}
	\label{eq:G_1j} \frac{\widehat{\mathcal{G}}_{1j}}{\sqrt{8/\pi}} &=&
\frac{ \delta_{1j}k^2 - k_1 k_j }{ \alpha_1 k_1^2 (k_2^2 + k_3^2) + \eta_b k^4
+ (\eta_c - \eta_b) k_1^2 k^2 }, \\
	\label{eq:G_2j} \frac{ \widehat{\mathcal{G}}_{2j} }{\sqrt{8/\pi}} &=&
\frac{ \delta_{2j} }{ (\eta_c - \eta_a) k_1^2 + \eta_a k^2} - k_2 k_j \frac{ (1
- \delta_{1j}) \frac{ \alpha_1 k_1^2 + (\eta_b - \eta_a) k^2 } { (\eta_c -
  \eta_a) k_1^2 + \eta_a k^2 } + 1 }{ \alpha_1 k_1^2 (k_2^2 + k_3^2) + \eta_b
k^4 + (\eta_c - \eta_b) k_1^2 k^2 }, \\
	\label{eq:G_3j} \frac{\widehat{\mathcal{G}}_{3j}}{\sqrt{8/\pi}} &=&
\frac{ \delta_{3j} }{ (\eta_c - \eta_a) k_1^2 + \eta_a k^2 } - k_3 k_j \frac{
(1 - \delta_{1j}) \frac{ \alpha_1 k_1^2 + (\eta_b - \eta_a) k^2 } { (\eta_c -
\eta_a) k_1^2 + \eta_a k^2 } + 1 }{ \alpha_1 k_1^2 (k_2^2 + k_3^2) + \eta_b k^4
+ (\eta_c - \eta_b) k_1^2 k^2 },
\end{eqnarray}
%
In this formulation, 
$\vec k = (k_1,k_2, k_3) = (k_x,k_y,k_z)$ is the wavenumber vector in the 
Fourier domain, $\delta_{ij}$ is the Dirac delta function and $i$ the imaginary 
unit.  In the isotropic limit case where $\eta_a = \eta_b = \eta_c = \eta$ and 
$\alpha_1 = 0$, we recover the Green's functions for isotropic fluids as 
expected.

\section{Response Function of a spherical particle in a Nematic Fluid}
\label{sec:Response_Leslie_Ericksen} 
%
In the low-Reynolds-number limit, the velocity of a particle moving in an
orthotropic fluid and the drag force exerted on it are related as
%
\begin{equation}
	\vec{F} = \overline{\overline{\zeta}} \cdot \vec{v}(\vec{x} = \vec{0})
= \overline{\overline{\zeta}} \cdot \vec v_0,
\label{eq:respof}
\end{equation}
%
where $\overline{\overline{\zeta}}$ is a tensorial response function,
also known as hydrodynamic resistance \citep{Microhydrodynamics}. 
The aim of this section is to obtain analytical expressions for the different
elements of $\overline{\overline{\zeta}}$. 
Given the singularity at the origin, the response function cannot be directly
calculated from equation \eqref{eq:velocity_greens} but it can be obtained by
regularizing the solution and assuming that the force applied by the sphere is
not a point force but a compact force distribution $\vec{F}(\vec{x})$
\citep{Levine.Lubensky.2001, Liron.Barta.1992}.  The velocity at the origin can
be obtained from the Fourier transform of the Green's function as
%
\begin{equation}
	\vec v_0 = \frac{1}{(2 \pi)^{3/2}} \iiint \frac{
\widehat{\overline{\overline{\mathcal{G}}}}}{8 \pi} \cdot \vec {\widehat
F}(\vec{k}) d^3 k = \overline{\overline{\gamma}} \cdot \vec F,
\label{eq:invrespof}
\end{equation}
%
where $ \vec{\widehat F}(\vec{k}) = \vec{F} \widehat{\mathcal{F}} (\vec{k})$ is
the Fourier transform of $\vec{F}(\vec{x})$, the function
$\widehat{\mathcal{F}} (\vec{k})$ is a regularization kernel that localized the
drag force in physical and/or Fourier space, and $\overline{\overline{\gamma}}$
is the hydrodynamic mobility. Note that equation \eqref{eq:invrespof} is just 
the inverse Fourier transform of $\vec{\widehat v}(\vec k)$ particularized at 
the origin, and that equations (\ref{eq:respof})-(\ref{eq:invrespof}) imply 
that $\overline{\overline{\zeta}} = {\overline{\overline{\gamma}}^{\,}}^{-1}$. 

\cite{Levine.Lubensky.2001} propose a volume localization approach to simplify
the calculation of $\overline{\overline{\gamma}}$ in the isotropic case by
considering that the particle radius limits the spectrum of allowed
fluctuations in the flow field. Hence, they only consider wavenumbers smaller
than $k_{max} = \pi/2 a$.  This approximation is written as
%
\begin{equation*}
	\widehat{\mathcal{F}}(k) = H \left( \frac{\pi}{2 a} - k \right),
\end{equation*}
%
where $H$ is the Heaviside function.  It leads to the following expression for
the inverse response function tensor
%
\begin{equation}
	\gamma_{ij} = \frac{1}{4 \sqrt{2 \pi} a} \int \limits_{\theta=0}^
{\theta=\pi} \sin{\theta} \left( \int \limits_{\varphi=0}^{\varphi=2 \pi}
\frac{k^2 \widehat{\mathcal{G}}_{ij}}{8 \pi} d \varphi \right) d \theta,
\label{eq:gamma_integrals}
\end{equation}
%
where $\theta$ and $\varphi$ are the inclination and azimuth angles in 
spherical coordinates, respectively. Note that $\widehat{\mathcal{G}}_{ij}$ is
inversely proportional to $k^2$, so that the whole expression is independent of
$k$.
While it holds a clear physical meaning, the sharp Fourier cut-off of the
Heaviside regularization function leads to a non-localized force distribution
in physical space and a velocity field with Gibbs oscillations in the radial
direction.  For this reason, the Gaussian regularization function
%
\begin{equation}
	\widehat{\mathcal{F}}(k) = e^{-a^2 k^2 / \pi}
\label{eq:gaussreg}
\end{equation}
%
is preferred in this study.  
This choice yields the same equation for $\gamma_{ij}$ and has the advantage of
providing a localized Gaussian force distribution and a smooth velocity field
in physical space (see \S \ref{sec:velos}). \textcolor{black}{This approach is
analogous to the regularized Stokeslet method \citep{Beale.Lai.2001,
Cortez.2001}.}

Owing to the symmetry of the Green's function \eqref{eq:G_1j}-\eqref{eq:G_3j}, 
we obtain that $\gamma_{ij} = 0$ for $i \neq j$, and both 
$\overline{\overline{\gamma}}$ and $\overline{\overline{\zeta}}$ are diagonal 
tensors. In particular,
%
\begin{equation}
	\vec{F} = \left[ \begin{array}{rcl} \zeta_{||} & 0 & 0 \\ 0 &
\zeta_\perp & 0 \\	
 0 & 0 & \zeta_\perp \end{array} \right] \cdot \vec v_0,
\label{eq:forceeq}
\end{equation}
%
where $\zeta_{||}$ is the response function in the direction parallel to the
nematic and $\zeta_{\perp}$ is the response function in all transverse
directions.  
These principal values can be determined independently by measuring the drag
force of a particle in the directions parallel and perpendicular to $\vec n$. 
For any other direction of motion, the drag force acting on a particle is not
parallel to $\vec v_0$.
In general, if $\vec v_0$ forms an angle $\beta_v$ with $\vec n$, the angle
between the drag force vector and the nematic director is $\beta_F =
\arctan[\tan(\beta_v) \zeta_{\perp} / \zeta_{||}]$ (see figure
\ref{fig:Sphere_Orthotropic}).

Armed with some tenacity, one can find an explicit analytical solution to the
integrals in equation \eqref{eq:gamma_integrals}, leading to the following
expressions for the diagonal components of the response function,
%
\begin{align}
	\zeta_{||} &= \frac 1 {\gamma_{11}}  = 
\frac{8 \pi a \eta_b B(\vec \eta)}{ 
{D_{+}(\vec \eta) \frac{ \arctan[C_{+}(\vec \eta)]}{C_{+}(\vec \eta) }
- D_{-}(\vec \eta) \frac{\arctan[C_{-}(\vec \eta)]}{C_{-}(\vec \eta)}} },
	\label{eq:response_function_11_analytic}
\\
	\zeta_{\perp} &= \frac 1 {\gamma_{22}} = \frac 1 {\gamma_{33}} = 
\frac{ 8 \pi a \eta_a \left( \frac{\eta_b}{\alpha_1} \right)^2 
C_{-}(\vec \eta)^8 \frac{E_{+}(\vec \eta)}{E_{-}(\vec \eta)}}
{\frac{\arctan \left(\sqrt{\eta_c /\eta_a-1}\right)}{\sqrt{\eta_c/\eta_a-1}} + 
\frac{1}{B(\vec \eta)} \frac{\eta_a}{\eta_b} \left[
\frac{\arctan[C_{-}(\vec\eta)]}{C_{-}(\vec\eta)} -
\frac{\arctan[C_{+}(\vec\eta)]}{C_{+}(\vec\eta)} \right]},
	\label{eq:response_function_22_analytic} 
\end{align}
%
where $B(\vec \eta)$, $C_{\pm}(\vec \eta)$, $D_{\pm}(\vec \eta)$ and
$E_{\pm}(\vec \eta)$ are non-dimensional functions of the viscosity vector
$\vec \eta = (\eta_a,\, \eta_b, \, \eta_c, \, \alpha_1)$.  These functions are
given in Appendix \ref{sec:functions_ap}.
Equations \eqref{eq:response_function_11_analytic} and
\eqref{eq:response_function_22_analytic} have a weak singularity at $\alpha_1
=0$ that can be removed by taking the limit $\alpha_1 \rightarrow 0$, leading
to
\begin{align} 
	\zeta_{||,\alpha_1=0} &= \frac{4 \pi a (\eta_c - \eta_b)}
{\frac{\eta_c}{\eta_b} \frac{ \arctan \left(\sqrt{\eta_c/\eta_b-1}\right)}
{\sqrt{\eta_c/ \eta_b-1}} - 1 },
\label{eq:response_function_11_alpha_1_0}
\\
	\zeta_{\perp,\alpha_1=0} &= \frac{8 \pi a (\eta_c-\eta_b)}
{ 1 - \frac{ \arctan \left(\sqrt{\eta_c/\eta_b-1} \right)}
{\sqrt{\eta_c/\eta_b-1}} + \frac{\eta_c - \eta_b}{\eta_a}
\frac{\arctan \left(\sqrt{\eta_c/\eta_a-1} \right)}{\sqrt{\eta_c/\eta_a-1}}}.
\label{eq:response_function_22_alpha_1_0} 
%
\end{align}

The isotropy point is of particular interest because diluted networks should
not be too far from it. Taylor expanding equations
\eqref{eq:response_function_11_analytic} and
\eqref{eq:response_function_22_analytic} around the point $\alpha_1 = 0$,
$\eta_a = \eta_b = \eta_c = \eta$, and keeping only the first order term we
obtain
%
\begin{align}
	\frac{\zeta_\parallel}{6 \pi a \eta} &\approx 1 + \frac{4}{35}
\frac{\alpha_1}{\eta} + \frac{4}{5} \left( \frac{\eta_b}{\eta} - 1 \right) +
\frac{1}{5} \left( \frac{\eta_c}{\eta} - 1 \right), 
\label{eq:tayl_isopar}
\\
	\frac{\zeta_\perp}{6 \pi a \eta} &\approx 1 + \frac{3}{70}
\frac{\alpha_1}{\eta} + \frac{1}{2} \left( \frac{\eta_a}{\eta} - 1 \right) +
\frac{1}{10} \left( \frac{\eta_b}{\eta} - 1 \right) + \frac{2}{5} \left(
\frac{\eta_c}{\eta} - 1 \right). 
\label{eq:tayl_isoper}
\end{align}
%
This Taylor expansion is consistent with previous results by
\cite{pokrovskii.tskhai.1986}, who studied the motion of a particle in a weakly
anisotropic fluid with $\eta_b = \eta_c$.

\begin{figure}
\hspace{7.3cm}
\includegraphics*[width=0.47\columnwidth,keepaspectratio]{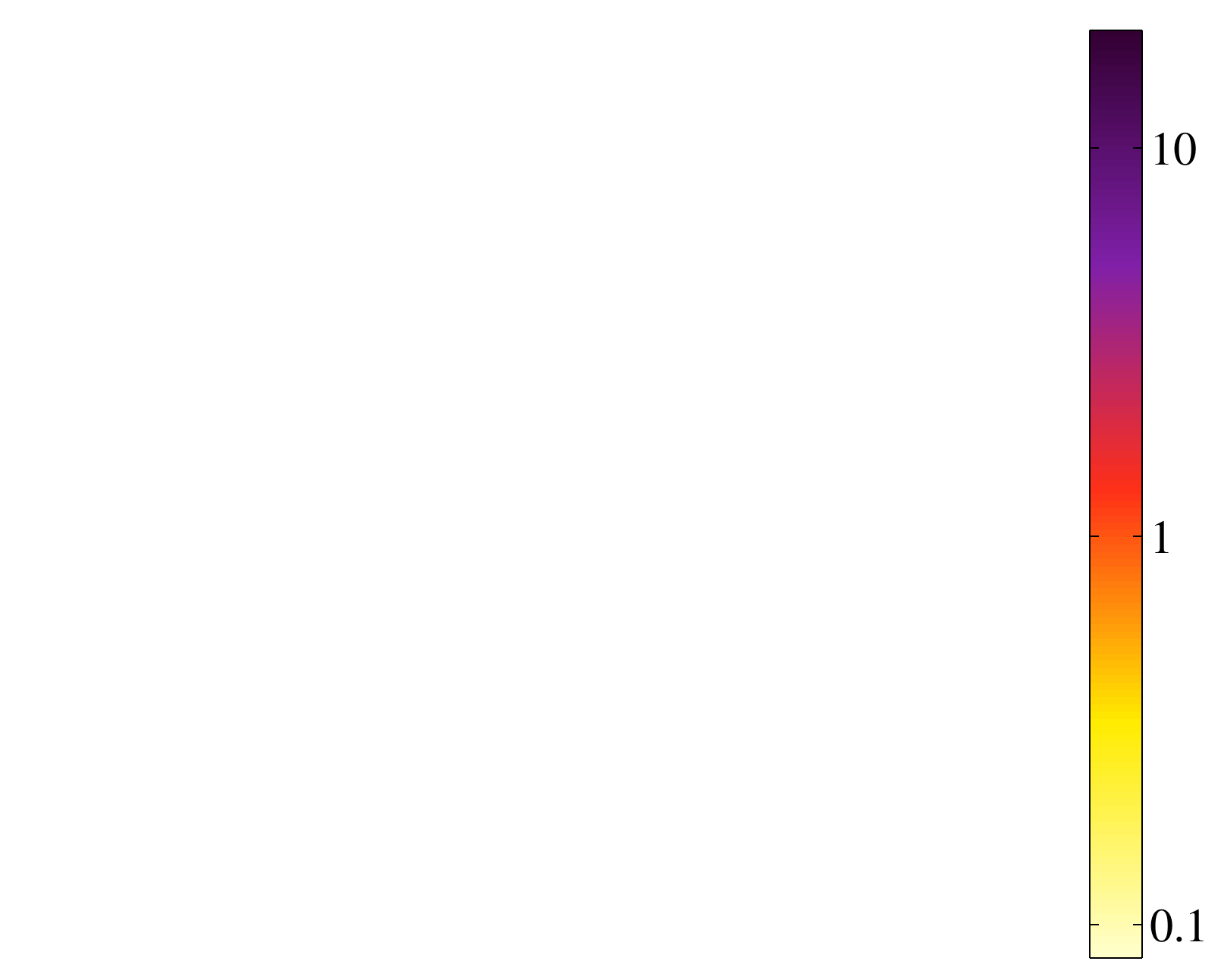}
\\[-6.05cm]
\includegraphics*[width=0.47\columnwidth,keepaspectratio]{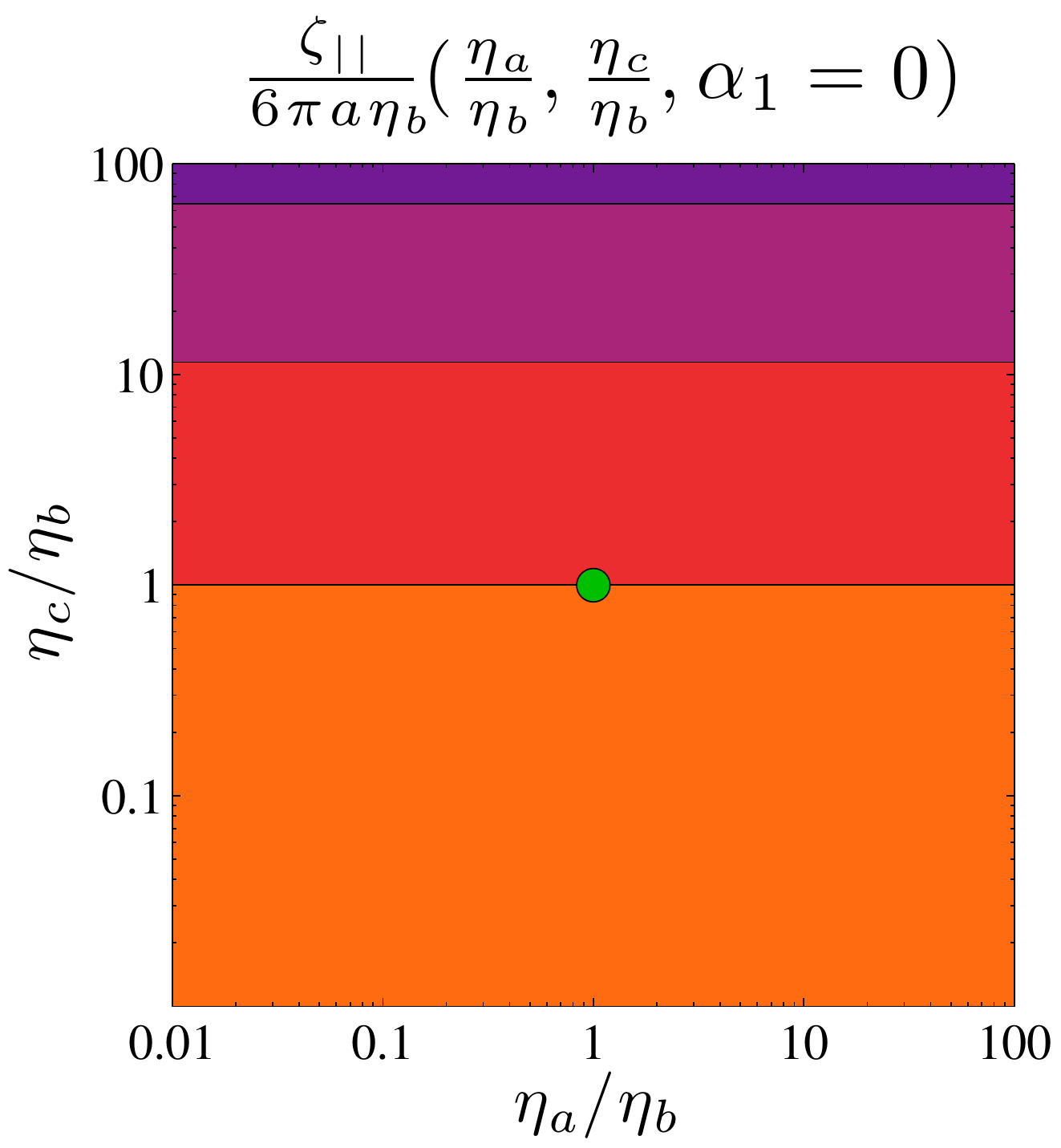}
\mylab{-0.43\textwidth}{0.47\textwidth}{(\aaa)}%
\includegraphics*[width=0.47\columnwidth,keepaspectratio]{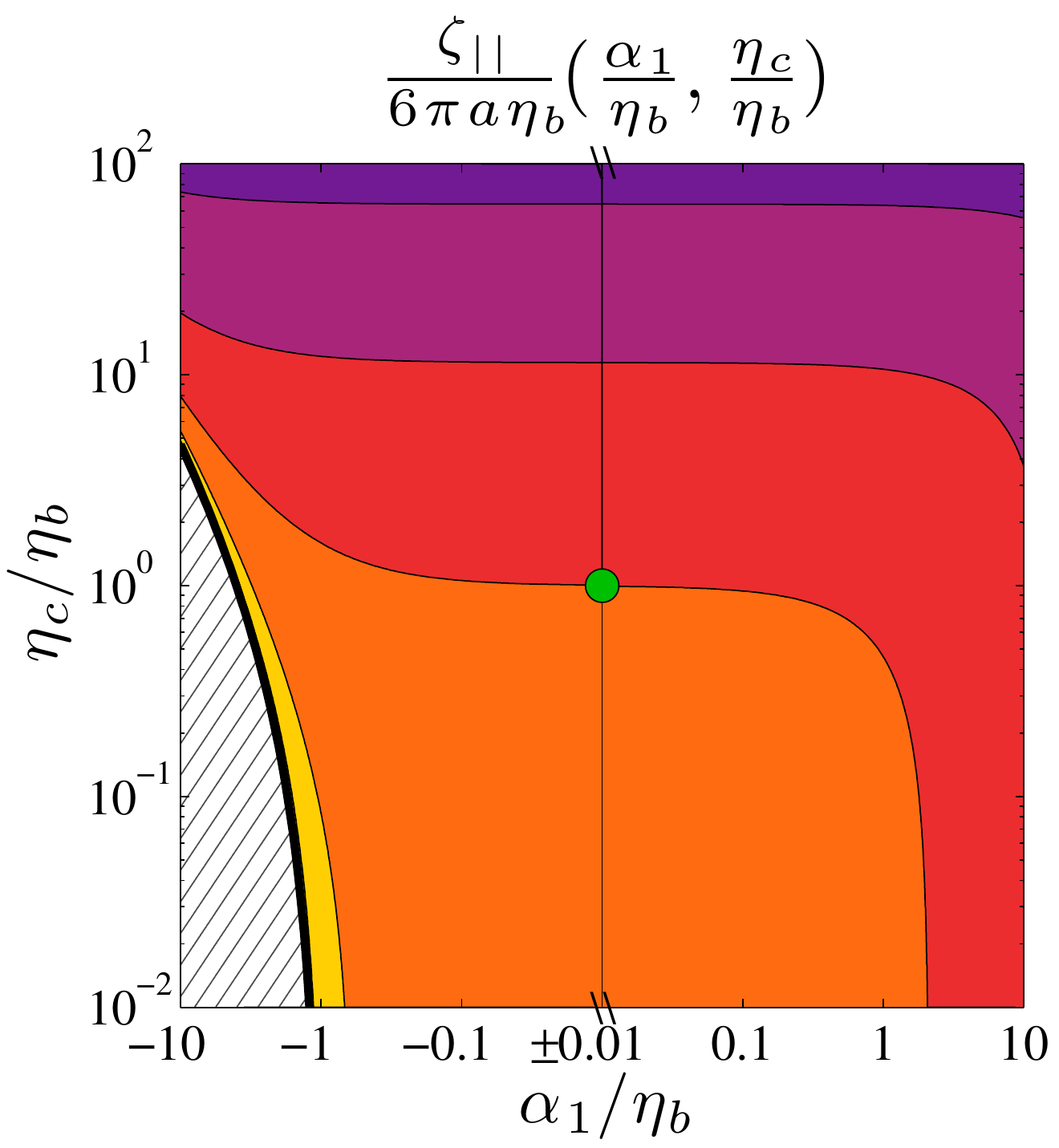}%
\mylab{-0.43\textwidth}{0.47\textwidth}{(\bbb)}%
	\caption{
Contour maps of the response function in the nematic direction, 
$\zeta_{||}$, obtained from eq. \eqref{eq:response_function_11_analytic}: 
(\aaa), $\zeta_{||}(\eta_a/\eta_b,\eta_c/\eta_b)$ for $\alpha_1=0$, showing
that this response function is independent of $\eta_a$; 
(\bbb), $\zeta_{||}(\alpha_1/\eta_b,\eta_c/\eta_b)$.  Note that the $x$-axis is
broken between $\alpha_1 = -0.01 \eta_b$ and $\alpha_1 = 0.01 \eta_b$.
The thick black curve indicates the locus where $\zeta_{||}=0$. The hatched
region to the left of this curve yields unphysical, complex-valued results for
the response function.
In both panels, the response functions have been normalized with $6 \pi a
\eta_b$.  The green circle represents the isotropic case with $\eta_a = \eta_b
= \eta_c$ and $\alpha_1 =0$.
} \label{fig:contour_zeta_11}
%
\end{figure}
%
\begin{figure}
\vspace{1.0cm}
\hspace{7.3cm}
\includegraphics*[width=0.47\columnwidth,keepaspectratio]{Figure_2c}
\\[-6.05cm]
\includegraphics*[width=0.47\columnwidth,keepaspectratio]{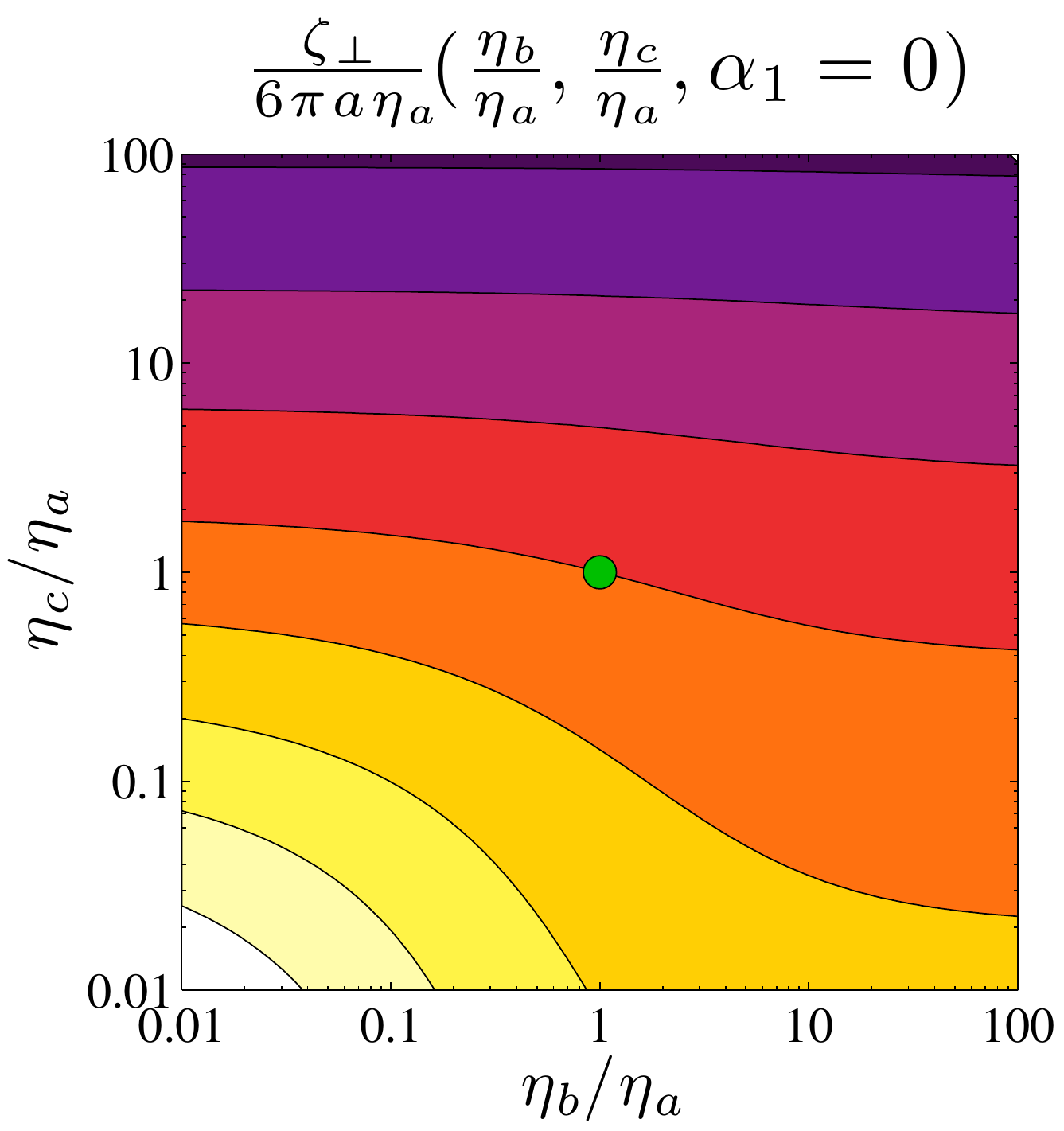}
\mylab{-0.43\textwidth}{0.47\textwidth}{(\aaa)}%
\includegraphics*[width=0.47\columnwidth,keepaspectratio]{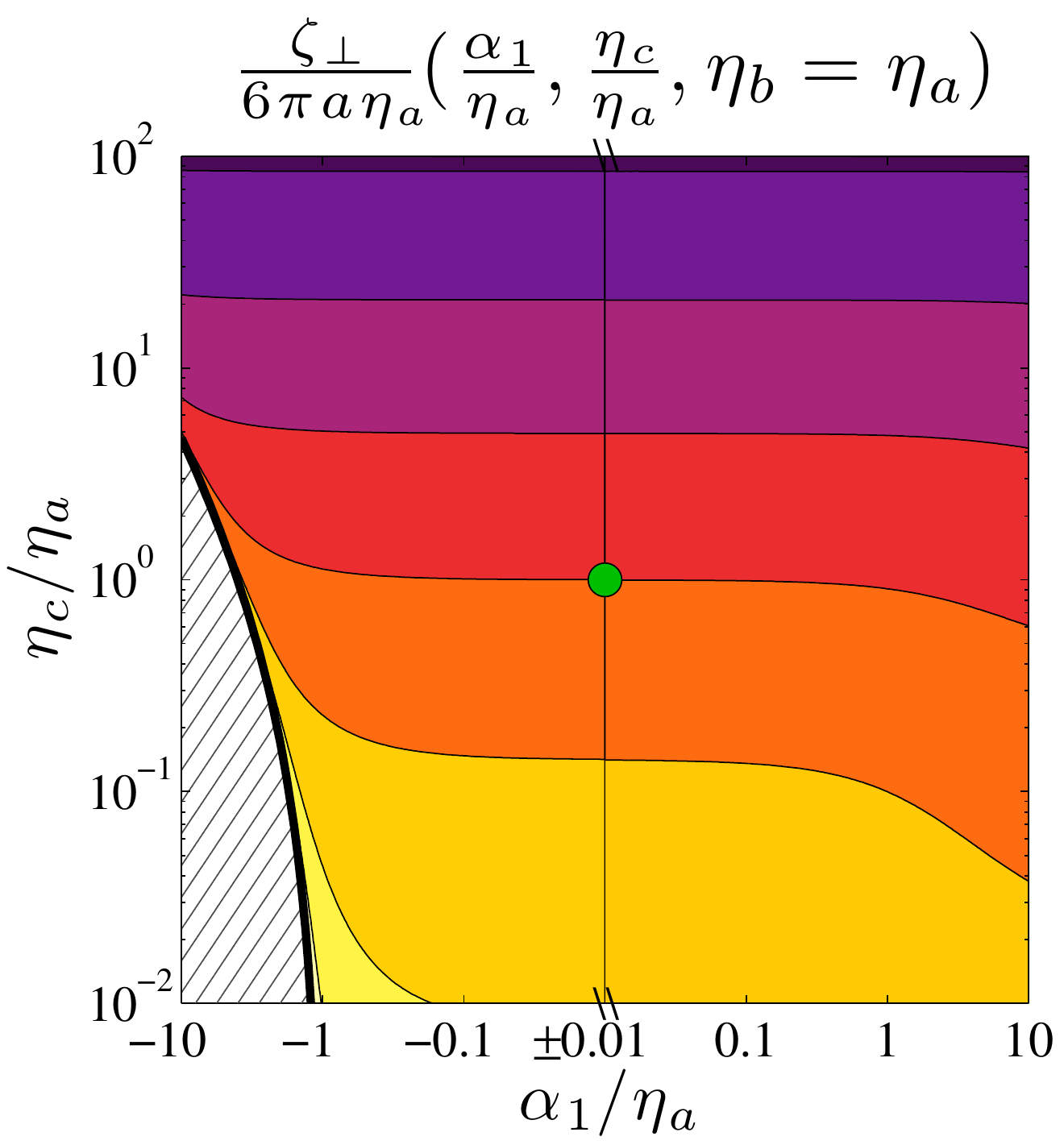}%
\mylab{-0.43\textwidth}{0.47\textwidth}{(\bbb)}%
	\caption{
Contour maps of the response function perpendicular to the 
nematic, $\zeta_{\perp}$, obtained from equation 
\eqref{eq:response_function_22_analytic}: 
(\aaa), $\zeta_{\perp}(\eta_b/\eta_a,\eta_c/\eta_a)$ for $\alpha_1=0$; 
(\bbb), $\zeta_{\perp}(\eta_b/\eta_a,\alpha_1/\eta_a)$ for $\eta_b = \eta_a$.
Note that the $x$-axis is broken between $\alpha_1 = -0.01 \eta_a$ and
$\alpha_1 = 0.01 \eta_a$.
The thick black curve indicates the locus where $\zeta_{\perp}=0$. The region
to the left of this curve yields unphysical, complex-valued results for the
response function.
In both panels, the response functions have been normalized with $6 \pi a
\eta_a$.  The green circle represents the isotropy point, $\eta_a = \eta_b =
\eta_c$ and $\alpha_1 =0$.
				 } \label{fig:contour_zeta_22}
%
\end{figure}

In general, both components of the response function increase with increasing
the viscosity coefficients but there are several aspects of this dependence
that deserve especial attention.
Figure \ref{fig:contour_zeta_11} displays the response function in the nematic
direction, $\zeta_{||}=\zeta_{11}$.  Interestingly, this component is
independent of the first Miesowicz viscosity $\eta_a$, as shown in figure
\ref{fig:contour_zeta_11}(\aaa) for $\alpha_1=0$, and in agreement with
\cite{Kneppe.Schneider.Schwesinger.1991}. This degeneracy can be explained by
realizing that the flow elicited by a sphere that translates parallel to the
nematic director is axially symmetric.  Working in polar cylindrical
coordinates $(r,\theta,x)$ and taking into account this symmetry, the momentum
balance equations \eqref{eq:fluid_equations_x}-\eqref{eq:fluid_equations_z} can
be rewritten as
%
\begin{align}
\partial_x p' &= \eta_b \nabla^2 u_x + \alpha_1 \partial_{xx}u_x, 
\label{eq:fluid_equations_xcyl} 
\\
	\partial_r p' &= \eta_c \left( \nabla^2 u_r - {u_r}/{r^2} \right) -
(\eta_c - \eta_a) \partial_r ( \nabla \cdot \vec v),
\label{eq:fluid_equations_rcyl}
\end{align}
%
where $\eta_c - \eta_a$ acts as a dilatational viscosity in the radial
direction that disappears when the velocity field is divergence-free.  
It is recognized that the resulting divergence-free, axially symmetric velocity
field has a stream function but this function is not pursued here for the sake
of consistency with the perpendicular problem, which is not axially symmetric.
Note also that the Green's function derived in \S
\ref{sec:Greens_Leslie_Ericksen} can be easily extended to networks with
significant volume fractions, where network compressibility leads to $\nabla
\cdot \vec v \ne 0$, and there is no stream function.  

Figure \ref{fig:contour_zeta_11}(\bbb) depicts the dependence of the parallel
response function on the remaining three viscosity coefficients by displaying
$\zeta_{||}/ (a \eta_b)$ as a function of $\alpha_1/\eta_b$ and $\eta_c/
\eta_b$.  This non-dimensionalization is chosen because the viscous shear
stresses in the $||$ direction are proportional to $\eta_b$ (see equation
\ref{eq:fluid_equations_x}).
In this scaling, the parallel response function shows a moderate dependence on
the ratio $\eta_c / \eta_b$.  In fact, the limiting values of
$\zeta_{||}(\alpha_1=0)$ for $\eta_c / \eta_b \rightarrow 0, \infty$ are 
$4\pi a \eta_b$ and $8 a \sqrt{\eta_b \eta_c}$, respectively, only differing by
a factor of order $(\eta_c / \eta_b)^{1/2}$.

The dependence of the parallel response function on $\alpha_1$ is even weaker,
especially in the case $\eta_c \gg \eta_b$, which is dominated by the viscous
stresses in the $\perp$ directions.  Near the isotropy point, the parallel
response function is relatively insensitive to $\alpha_1$, consistent with the
small factor multiplying this coefficient in the Taylor expansion of equation
\eqref{eq:tayl_isopar}.
Figure \ref{fig:contour_zeta_11}(\bbb) also considers $\alpha_1 <0$ as this
viscosity coefficient has been reported to be negative in some nematic liquid
crystals  \citep{Kneppe.Schneider.1981}.
Our analysis indicates that lowering $\alpha_1$ below zero causes the response
function to decrease.  This decrease is moderate everywhere but near the curve
%
\begin{equation}
\frac{\eta_c} {\eta_b} = 1 - \frac {\alpha_1} {\eta_b} - 2
\sqrt{\frac{-\alpha_1}{\eta_b}},\;\;\; \alpha_1 < -\eta_b,
\label{eq:singualp}
\end{equation}
%
where the function $B(\vec \eta)$ that multiplies the right-hand side of
equation \eqref{eq:response_function_11_analytic} becomes zero and so does the
parallel response function (see Appendix \ref{sec:functions_ap}).  The region
to the left of this curve in the $(\alpha_1/\eta_b,\eta_c/\eta_b)$ plane
produces complex values of $\zeta_{||}$ and, thus, is physically impossible.

The transverse response function, $\zeta_{\perp} = \zeta_{22}$ (equation
\ref{eq:response_function_22_analytic}), has a more complicated dependence on
the viscosity coefficients than its parallel counterpart because the flow
elicited by $\vec v_0$ is not axially symmetric.
In order to analyze this dependence, it is convenient to plot $\zeta_{\perp}/
(a \eta_a)$ as a function of $\eta_b/\eta_a$, $\eta_c/ \eta_a$ and $\alpha_1 /
\eta_a$.  
This representation allows us to evaluate the relative importance of the
different sources of anisotropy in the Leslie-Ericksen constitutive equation
\eqref{eq:Leslie_Eriksen}, namely bending of the fluid with respect to the
nematic and anisotropy in the stress-strain relationship.

Figure \ref{fig:contour_zeta_22}(\aaa) displays the transverse response
function in the case $\alpha_1=0$.  The plot reveals that $\zeta_{\perp}/ (a
\eta_a)$ is more sensitive to $\eta_c / \eta_a$ than to $\eta_b / \eta_a$.
This is especially true in the limit $\eta_c \gg \eta_a$, in which the bending
stresses become dominant in the equations of motion, and the response function
has a concise limit that is independent of the second Miesowicz viscosity,
$$\zeta_\perp \rightarrow 16 a \eta_a \sqrt{\eta_c / \eta_a}.$$
The response function also becomes independent of $\eta_b$ in the limit $\eta_b
\gg \eta_a$, in which the anisotropic momentum diffusivity dominates the flow
dynamics, and 
$$ \left.{\zeta_{\perp} \rightarrow 8\pi a \eta_a  \sqrt{\eta_c / \eta_a -1 }}
\middle/ \right.{\arctan\left(\sqrt{\eta_c / \eta_a-1}\right)}.$$
The $\alpha_1$-dependence of $\zeta_\perp$ is similar to that of $\zeta_{||}$,
as shown in figure \ref{fig:contour_zeta_22}(\bbb) for the case $\eta_a =
\eta_b$.
The transverse response function is fairly insensitive to $\alpha_1$ for the
most part of the parameter space, particularly near the isotropy point and for
large $\eta_c / \eta_a$.  Much like the $||$ case, the effect of $\alpha_1$
only becomes strong near the curve defined by equation \eqref{eq:singualp},
where the transverse response function is zero.  In the scaling used to plot
figure \ref{fig:contour_zeta_22}(\bbb), this curve has an asymptote at
$(\alpha_1 / \eta_a,\, \eta_c/\eta_a) = (-\eta_b / \eta_a, 0)$.  Thus, as
$\eta_b$ increases, the asymptote moves towards larger negative values of
$\alpha_1/\eta_a$ and the influence of $\alpha_1$ becomes less important.

\subsection{Pseudo-isotropic conditions } \label{sec:Two_viscosities}
%
The analysis of the equations of motion
(\ref{eq:fluid_equations_x})-(\ref{eq:vorti}) suggests that the anisotropy in
the Leslie-Ericksen constitutive relation (\ref{eq:Leslie_Eriksen}) may
influence the response function by means of two separate mechanisms.
One is the anisotropy of the stress-strain relationship, which is proportional
to the viscosity difference $\eta_b - \eta_a$, while the other is the
resistance to the bending of the fluid, which is proportional to the viscosity
difference $\eta_c - \eta_a$. 
The aim of this section is to dissect the effects of the two anisotropy
mechanisms.
We first study the response function in a rotationally pseudo-isotropic fluid 
where anisotropy comes exclusively from the anisotropic momentum diffusivity, 
and which corresponds to the horizontal line $\eta_c = \eta_a$ in figures 
\ref{fig:contour_zeta_11}(\aaa) and \ref{fig:contour_zeta_22}(\aaa).  
We then consider a strain pseudo-isotropic fluid, where the anisotropy comes 
solely from the resistance to bending of the fluid with respect to the nematic, 
and which corresponds to the vertical line $\eta_b = \eta_a$ in figures 
\ref{fig:contour_zeta_11}(\aaa) and \ref{fig:contour_zeta_22}(\aaa). 

The response functions derived in this section also provide with distinct
two-viscosity benchmarks to test the accuracy of existing PTM methods when
applied to anisotropic fluids (see \S \ref{sec:DPTM}).
This assessment needs to be performed on two-viscosity fluids as only two of
the principal components of the response functions are independent in a nematic
fluid (see equation \ref{eq:forceeq}).
This limitation can be potentially circumvented by considering two-point
response functions but such effort is beyond the scope of this paper.

An aspect of particular interest that is evaluated on pseudo-isotropic fluids
is the accuracy of the directional effective viscosity approximation.
Based on the principal components of the response function, one can define two
effective directional viscosities $\eta_{|| , \perp}^{\mbox{\scriptsize eff}} =
\zeta_{|| , \perp} / (6 \pi a)$, similar to \cite{Stark.Ventzki.2001}. This
phenomenological approach has been previously used by experimentalists to
characterise different types of nematic fluids
\citep{Loudet.2004,Hasnain.2006,Rogers.Waigh.Lu.2008,DelAlamo.2008,He.Mak.Liu.Tang.2008}
but it has not been justified yet.

\subsubsection{Rotationally Pseudo-Isotropic Fluid
($\boldsymbol {\eta_a =\eta_c}$)} \label{subsec:etaaeqetac}

\begin{figure}
\vspace{0.5cm}
\includegraphics*[width=0.5\columnwidth,keepaspectratio]{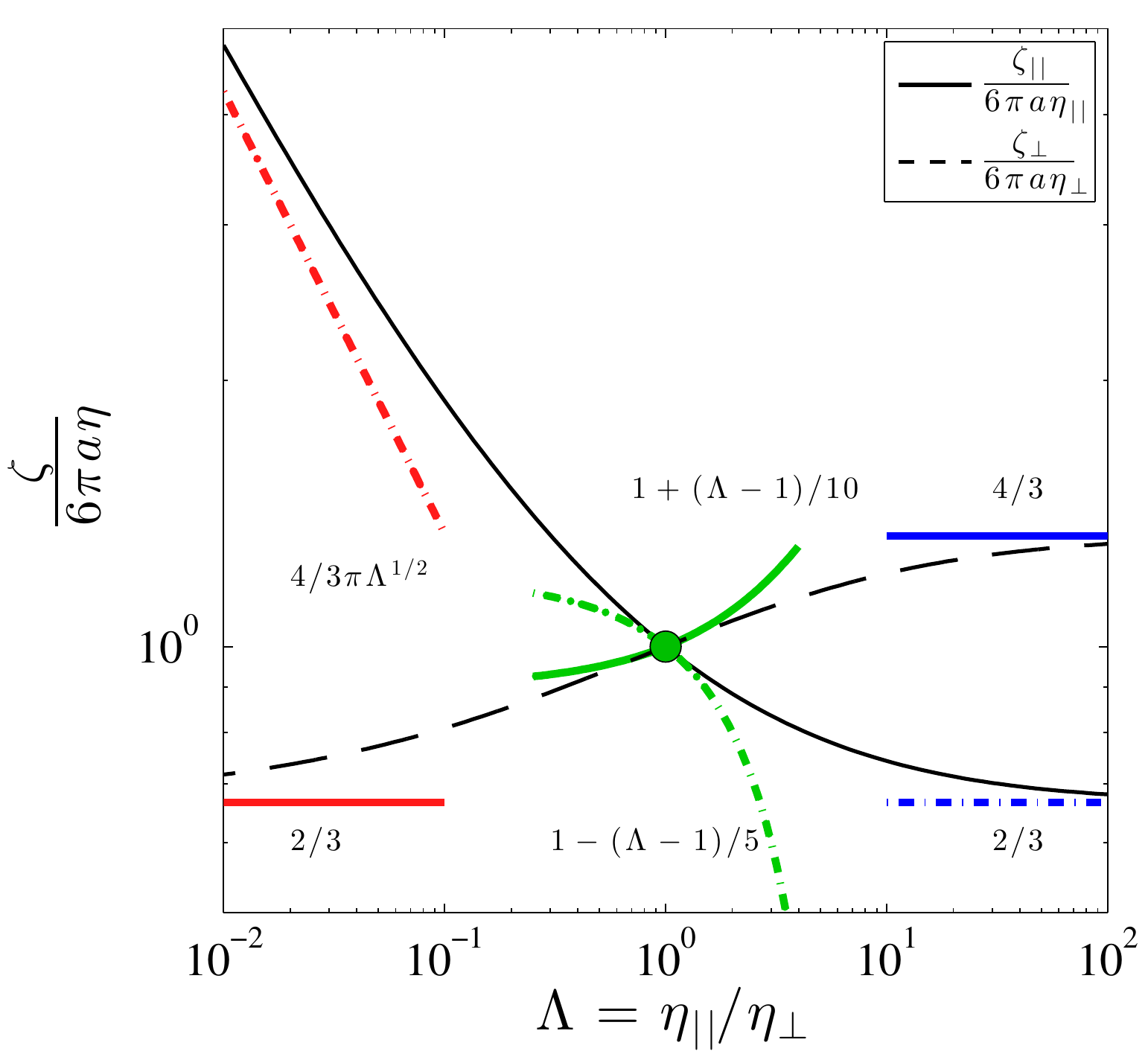}
\mylab{-0.41\textwidth}{0.47\textwidth}{(\aaa)}
\includegraphics*[width=0.5\columnwidth,keepaspectratio]{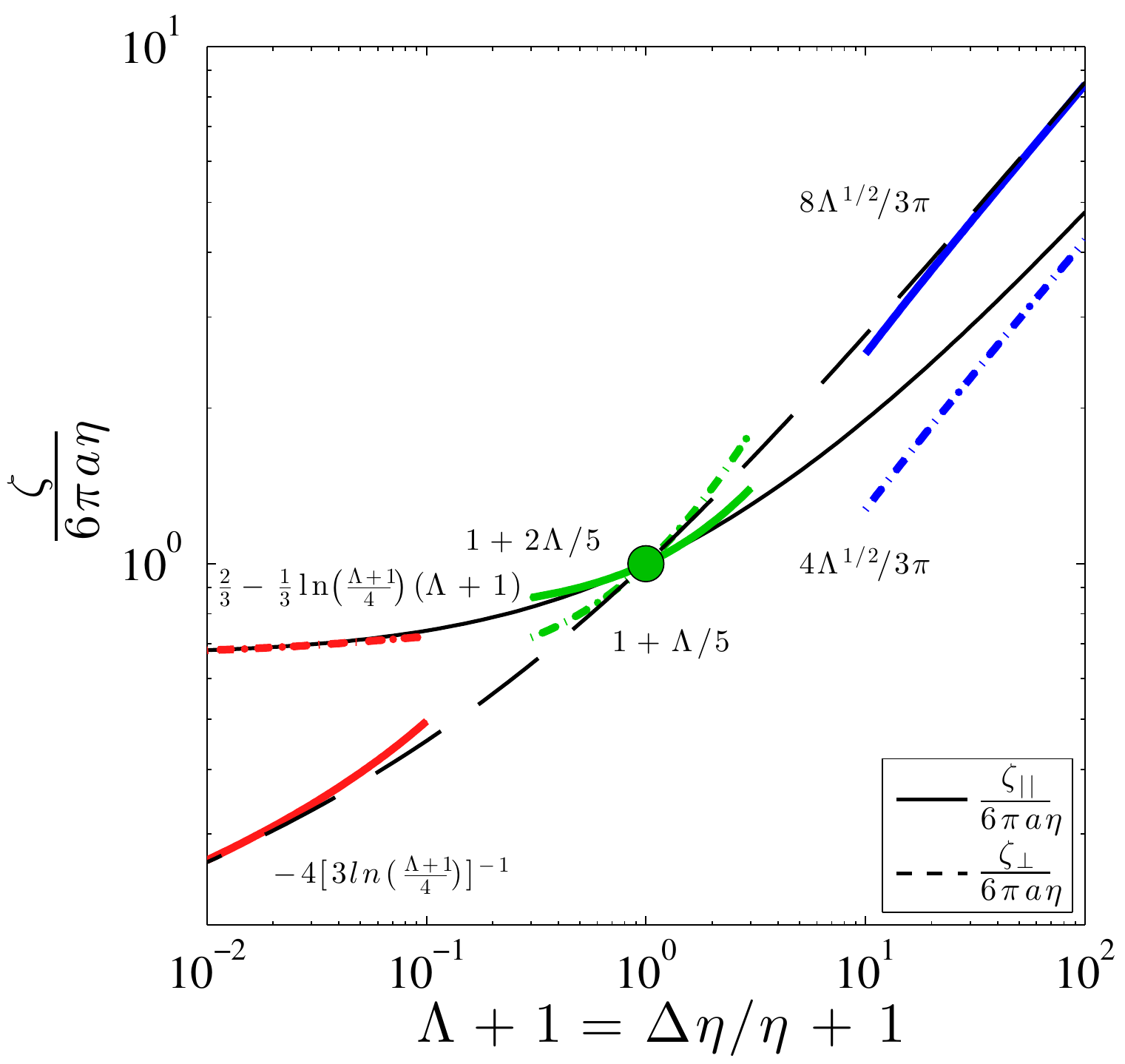}
\mylab{0.6\textwidth}{0.50\textwidth}{(\bbb)}
	\caption{
Principal components of the response function (\aaa) in a 
rotationally pseudo-isotropic fluid, $\alpha_1=0$, $\eta_b=\eta_{||}$ and
$\eta_a = \eta_c = \eta_\perp$, represented as a function of the viscosity
ratio $\Lambda = \eta_{||} / \eta_{\perp}$ and (\bbb) in a strain 
pseudo-isotropic fluid $\alpha_1=0$, $\eta_a=\eta_b=\eta$ and 
$\eta_c = \Delta \eta + \eta$, represented as a function of the viscosity ratio 
$\Lambda = \Delta \eta / \eta$: \solid, $\zeta_{||} / 6 \pi a \eta_{||}$: 
\dashed, $\zeta_\perp / 6 \pi a \eta_{\perp}$. The green circle marks the 
isotropy point: (\aaa) $\Lambda = 1$ and (\bbb) $\Lambda = 0$. The thick color 
segments mark the asymptotic behavior of the response function for (\aaa) 
$\Lambda \ll 1$ (red), $\Lambda \approx 1$ (green) and $\Lambda \gg 1$ (blue), 
and (\bbb) $\Lambda \to -1$ (red), $\Lambda \approx 0$ (green) and 
$\Lambda \gg 1$ (blue).
}
	\label{fig:zeta_2vis}
%
\end{figure}

Consider a fluid where $\alpha_1$ is very small and $\alpha_2 = \alpha_5$.
Under these conditions, the last term in the transverse components of the
momentum balance equation (equations
\ref{eq:fluid_equations_y}-\ref{eq:fluid_equations_z}) disappears, yielding the
rotationally isotropic pseudo-Stokes problem
%
\begin{equation}
\label{eq:rotiso} \nabla p' = \left[ \begin{array}{rcl} \eta_{||} & 0 & 0 \\ 0
& \eta_\perp & 0 \\	
 0 & 0 & \eta_\perp \end{array} \right] \cdot \nabla^2 \vec{v},
\end{equation}
where $\eta_{||} = \eta_b$ denotes the viscosity parallel to the director and
$\eta_\perp = \eta_a$ is the viscosity in the transverse directions. In this
case, the directionality of the problem is solely due to the anisotropic
diffusion of momentum along different directions (i.e. anisotropy in the
stress-strain relationship).

Taking the limit $\eta_a \rightarrow \eta_c$ on equations
\eqref{eq:response_function_11_analytic}-\eqref
{eq:response_function_22_analytic}, and introducing the anisotropy ratio
$\Lambda = \eta_{||} /\eta_\perp$, we find that
%
\begin{align}
	\label{eq:zeta_1} 
\frac{\zeta_{||}}{6 \pi a \eta_{||}} &= 
\frac{\frac{2}{3} (1 - \Lambda)}
{ -\Lambda + \frac{\arctan \left( \sqrt{ {1/\Lambda} -1 } \right)}{ \sqrt{
1/\Lambda -1 } } }, \\
	\label{eq:zeta_2} 
\frac{\zeta_\perp}{6 \pi a \eta_\perp} &= 
\frac{ \frac{4}{3} (1 - \Lambda)}
{ -\Lambda + 2 - \frac{\arctan \left( \sqrt{ 1/\Lambda
- 1 } \right)}{ \sqrt{ 1/\Lambda -1 } } }.
\end{align}
%
These response functions are plotted in figure \ref{fig:zeta_2vis}(\aaa)
together with their tangent lines at the isotropy point $\Lambda = 1$ and their
asymptotes for highly anisotropic conditions ($\Lambda \gg 1$ and $\Lambda \ll
1$).  This figure shows 1D cuts of figures \ref{fig:contour_zeta_11}(\aaa) and
\ref{fig:contour_zeta_22}(\aaa) along the horizontal line $\eta_c = \eta_a$.

The behavior of the $||$ and $\perp$ response functions near the isotropy point
is given by their Taylor expansion
%
\begin{align*}
	\frac{\zeta_{||}}{6 \pi a \eta_{||}} &= 1 - \frac{1}{5} (\Lambda - 1) +
\frac{27}{175}(\Lambda - 1)^2 + O[(\Lambda - 1)^3], \\
	\frac{\zeta_\perp}{6 \pi a \eta_\perp} &= 1 + \frac{1}{10} (\Lambda -
1) - \frac{33}{700} (\Lambda - 1)^2 + O[(\Lambda - 1)^3],
\end{align*}
%
where the leading-order terms correspond to the expected isotropic result
$\zeta = 6\pi a \eta$, and the higher-order terms are the corrections
introduced by the anisotropy of the fluid.  The numerical values of the
coefficients in these expansions indicate that the anisotropic corrections are
weaker in the transverse response function than in the parallel one, consistent
with the weaker variation of $\zeta_\perp$ observed in figure
\ref{fig:zeta_2vis}(\aaa).

In order to determine whether the effects of anisotropy are bounded in a
rotationally pseudo-isotropic fluid, we consider the asymptotic behavior of the
response function for $\Lambda \gg 1$, where
%
\begin{align*}
	\frac{\zeta_{||}}{6 \pi a \eta_{||}} \simeq \frac{2}{3},\;\;\;\; 
	\frac{\zeta_\perp}{6 \pi a \eta_\perp} \simeq \frac{4}{3}.
\label{eq:rotlim1}
\end{align*}
%
Interestingly, these expressions are equivalent to those of a prolate ellipsoid 
moving along its principal axes in an isotropic liquid 
\citep{Microhydrodynamics}.  
It is important to note that the two components of the response function are 
decoupled from each other, so that increasing $\eta_{||}$ does not affect 
$\zeta_{\perp}$ and vice versa.  
Thus, the ratio of effective diffusivities becomes proportional to the ratio of
actual viscosities of the fluid for $\Lambda \gg 1$, 
%
\begin{align*}
\frac{\eta^{\mbox{\scriptsize eff}}_{||}} { \eta^{\mbox{\scriptsize eff}}_\perp} 
=
	\frac{\zeta_{||}}{\zeta_\perp} \simeq \frac 1 2
\frac{\eta_{||}}{\eta_\perp},
\end{align*}
%
although these ratios differ by a factor of two.

The second asymptotic limit is $\Lambda \ll 1$, which results in
%
\begin{align*}
	\frac{\zeta_{||}}{6 \pi a \eta_{||}} \simeq \frac{4}{3 \pi
\sqrt{\Lambda}}, \;\;\;\;\;
	\frac{\zeta_\perp}{6 \pi a \eta_\perp} \simeq \frac{2}{3},
%
%
\end{align*}
%
It can be seen that, in this limit, the $||$ component of the response function
not only depends on $\eta_{||}$ but also on $\eta_\perp$, so that the ratio of
effective diffusivities,
%
\begin{align*}
\frac{\eta^{\mbox{\scriptsize eff}}_{||}} { \eta^{\mbox{\scriptsize
eff}}_\perp} =
\frac{\zeta_{||}}{\zeta_\perp} \simeq \frac 2 {\pi}
\sqrt{\frac{\eta_{||}}{\eta_\perp}} \gg \frac{\eta_{||}}{\eta_\perp},
\end{align*}
%
becomes a rather poor measure of the ratio of actual viscosities of the fluid.

\subsubsection{Strain Pseudo-Isotropic Fluid
($\boldsymbol {\eta_a =\eta_b}$)} \label{subsec:etaaeqetab}

Consider a fluid where $\alpha_1$ and $\alpha_3$ are very small and $\alpha_2 =
-\alpha_5$. 
Under these conditions, the equations of motion
\eqref{eq:fluid_equations_x}-\eqref{eq:fluid_equations_z} become
%
\begin{equation}
\nabla p' = \eta \nabla^2 \vec{v} + \Delta \eta \partial_{||} (\vec \omega
\times \vec n),
\label{eq:pseudostrain}
\end{equation}
%
where $\eta = \eta_a$ and $\Delta \eta = \eta_c - \eta_a$.  
The Laplacian term on the right hand side of equation \eqref{eq:pseudostrain}
can be seen as coming from isotropic momentum diffusion with viscosity $\eta$.
On the other hand, the last term in the equation represents the resistance of
the fluid to bending with respect to the nematic, and this is the only source
of anisotropy in the strain pseudo-isotropic condition.  The coefficient
$\Delta \eta$ can be interpreted as a rotational viscosity.

Taking the limit $\eta_b \rightarrow \eta_a$ on equations
\eqref{eq:response_function_11_analytic}-\eqref
{eq:response_function_22_analytic}, and introducing the anisotropy ratio
$\Lambda = \Delta \eta / \eta$, we find that
%
\begin{align}
	\label{eq:zeta_1_C} 
\frac{\zeta_{||}}{6 \pi a \eta } &= \frac{\frac{2}{3} \Lambda} {(\Lambda + 1)
\frac{\arctan \left( \sqrt{ \Lambda } \right)}{ \sqrt{ \Lambda } } - 1}, \\
	\label{eq:zeta_2_C} 
\frac{\zeta_\perp}{6 \pi a \eta} &= \frac{\frac{4}{3} \Lambda } { (\Lambda - 1)
\frac{\arctan \left( \sqrt{ \Lambda } \right)} { \sqrt{ \Lambda } } + 1}.
\end{align}
%
Note that, in this case, the anisotropy ratio ranges between $\Lambda = -1$
($\eta_c \ll \eta_a$) and $\Lambda = \infty$ ($\eta_c \gg \eta_a$), and the
isotropy point corresponds to $\Lambda = 0$.

Similar to figure \ref{fig:zeta_2vis}(\aaa), the parallel and perpendicular 
response
functions for a strain pseudo-isotropic fluid are plotted in figure
\ref{fig:zeta_2vis}(\bbb), along with their tangent lines at the isotropy point 
and their asymptotes for highly anisotropic conditions.
In this case, the plotted curves are one-dimensional sections of figures
\ref{fig:contour_zeta_11}(\aaa) and \ref{fig:contour_zeta_22}(\aaa) along the 
vertical line $\eta_a = \eta_b$.

The behavior of the response function near the isotropy point is given by its
Taylor expansion,
%
\begin{align*}
	\frac{\zeta_{||}}{6 \pi a \eta} &= 1 + \frac{1}{5} \Lambda -
\frac{8}{175} \Lambda^2 + O[\Lambda^3] , \\
	\frac{\zeta_\perp}{6 \pi a \eta} &= 1 + \frac{2}{5} \Lambda -
\frac{17}{175} \Lambda^2 + O[\Lambda^3].
\end{align*}
%
which shows that, in contrast to the rotationally pseudo-isotropic condition, 
the anisotropy corrections are stronger in the transverse direction than in the
parallel direction.  
\textcolor{black}{This result is consistent with the intuitive notion that
particle motion in the $\perp$ direction bends the nematic more than particle
motion in the $||$ direction.}

In the limit $\Lambda \gg 1$, the asymptotic behavior of the response function
is
%
\begin{equation*}
	\frac{\zeta_{||}}{6 \pi a \eta} \simeq \frac{4}{3 \pi} \sqrt{\Lambda} ,
\;\;\;\;\;
	\frac{\zeta_\perp}{6 \pi a \eta} \simeq \frac{8}{3 \pi} \sqrt{\Lambda},
%
%
\end{equation*}
%
whereas, in the limit $\Lambda \to -1$, we have
%
\begin{align*}
	\frac{\zeta_{||}}{6 \pi a \eta} & \simeq \frac{2}{3} - \frac{2}{3}
\left[ 1 + \frac{1}{2} \ln \left( \frac{ \Lambda + 1 }{4} \right) \right]
\left( \Lambda + 1 \right), \nonumber\\
	\frac{\zeta_\perp}{6 \pi a \eta} & \simeq -\frac{4}{3} \frac{1} {\ln
\left( \frac{ \Lambda + 1 }{4} \right) + 1}.
%
%
\end{align*}

Comparing these asymptotic responses with those in figure
\ref{fig:zeta_2vis}(\aaa) reveals that the effects of anisotropy in the
stress-strain relationship are fundamentally different from those arising from
nematic bending.
In rotationally pseudo-isotropic fluids, the response function becomes
independent of the anisotropy ratio for high levels of anisotropy, leading to
horizontal asymptotes in figure \ref{fig:zeta_2vis}(\aaa), and rendering upper
bounds to the effect of anisotropy.
The only exception to this behavior is $\zeta_{||}(\eta_{||} \ll \eta_\perp)$,
which has a square-root dependence on $\eta_\perp$ due to the fact that the
flow is still infinitesimally deflected in the $\perp$ direction by the sphere
in this limit, as will be shown in \S \ref{sec:vlongi}.

In contrast, the strain pseudo-isotropic response function remains dependent on
both $\eta$ and $\Delta \eta$ even at high levels of anisotropy, suggesting a
strong interaction between the anisotropic stresses caused by bending and the
diffusion of momentum by strain.
These two mechanisms only appear to decouple from each other for
$\zeta_{||}(\Delta \eta \rightarrow -\eta)$, which is consistent with the
independence of the parallel response with respect to the first Miesowicz
coefficient (figure \ref{fig:contour_zeta_11}\aaa), as $\Delta \eta = -\eta_a$
in this limit. These results inevitably call for caution in employing the
concept of effective directional viscosities. 
This notion may be qualitatively useful in rotationally pseudo-isotropic fluids
where the effect of anisotropy can be partially separated in different
directions.  
However, its usefulness becomes limited in fluids with resistance to bending,
where the anisotropy appears to act in all directions concurrently and in a
non-trivial manner.

\subsection{ Comparison with numerical results } 
\label{sec:Comparison}

This section uses data from existing numerical simulations to validate the
analytical procedures employed to determine $\overline{\overline{\zeta}}$.
Owing to the relatively high number of different viscosities involved in the
problem, it is difficult to find simulation studies that cover a significant
part of the parameter space.  The most comprehensive set of simulations is
\cite{Heuer.Kneppe.Schneider.1992}, who compute the steady creeping flow of a 
nematic liquid of uniform director around a sphere using finite differences and 
a relaxation time integrator.  They provide empirical formulae for both 
$\zeta_{||}$ and $\zeta_{\perp}$ that fit their simulation results near the 
isotropy point.

\begin{figure}
\vspace{0.5cm}
\includegraphics*[width=0.49\columnwidth,keepaspectratio]{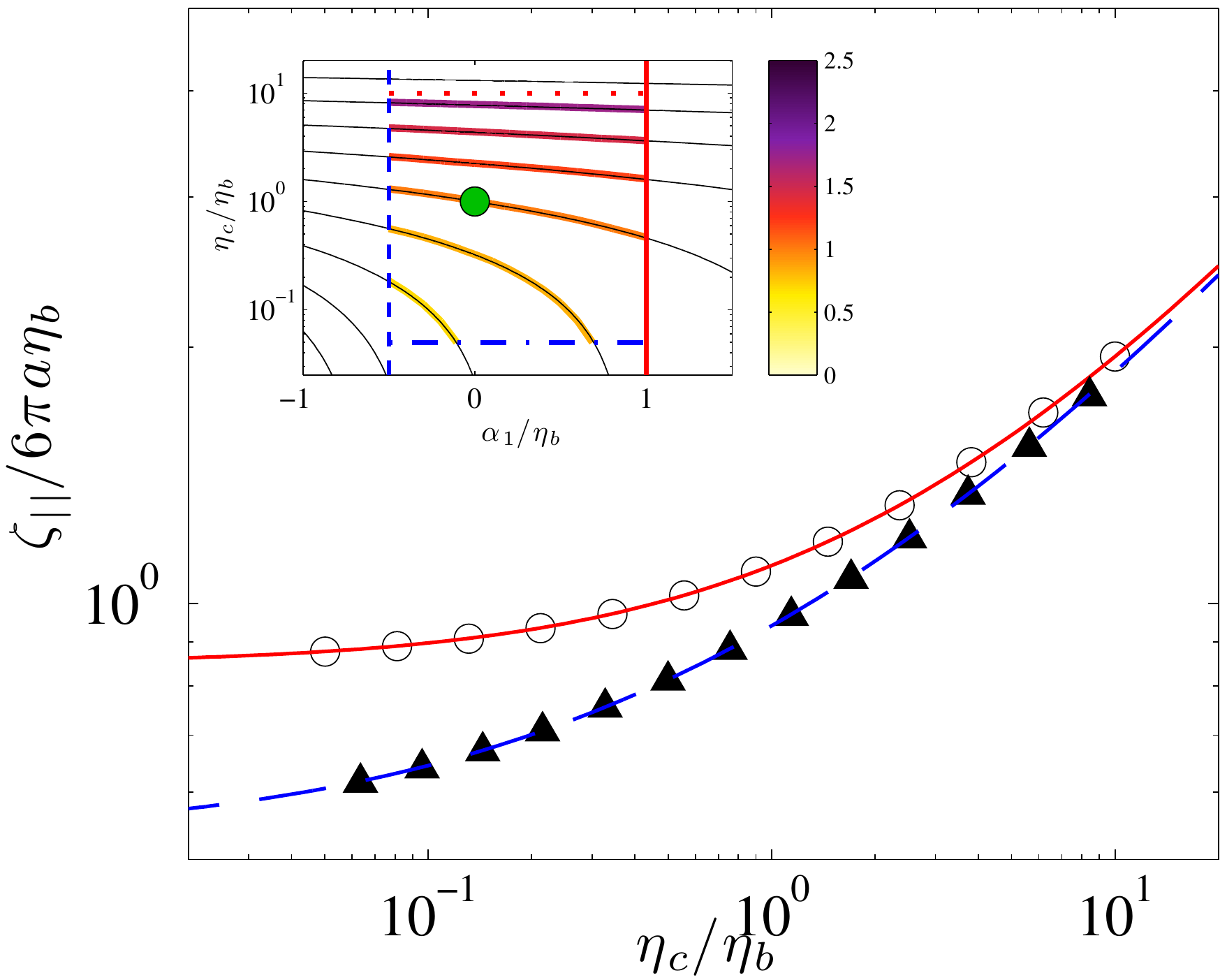}
\mylab{-0.48\textwidth}{0.41\textwidth}{(\aaa)}
\includegraphics*[width=0.49\columnwidth,keepaspectratio]{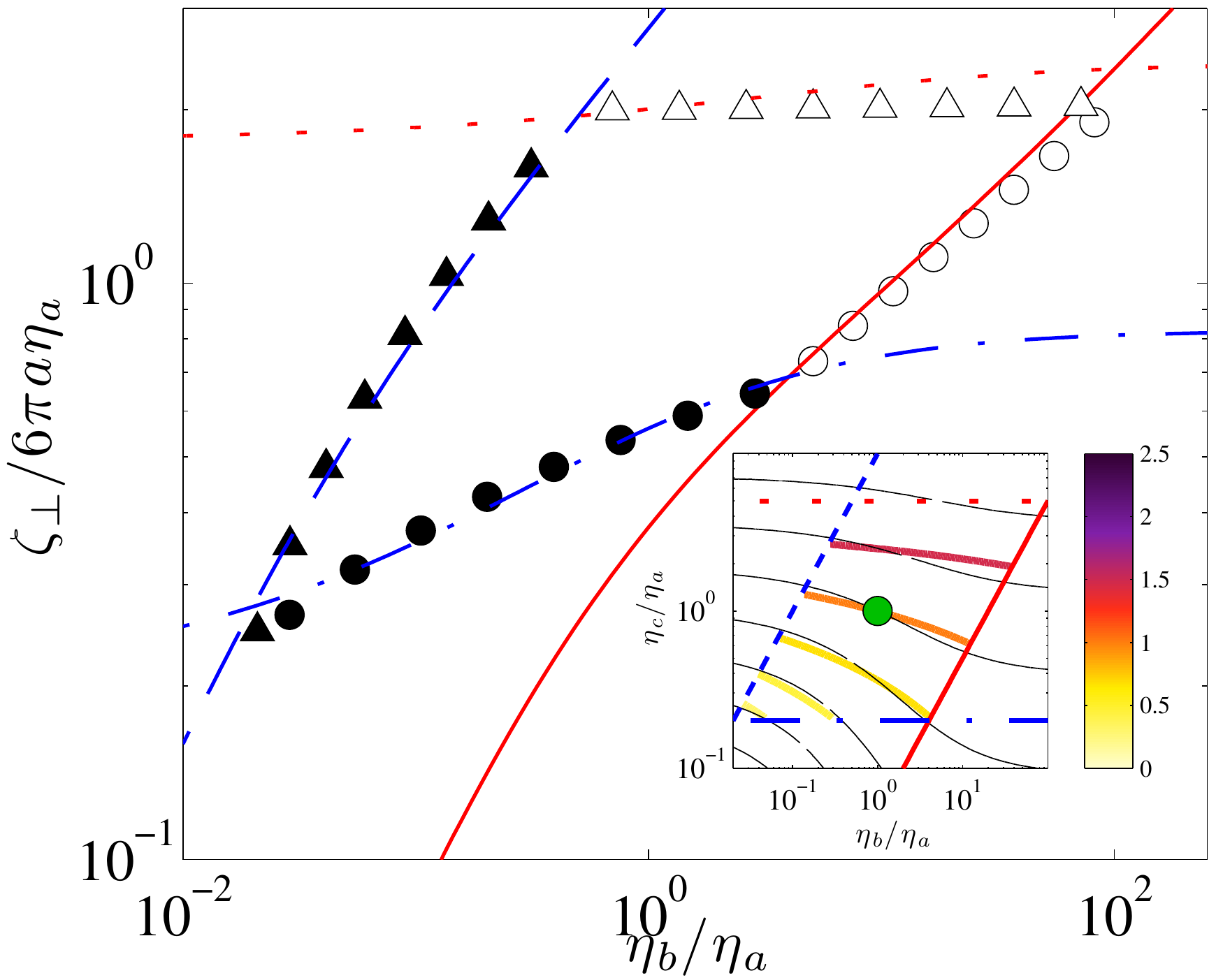}
\mylab{-0.48\textwidth}{0.41\textwidth}{(\bbb)}
	\caption{
Comparison between our analytical expression for the response 
function (lines) and numerical simulations by 
\citep{Heuer.Kneppe.Schneider.1992} (symbols). The two line plots on each figure
represent the lower and upper limits of the parameter ranges provided in 
\citep{Heuer.Kneppe.Schneider.1992}.
These limits and the entire parameter ranges are represented in the inset
contour maps, where the thick colour contours come from the simulations, the
thin black contours come from equations \eqref{eq:response_function_11_analytic}
and \eqref{eq:response_function_22_analytic}, and the green circle denotes the
isotropic case.
(\aaa) Response function in the nematic direction, $\zeta_{||}$, normalized
with $a \eta_b$ and represented as a function of $\eta_c / \eta_b$ and
$\alpha_1 / \eta_b$:
\textcolor{blue}{\dashed} and \solidtrian, $\alpha_1 = -0.5 \eta_b$; 
\textcolor{red}{\solid} and \circle, $\alpha_1 = \eta_b$; 
\textcolor{blue}{\chndot}, $\eta_c = 0.05 \eta_b$; 
\textcolor{red}{\dotted}, $\eta_c = 10 \eta_b$. 
(\bbb) Response function perpendicular to the nematic direction,
$\zeta_{\perp}$, computed for $\alpha_1 = 0$, normalized with $a \eta_a$ and
represented as a function of $\eta_b / \eta_a$ and $\eta_c / \eta_a$: 
\textcolor{blue}{\dashed} and \solidtrian, $\eta_c = 10 \eta_b$; 
\textcolor{red}{\solid} and \circle, $\eta_b = 0.05 \eta_c$; 
\textcolor{blue}{\chndot} and \solidcircle, $\eta_c = 0.2 \eta_a$; 
\textcolor{red}{\dotted} and \trian, $\eta_c = 5 \eta_a$. 
}
\label{fig:Comparison_Heuer}
%
\end{figure}

Figure \ref{fig:Comparison_Heuer}(\aaa) compares the analytical expression for
$\zeta_{||}$ (\ref{eq:response_function_11_analytic}) with Heuer \etal's
empirical fit in the parameter range where these authors find the fit to be
accurate, that is $0.05 < \eta_c/\eta_b < 10$ and $-1/2 < \alpha_1/\eta_b < 1$
\citep{Heuer.Kneppe.Schneider.1992}.  Both results are in an excellent
agreement, not only around the isotropy point but in the whole domain,
including its boundaries $\alpha_1 = -\eta_b / 2$ and $\alpha_1 = \eta_b$.

Figure \ref{fig:Comparison_Heuer}(\bbb) displays a similar comparison for
$\zeta_{\perp}$, this time in the domain $0.2 < \eta_c/\eta_b < 5$, $0.2<
\eta_c/\eta_a < 5$ and $\alpha_1 =0$, which is the range of accuracy of Heuer
\etal's  empirical fit to their simulation data
\citep{Heuer.Kneppe.Schneider.1992}.  The agreement is again perfect near the
isotropy point and for low values of $\eta_b/\eta_a$ and $\eta_c/\eta_a$
(bottom left corner of figure \ref{fig:Comparison_Heuer}(\bbb)).
However, it worsens slightly for high values of $\eta_b/\eta_a$ and
$\eta_c/\eta_a$ (top right corner of the figure), where the simulation results
seem to underestimate $\zeta_\perp$.  These differences are however small and
remain below $12\%$.
The observed divergence may be due to imprecisions in Heuer \etal's fit to
their own simulation results far from the isotropy point.
This hypothesis is supported by comparing their simulation results to the
prediction from the empirical fit for the one case in their paper where both
are available, \ie $(\eta_b / \eta_a ,\, \eta_c / \eta_a, \alpha_1/\eta_a) =
(1/3,\, 10/3, \,0) $.  For this combination of viscosities, the simulation
yields $\zeta_{\perp} / 6 \pi a \eta_a= 5.28 $, which is in excellent agreement
with our calculation of $\zeta_\perp / 6 \pi a \eta_a  = 5.27$.  However, the
empirical fit predicts $\zeta_\perp / 6 \pi a \eta_c = 5.47$, which differs by
$4\%$ from the simulation result.

\section{Assessing Particle Tracking Microrheology of Nematic Bio-Polymer
Networks}
\label{sec:DPTM} 

This section employs the expressions for the response function derived in \S
\ref{sec:Two_viscosities} to evaluate the accuracy of existing PTM formulae
when applied to pseudo-isotropic nematic fluids.  
This analysis allows us to separately determine the ability of those formulae
to estimate strain directional viscosities or bending viscosities.

PTM determines the viscosity of a fluid from the measured response function of
embedded microparticles.  Whether the motion of the particles is spontaneously
driven by the thermal excitation of the fluid (passive PTM) or externally
forced (active PTM), the fluid viscosity is estimated by fitting the measured
response function to a model for the drag of a canonical particle.
The simplest possible model is the Stokes's formula
$ 
	\zeta = 6 \pi a \eta,
$
which is valid for isotropic fluids \citep{Mason.1995,Mason.1997,Mason.2000}.
A second, more recent approach that is applied to nematic biogels and liquid
crystals consists of applying Stokes's equation separately in the two principal
directions of the motion to obtain two effective viscosity coefficients
\citep{Stark.Ventzki.2001,
Loudet.2004,Hasnain.2006,Rogers.Waigh.Lu.2008,DelAlamo.2008,
He.Mak.Liu.Tang.2008},
%
\begin{equation}
	\label{eq:Directional_Friction_Coefficients} \zeta_{||, \perp} = 6 \pi
a \eta_{||, \perp}^{\mbox{\scriptsize eff}}.
\end{equation} 

\begin{figure} 
\includegraphics*[width=0.45\columnwidth,keepaspectratio]{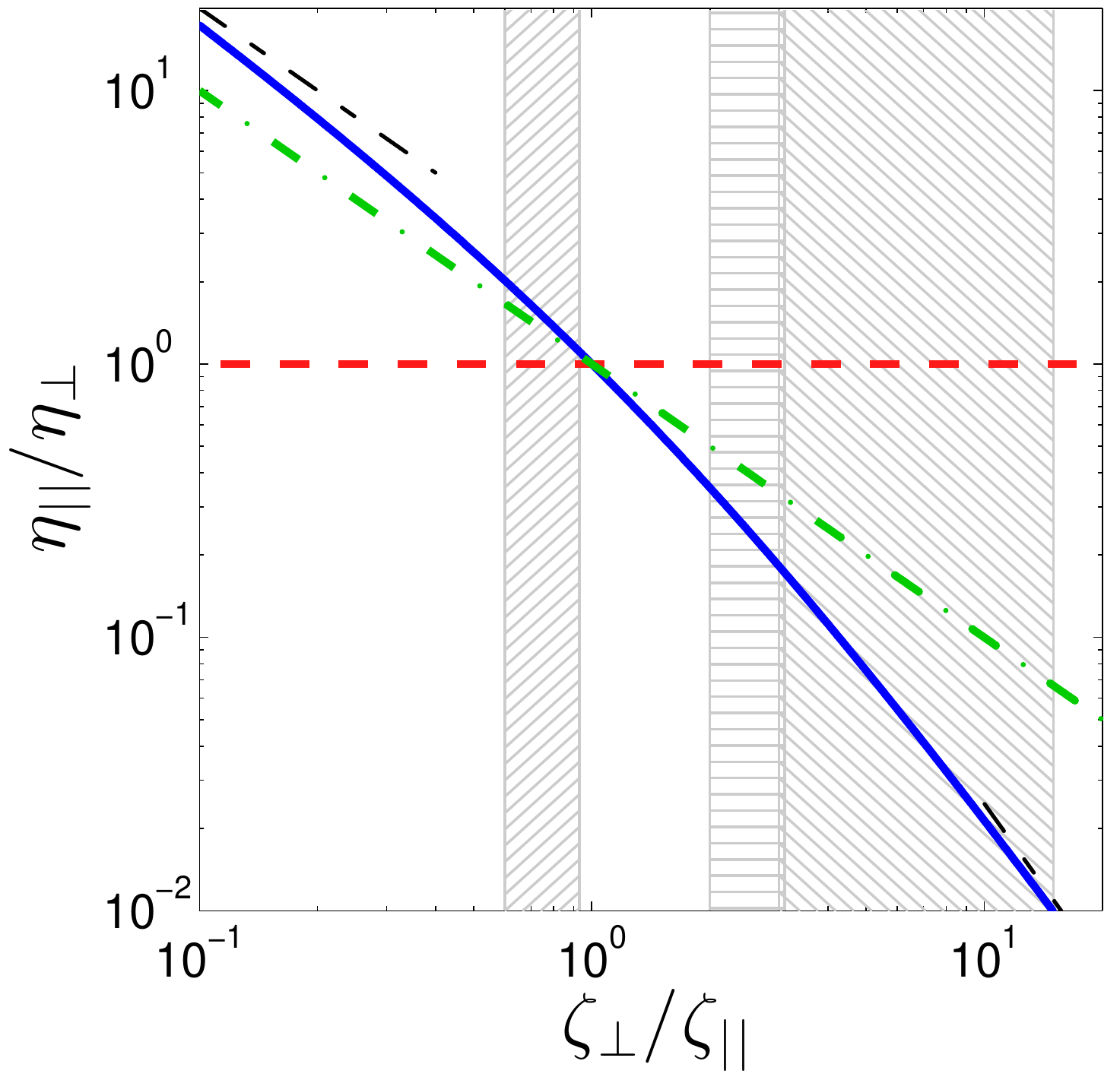}
\mylab{-0.36\textwidth}{0.1\textwidth}{(\aaa)}
\includegraphics*[width=0.45\columnwidth,keepaspectratio]{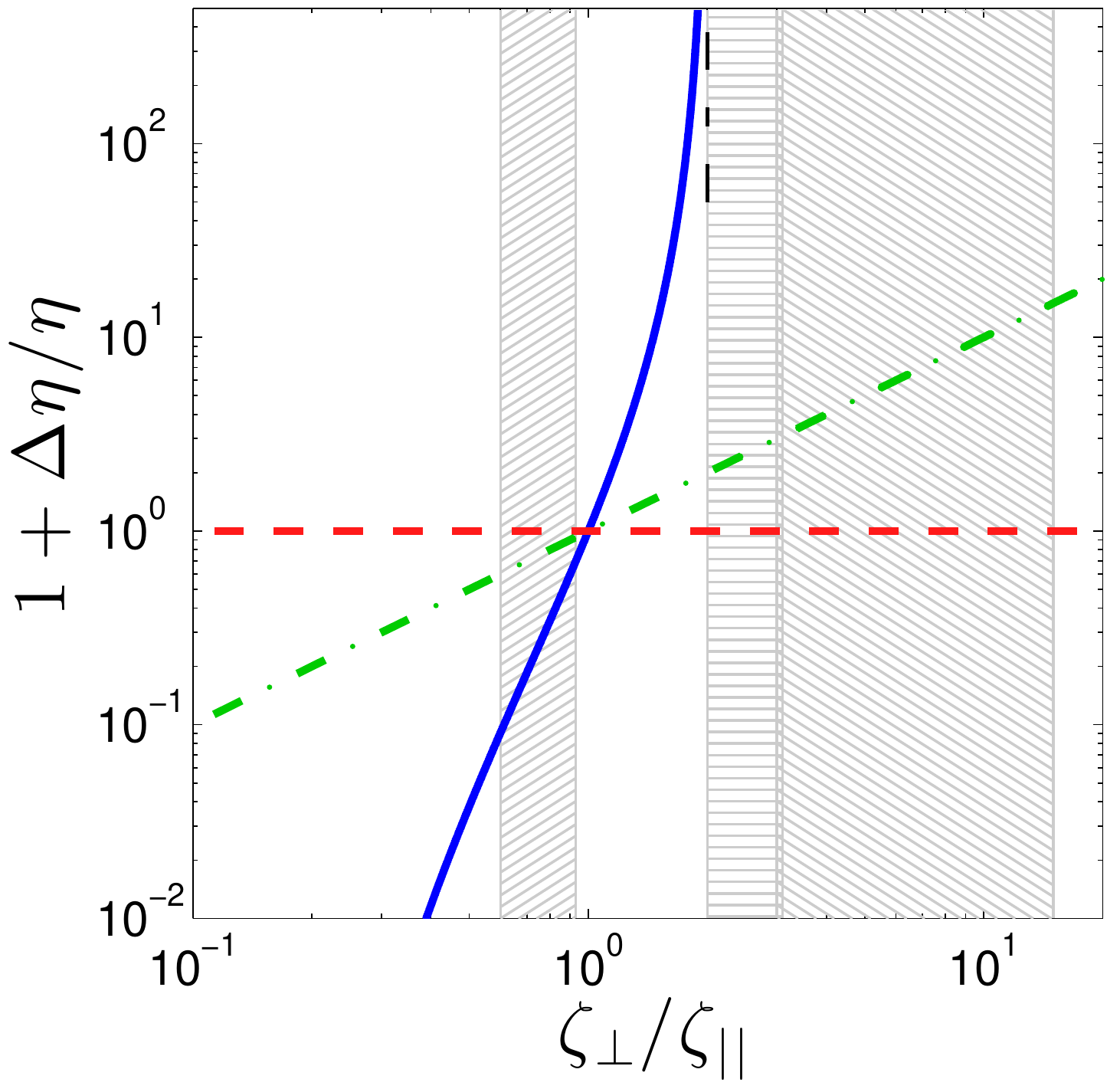}
\mylab{-0.36\textwidth}{0.1\textwidth}{(\bbb)}
\caption{
(\aaa) Ratio of viscosity coefficients,
$\eta_{||}/\eta_{\perp}$, as a function of the ratio of directional response
functions, $\zeta_{\perp}/\zeta_{||}$. 
Each curve is obtained by using a different model for the response function:
\textcolor{red}{\dashed}, isotropic Stokes's formula; 
\textcolor{verde}{\chndot}, effective viscosity approach (equation
\ref{eq:Directional_Friction_Coefficients});
\textcolor{blue}{\solid}, rotationally pseudo-isotropic fluid (see \S
\ref{subsec:etaaeqetac}); \textcolor{black}{\bdash \bdashort \spacce{1}
\bdash}, asymptotes.
The hatched areas represent experimental ranges of $\zeta_{\perp}/\zeta_{||}$
for biopolymer networks:
left hatched, sheared DNA solutions \citep{Hasnain.2006}; horizontally hatched,
nematic reconstituted actin networks \citep{He.Mak.Liu.Tang.2008}; right
hatched, cytoplasm of live bovine aortic endothelial cells
\citep{DelAlamo.2008}. (\bbb) Same as panel (\aaa) for a strain
pseudo-isotropic fluid (see \S \ref{subsec:etaaeqetab}) where the relevant
ratio of viscosities is $\Delta \eta / \eta$} 
\label{fig:PTM_1}
%
%
\end{figure}

From the statistical mechanics perspective, this approach has been justified on
the grounds that the motion of the particle in one principal direction is
uncorrelated from its motion in the second principal direction
\citep{DelAlamo.2008}.
\textcolor{black}{
From the point of view of fluid mechanics, the effective viscosities were
proposed as friction coefficients that quantify overall directional diffusivity
in a nematic fluid \citep{Stark.Ventzki.2001}.  However, their relation to the
actual directional viscosities of the fluid has not been characterized yet.}
Figure \ref{fig:PTM_1} sheds light into this question with the aid of
experimental data coming from reconstituted biopolymer gels
\citep{Hasnain.2006,He.Mak.Liu.Tang.2008} and live cells \citep{DelAlamo.2008}.
The figure displays the ratio of viscosity coefficients in pseudo-isotropic
fluids as a function of the ratio  $\zeta_\perp/\zeta_{||}$ calculated in this
study, together with the predictions from the isotropic Stokes's law and the
effective viscosity approach \eqref{eq:Directional_Friction_Coefficients}. The
range of values of $\zeta_\perp/\zeta_{||}$ measured in experiments is included
in the plots.  In passive PTM experiments, this ratio is equal to the ratio of
mean squared displacements of the particles $\bra \Delta x^2\ket_{\perp} / \bra
\Delta x^2\ket_{||}$ \citep{DelAlamo.2008}.

A first interesting observation that can be made from comparing figures
\ref{fig:PTM_1}(\aaa) and \ref{fig:PTM_1}(\bbb) is that biopolymer networks, 
both reconstituted and inside live cells, behave closer to a rotationally
pseudo-isotropic fluid than to a strain pseudo-isotropic one.  In fact, figure
\ref{fig:PTM_1}(\bbb) reveals that most of the reported values of
$\zeta_\perp/\zeta_{||}$ for these materials are disallowed by the strain
pseudo-isotropic model.
In this model,  $\zeta_\perp/\zeta_{||}$ has a vertical asymptote at 2 and,
thus, fluids where $\zeta_\perp/\zeta_{||}$ is measured to be $> 2$ are
necessarily strain anisotropic.
The only exception is the sheared DNA solution \citep{Hasnain.2006}, which 
shows a nearly isotropic behavior that can be in principle consistent with 
both the rotationally pseudo-isotropic and the strain pseudo-isotropic 
conditions.

If, in consonance with the previous observation, one assumes that the rheology
of nematic biopolymer networks is relatively close to rotationally
pseudo-isotropic, figure \ref{fig:PTM_1}(\aaa) can be used to estimate the 
accuracy of previous response function models when applied to those 
biomaterials.
Note that this estimation is not expected to be very precise as we merely
proved that these fluids are not strain pseudo-isotropic, which is not exactly
equivalent to proving that they are rotationally pseudo-isotropic.
Nevertheless, we proceed with this estimation for lack of a better measure.
Comparing the different curves in figure \ref{fig:PTM_1}(\aaa) within the range 
of experimental values of $\zeta_\perp/\zeta_{||}$ obtained from the literature
indicates that previous models render errors of up to one order (effective
viscosity) or two orders of magnitude (isotropic Stokes).  
In both cases, the actual anisotropy of the fluid, given by the ratio
$\eta_\perp / \eta_{||}$, is underestimated, with the effective viscosity
approach predicting $0.8 \lesssim \eta_{\perp}/\eta_{||} \lesssim 9$, and the
present results predicting $0.5 \lesssim \eta_{\perp}/\eta_{||} \lesssim 100$.
Thus, researchers should exercise caution when interpreting directional PTM
data.

\section{Influence of Anisotropy on the Far Velocity Field} \label{sec:velos}
 
Anisotropy does not alter the $r^{-1}$ decay of the velocity far away from the
sphere but it does modify the dependence of $\vec v$ on the azimuth and
inclination angles. 
The velocity in the far field of the sphere can be calculated from equations
\eqref{eq:velocity_greens} and (\ref{eq:forceeq}), yielding
%
\begin{equation*}
	\label{eq:Far_Field_v} \vec{v}(\vec{x}) = \frac{\overline{ \overline{
\mathcal{G} (\vec x) } }} {8 \pi} \cdot \left[ \begin{array}{rcl} \zeta_{||} &
0 & 0 \\ 0 & \zeta_\perp & 0 \\ 0 & 0 & \zeta_\perp \end{array} \right] \cdot
\vec{v}_0,
\end{equation*}
%
where $\vec{v}_0$ is the velocity of the particle.  Similar to \S
\ref{sec:Response_Leslie_Ericksen}, the Green's function is obtained by Fourier
transforming equations \eqref{eq:G_1j}-\eqref{eq:G_3j} after regularizing with
the Gaussian \eqref{eq:gaussreg}.
This section focuses on the two basic flow configurations in which $\vec{v}_0$
is parallel and perpendicular to $\vec n$.  All other possible configurations
are linear combinations of these two basic flows.

\subsection{Velocity field caused by a sphere moving parallel to the nematic
director ($\vec v_0 \, ||\, \vec n$)} 
\label{sec:vlongi}

\begin{figure}
\vspace{0.5cm}
\includegraphics*[width=0.5\columnwidth,keepaspectratio]{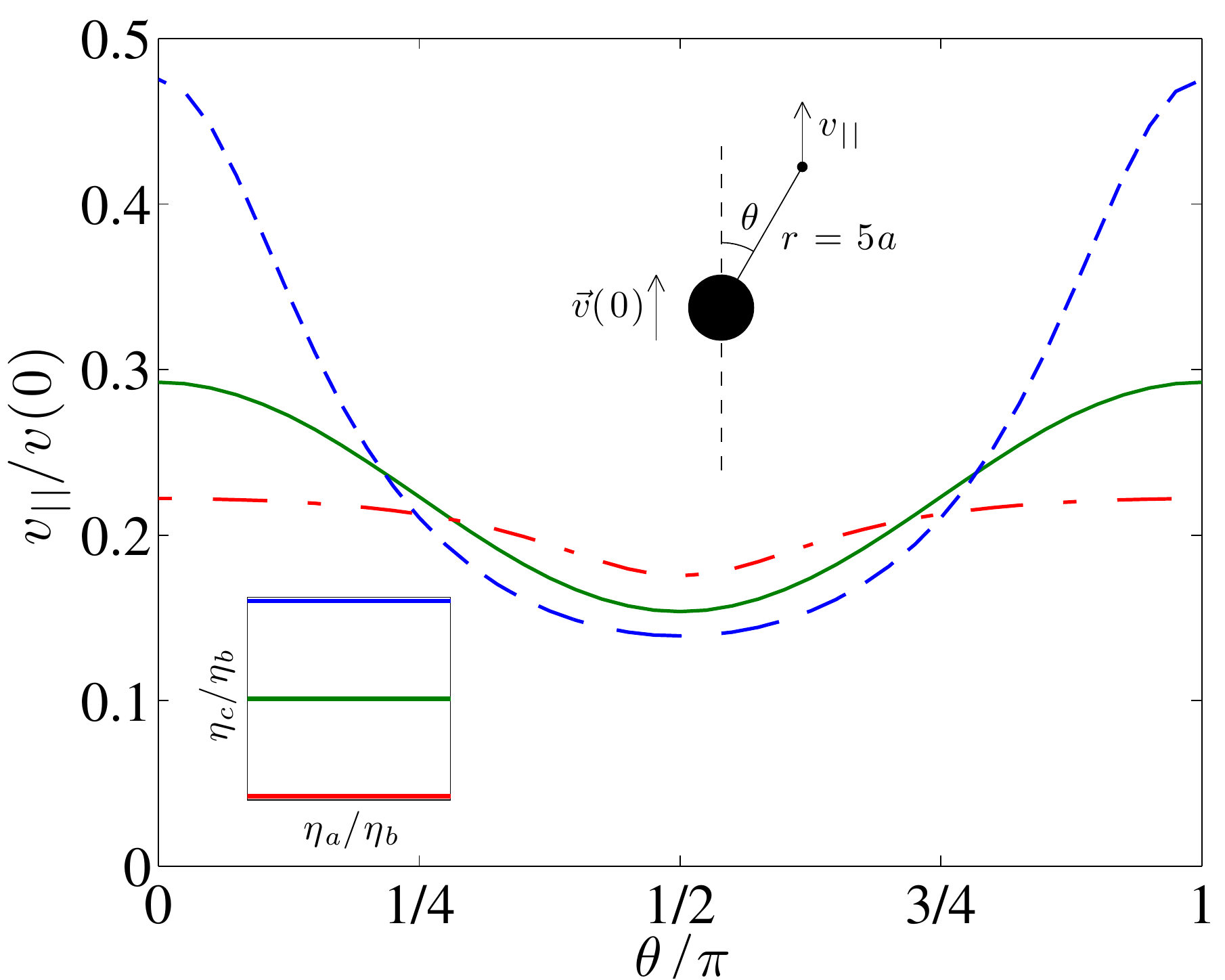}
\mylab{-0.45\textwidth}{0.41\textwidth}{(\aaa)}%
\includegraphics*[width=0.5\columnwidth,keepaspectratio]{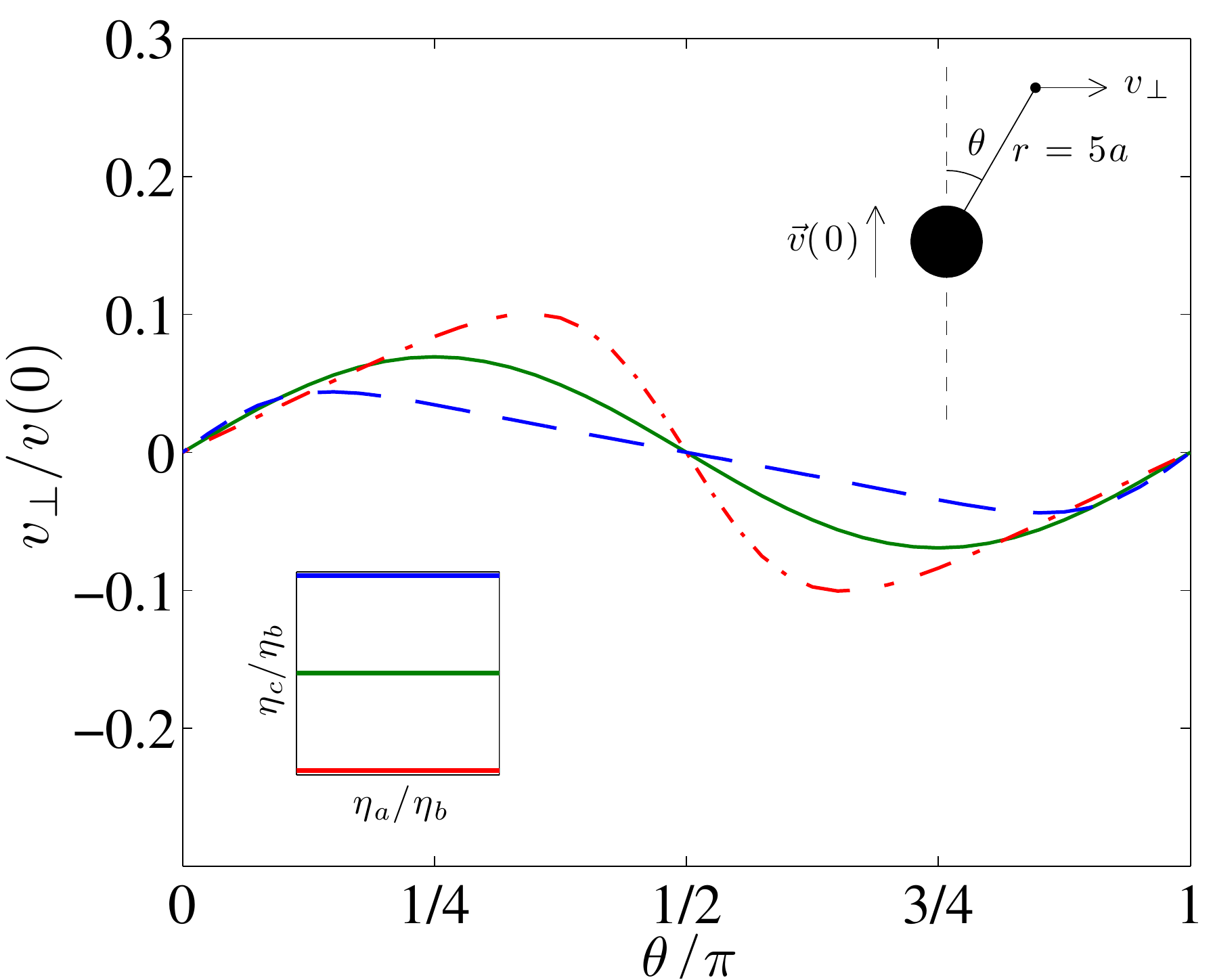}
\mylab{0.57\textwidth}{0.44\textwidth}{(\bbb)}%
	\caption{
Longitudinal ($v_{||}$, panel \aaa) and transverse ($v_\perp$, panel \bbb) flow
velocity components elicited by a sphere moving parallel to the nematic
director.
The velocities are calculated at a distance $r=5a$ away from the sphere centre,
represented as a function of the inclination angle $\theta$.
Each line type corresponds to a different value of $\eta_c / \eta_b$, as
indicated in the inset map at the lower left corner of each figure:
\textcolor{blue}{\dashed}, $\eta_c = 10 \eta_b$; \textcolor{verde}{\solid},
$\eta_c = \eta_b$; \textcolor{red}{\chndot}, $\eta_c = \eta_b / 10$.
				 } \label{fig:velopar}
%
\end{figure}

Figure \ref{fig:velopar} displays velocity profiles around a sphere that
translates parallel to the nematic director (\ie northwards) and experiences a
resistance equal to $\zeta_{||}$. 
The velocity components parallel and perpendicular to the director, $v_{||}$
and $v_\perp$ respectively, are shown.  These components fully define the
axially symmetric velocity field elicited by the sphere. 
For illustrative purposes, we have represented $v_{||}$ and $v_\perp$ at a
distance $r=5a$ from its centre but similar results are obtained at other
radial locations.

As noted above, the flow generated by $\vec v_0 \, || \, \vec n$ is independent
of the first Miesowicz viscosity, so we restrict our analysis to the effect of
the other two Miesowicz coefficients.  
For simplicity, we focus on the case $\alpha_1 = 0$ in which the equations
governing the flow
\eqref{eq:fluid_equations_xcyl}-\eqref{eq:fluid_equations_rcyl}  become
%
\begin{align}
\partial_x p' &= \eta_b \nabla^2 u_x,
\label{eq:pseudostokes_xcyl} 
\\
\partial_r p' &= \eta_c \left( \nabla^2 u_r - {u_r}/{r^2} \right).
\label{eq:pseudostokes_rcyl}
\end{align}
%
These equations are similar to Stokes's equations but have different viscosity
coefficients in the axial and radial directions. The resulting flow is
therefore rotationally pseudo-isotropic, and $\eta_b$ and $\eta_c$ represent
the viscosities in the $||$ and $\perp$ directions, respectively (see equation
\ref{eq:rotiso}).
Curiously enough, invoking axial symmetry allows us to recast the very same
equations as
%
\begin{align}
	\partial_x p' &= \eta_b \nabla^2 u_x , 
\label{eq:pseudorot_xcyl} \\
\partial_r p' &= \eta_b (\nabla^2 u_r - u_r / r^2) + (\eta_c - \eta_b)
\partial_x \omega_\theta,
\label{eq:pseudorot_rcyl}
\end{align} 
%
which are now consistent with strain pseudo-isotropic conditions with $\eta_b =
\eta$ and $\eta_c = \eta + \Delta \eta$ (see equation \ref{eq:pseudostrain}).
The equivalence of these two formulations implies that the flow generated by a
sphere moving parallel to the director is both strain pseudo-isotropic
and rotationally isotropic. 
However, it is important to note that this conjunction arises exclusively from
the flow geometry and not from the fluid properties, and that it does not mean
that the flow is isotropic.  In this configuration, the bending stresses are
still different than zero and momentum diffuses differently in the $||$ and
$\perp$ directions unless $\eta_c = \eta_a$ and $\eta_b = \eta_a$,
respectively.
This is possible because the axial symmetry of the flow imposes a connection
between the two types of anisotropic stresses, which is reflected by the
equivalence $\eta_{||} = \eta$ and $\eta_\perp = \eta + \Delta \eta$ that
follows from comparing equations
\eqref{eq:pseudostokes_xcyl}-\eqref{eq:pseudostokes_rcyl} with equations
\eqref{eq:pseudorot_xcyl}-\eqref{eq:pseudorot_rcyl}.
Consistent with this idea, the change of variables 
$(\eta_{||},\eta_\perp) = (\eta, \eta + \Delta \eta)$
transforms the rotationally pseudo-isotropic $\zeta_{||}$ \eqref{eq:zeta_1}
into the strain pseudo-isotropic one \eqref{eq:zeta_1_C}.  Of course, the same
does not happen for $\zeta_\perp$ (equations \ref{eq:zeta_2} and
\ref{eq:zeta_2_C} ) because the flow elicited by $\vec v_0 \, \perp \, \vec n$
is not axially symmetric.

\begin{figure}
\centering
\includegraphics*[width=0.4\columnwidth,keepaspectratio]{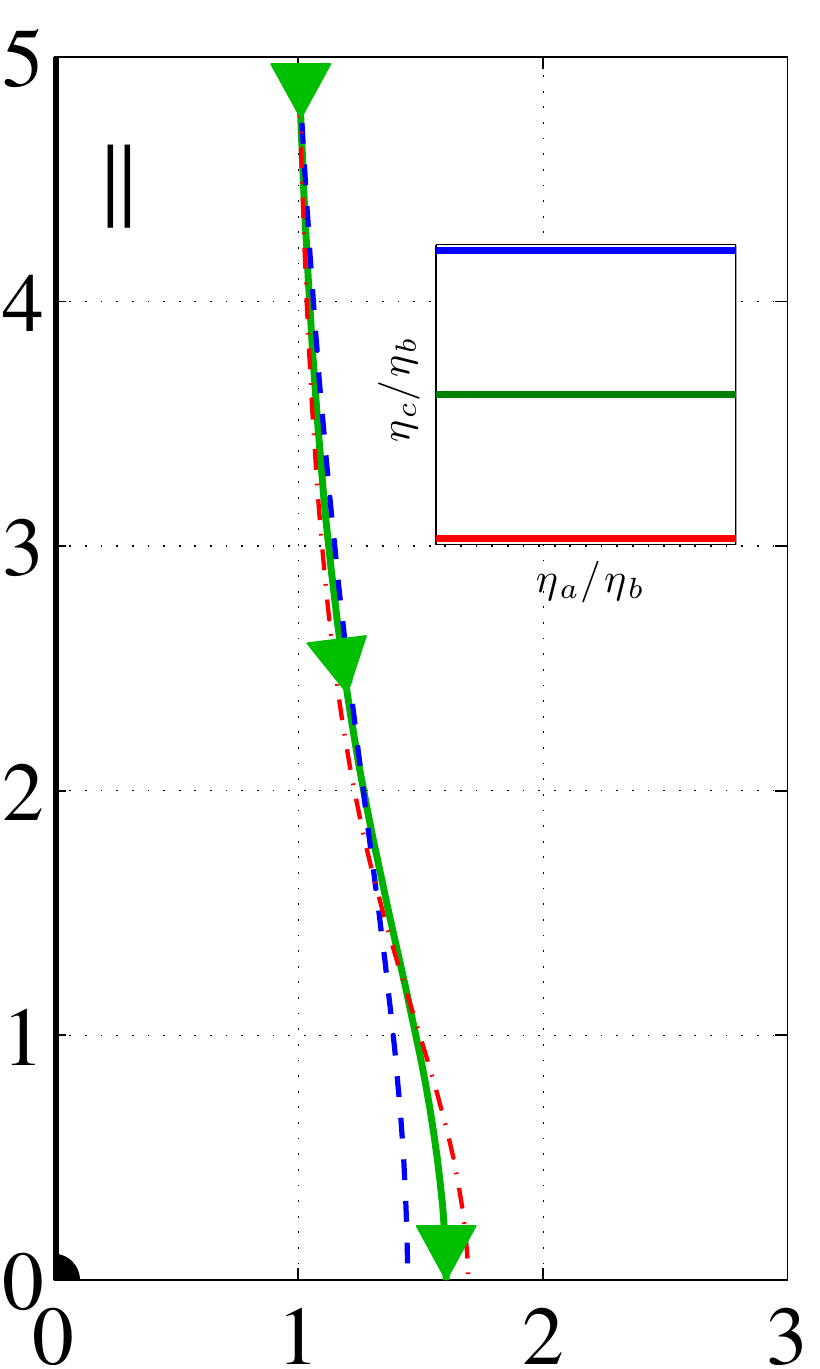}
	\caption{
Streamlines on the meridional plane of a sphere that translates parallel to the
nematic director (thick black vertical axis).  
The arrows indicate the flow direction in a reference frame moving with the
sphere.  The spatial coordinates are normalized with the sphere radius.
One streamline is plotted for each value of $\eta_c / \eta_b$ considered in
figure \ref{fig:velopar}:
\textcolor{blue}{\dashed}, $\eta_c = 10 \eta_b$; \textcolor{verde}{\solid},
$\eta_c = \eta_b$; \textcolor{red}{\chndot}, $\eta_c = \eta_b / 10$.
				 } \label{fig:lindecor_par}
%
\end{figure}

Probably due to this enhanced pseudo-isotropy, the velocity profiles in the
$\vec v_0 \, ||\, \vec n$ configuration conserve many isotropic features and
the streamlines are not substantially affected by the viscosity ratio $\eta_c /
\eta_b$.
In particular, figure  \ref{fig:velopar}(\aaa) shows that, regardless of the
value of $\eta_c /\eta_b$, the longitudinal velocity component has a single
maximum on the polar axis ($\theta=0$) and a single minimum on the equatorial
plane ($\theta = \pi/2$).  The viscosity ratio $\eta_c / \eta_b$, however, has 
a marked effect on the peak-to-valley variation of $v_{||}$.  For $\eta_c /
\eta_b \ll 1$, the spatial distribution of $v_{||}$ is nearly uniform.
However, for $\eta_c \gg \eta_b$, there is a significant enhancement of the
flow velocity on the polar axis, which is accompanied by a slight velocity
damping on the equatorial plane.

The transverse velocity component is shown in figure \ref{fig:velopar}(\bbb).
Owing to the symmetry of this flow configuration, $v_\perp$ is zero both on the
polar axis and the equatorial plane.
We find that the ratio $\eta_c / \eta_b$ influences both the maximum value of
$v_\perp$ and its azimuthal location, which is  $\theta = \pi/4$ in the
isotropic case.  For $\eta_c < \eta_b$ the peak value of $v_\perp$ increases
and its location is displaced towards the equator, whereas the opposite happens
for $\eta_c > \eta_b$.

Figure \ref{fig:lindecor_par} displays the flow streamlines obtained by
integrating the far velocity field after subtracting the speed of the sphere.  
In order to illustrate the effect of anisotropy, the integration is performed
for different values of $\eta_c / \eta_b$ from a common initial point far
upstream of the sphere. 
It is found that anisotropy only has a modest influence on the flow streamlines
when the sphere moves parallel to the nematic.
Consistent with $\eta_b$ and $\eta_c$ acting as $\eta_{||}$ and $\eta_\perp$, 
respectively, in equations 
\eqref{eq:pseudostokes_xcyl}-\eqref{eq:pseudostokes_rcyl}
and with the observed decrease of $v_\perp$ with $\eta_c / \eta_b$ (figure
\ref{fig:velopar}\bbb),
fluid particles approaching the sphere are deflected somewhat more in the
$\perp$ direction for $\eta_c < \eta_b$ than in isotropic conditions, and the
opposite happens for $\eta_c > \eta_b$. 
However, the level of deflection in this axially symmetric configuration is
considerably smaller than when the sphere is moving perpendicular to the
nematic (figure \ref{fig:lindecor_per}).  
%

\subsection{Velocity field caused by a sphere moving perpendicular to the 
nematic director ($\vec v_0 \perp \vec n$)} 
\label{sec:vperpe}

The flow generated by a sphere moving perpendicular to the nematic director
(\eg eastwards) is more complex than its $\vec v_0\, ||\, \vec n$ counterpart
because axial symmetry is broken, the velocity field is three-dimensional and
all four viscosity coefficients affect the flow.
In this section, we study this dependence by plotting velocity profiles and
streamlines for all of the possible combinations of the viscosity ratios defined
by $\eta_b / \eta_a$ and $\eta_c / \eta_a \ll, =, \gg 1$ ($0.1,\, 1,\, 10$),
and consistent with the scaling of the response function chosen to plot figure
\ref{fig:contour_zeta_22}.
For simplicity, we restrict our analysis to $\alpha_1 = 0$ similar to the $\vec
v_0\, ||\, \vec n$ configuration.

\begin{figure*}
%
\vspace{0.5cm}
\includegraphics*[width=0.329\columnwidth,keepaspectratio]{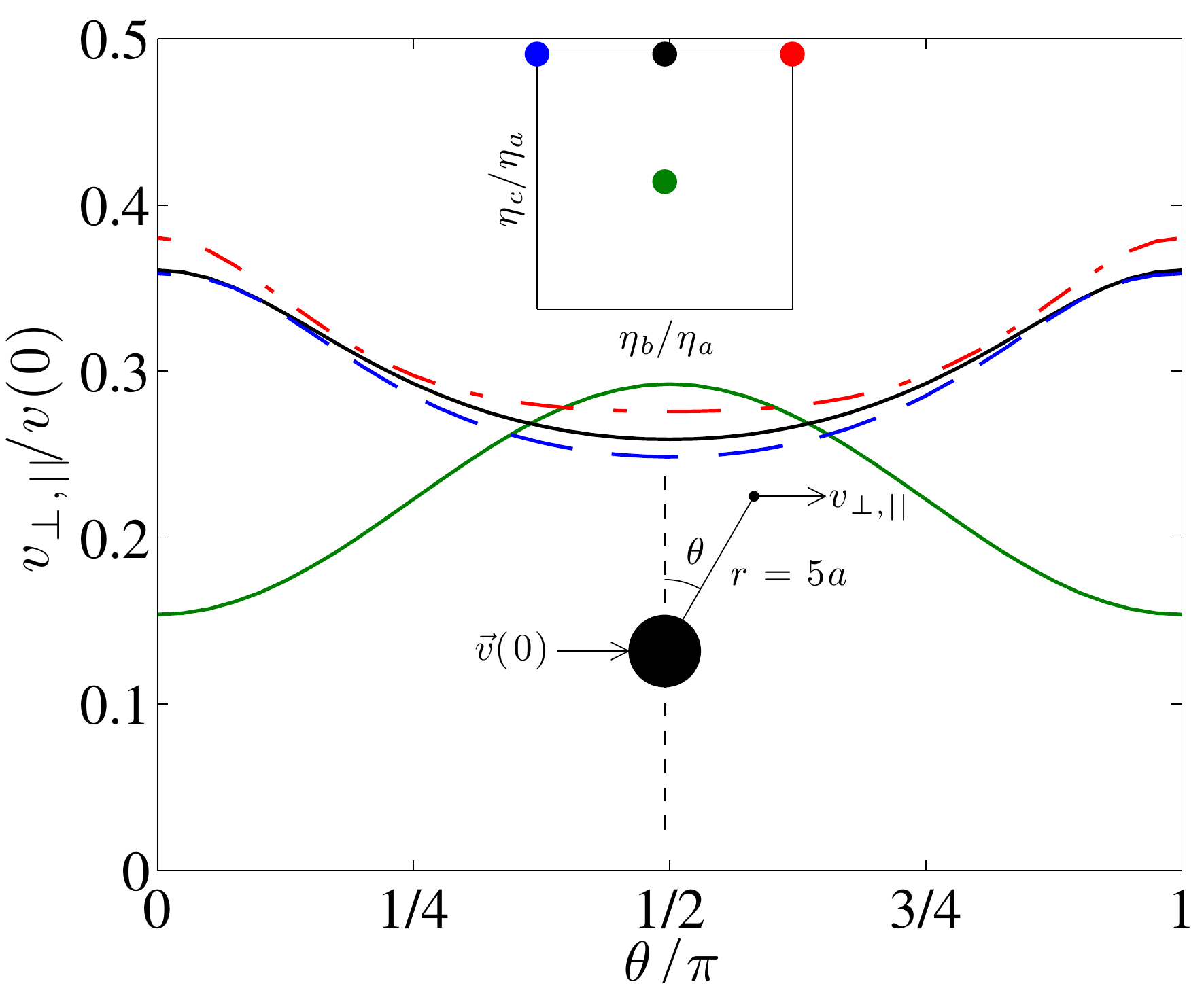}
\mylab{-0.3\textwidth}{0.275\textwidth}{(\aaa)}%
\includegraphics*[width=0.329\columnwidth,keepaspectratio]{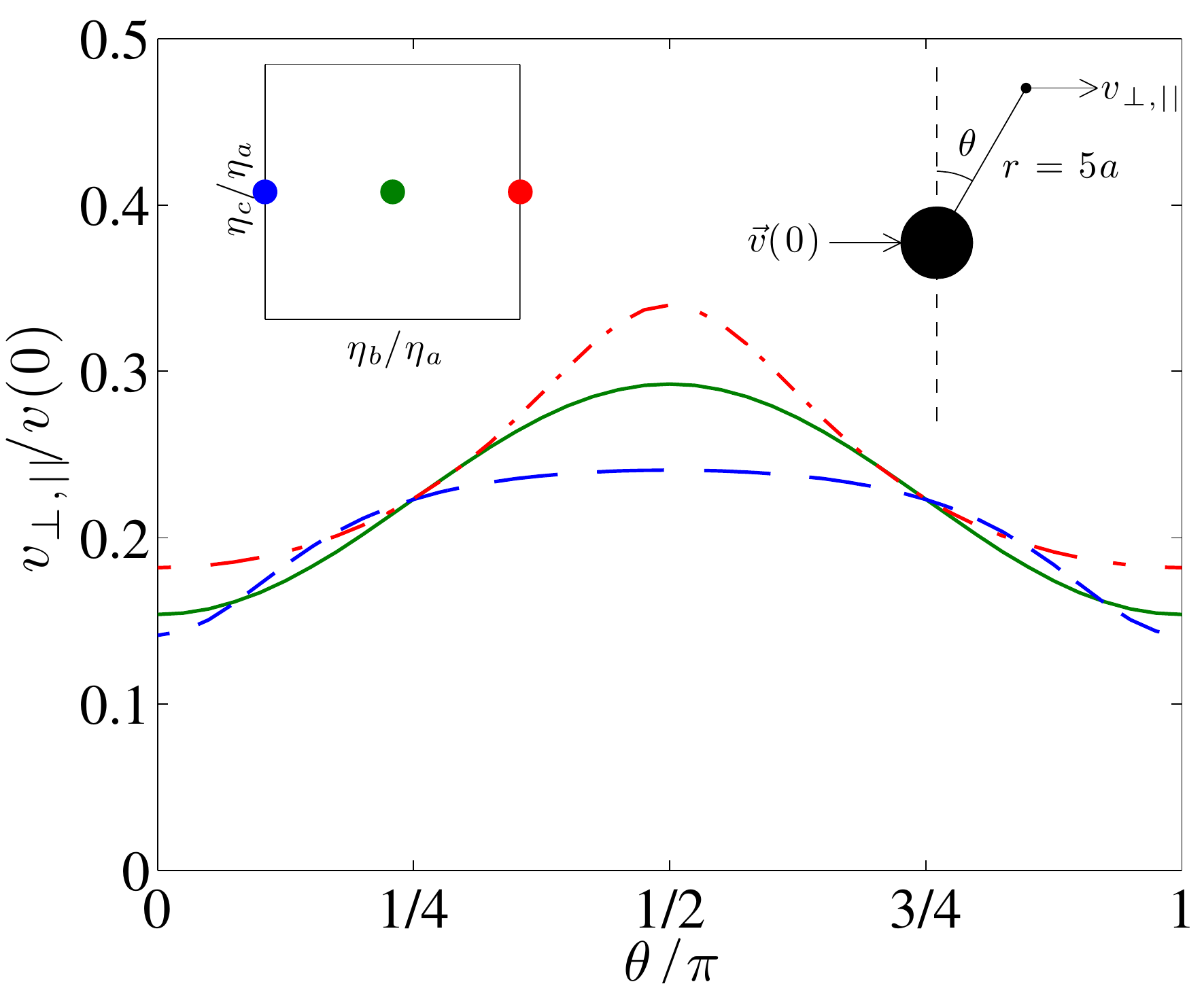}
\mylab{-0.3\textwidth}{0.275\textwidth}{(\bbb)}%
\includegraphics*[width=0.329\columnwidth,keepaspectratio]{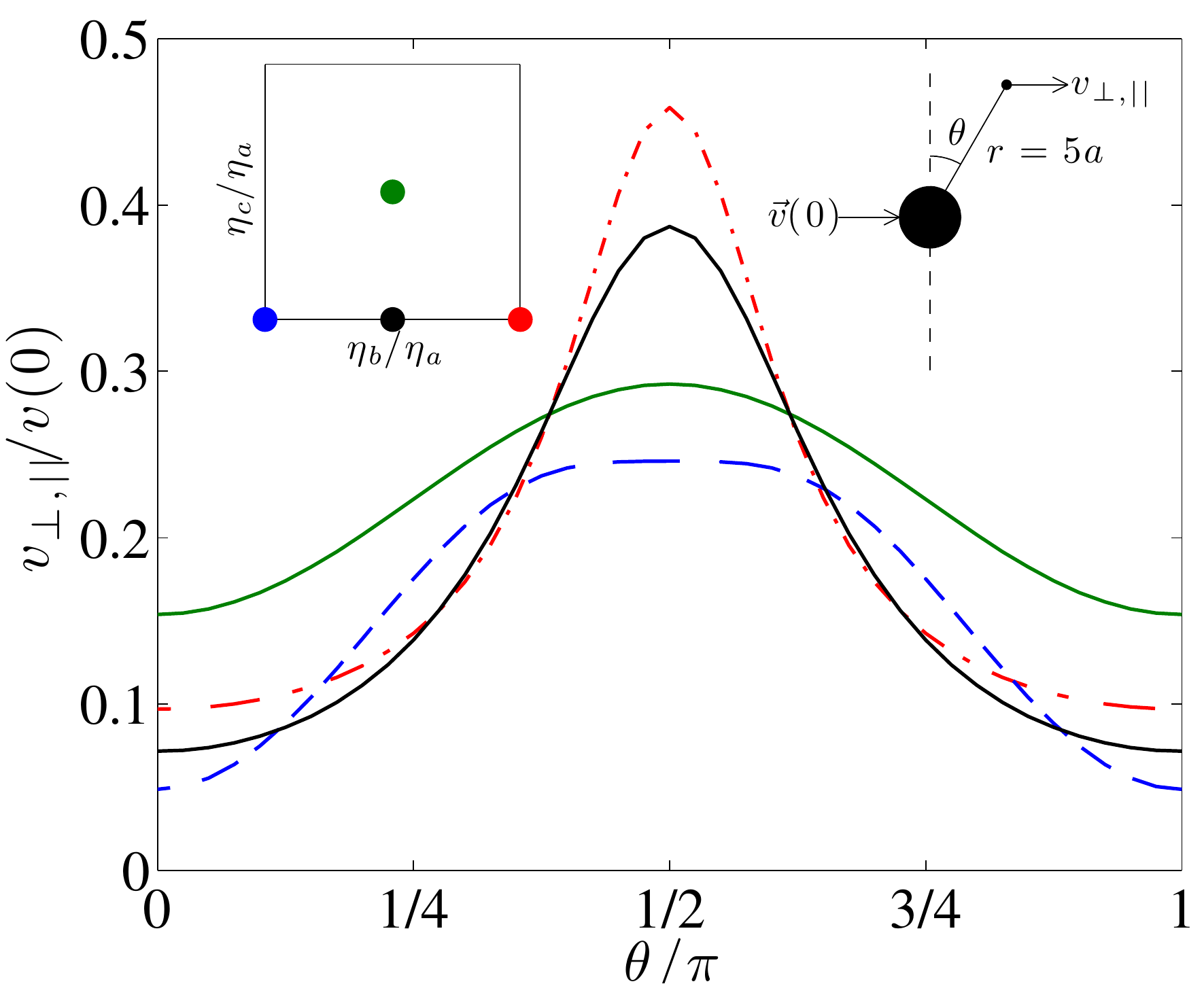}
\mylab{0.71\textwidth}{0.30\textwidth}{(\ccc)}%
\\
\includegraphics*[width=0.329\columnwidth,keepaspectratio]{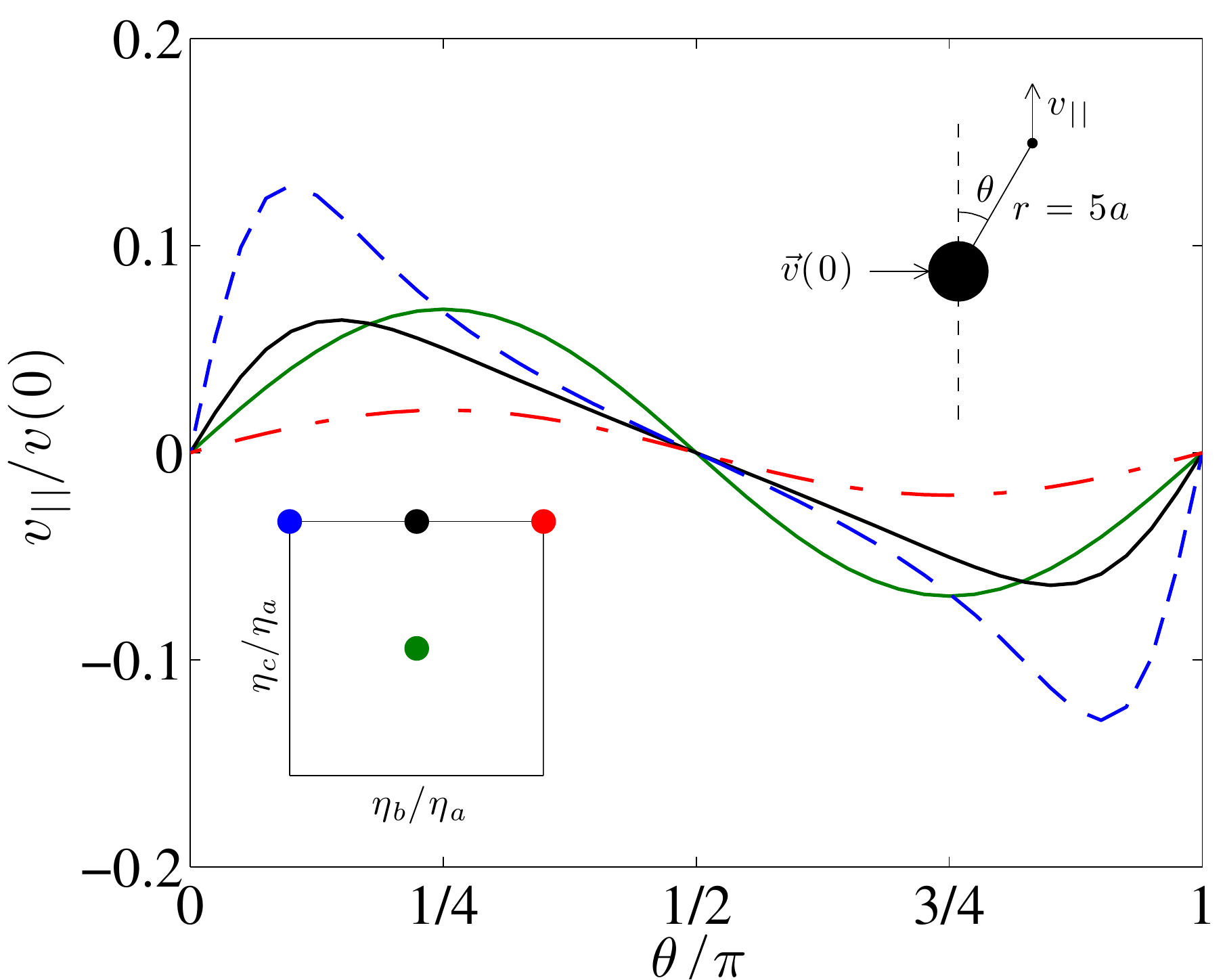}
\mylab{-0.3\textwidth}{0.275\textwidth}{(\ddd)}%
\includegraphics*[width=0.329\columnwidth,keepaspectratio]{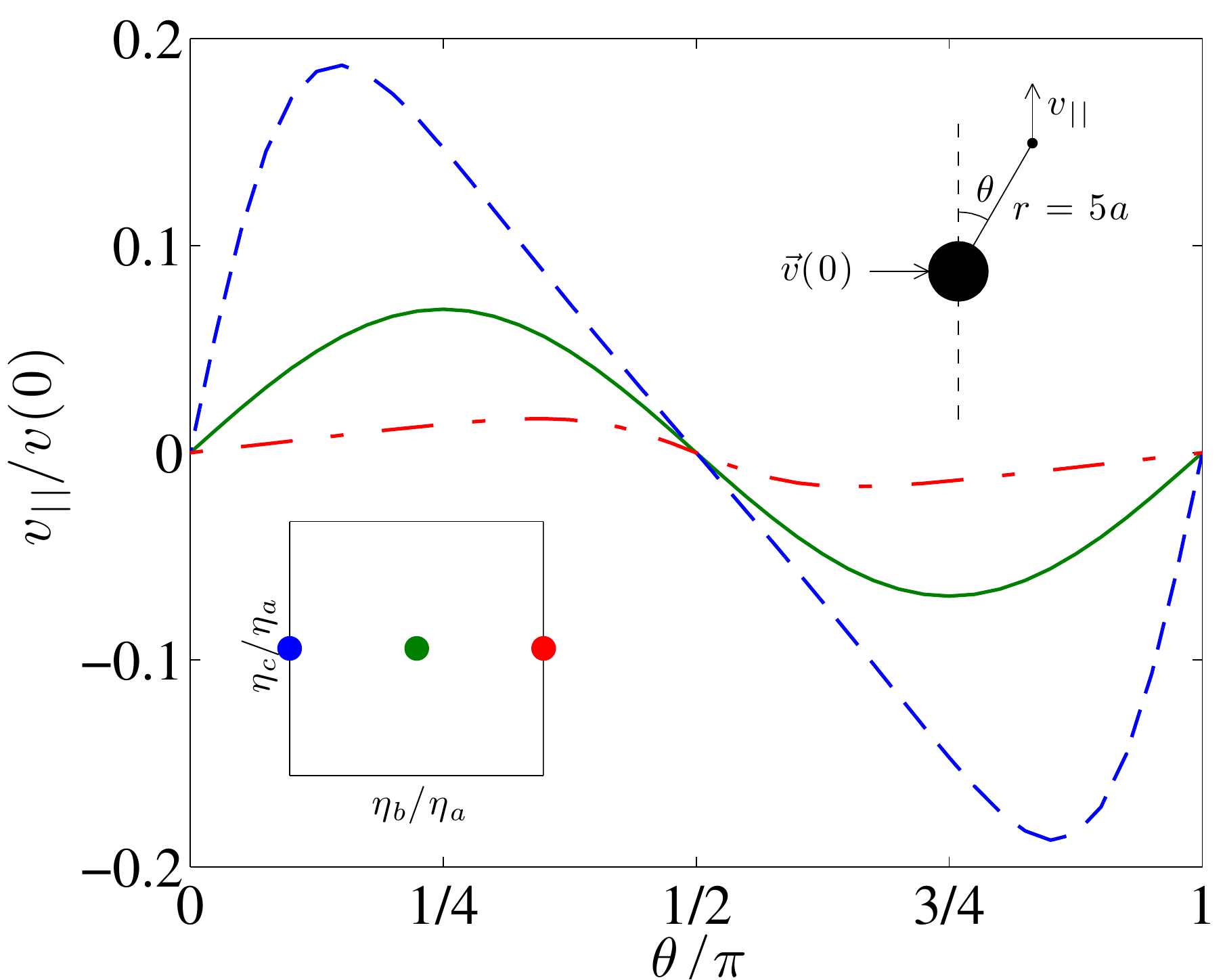}
\mylab{-0.3\textwidth}{0.275\textwidth}{(\eee)}%
\includegraphics*[width=0.329\columnwidth,keepaspectratio]{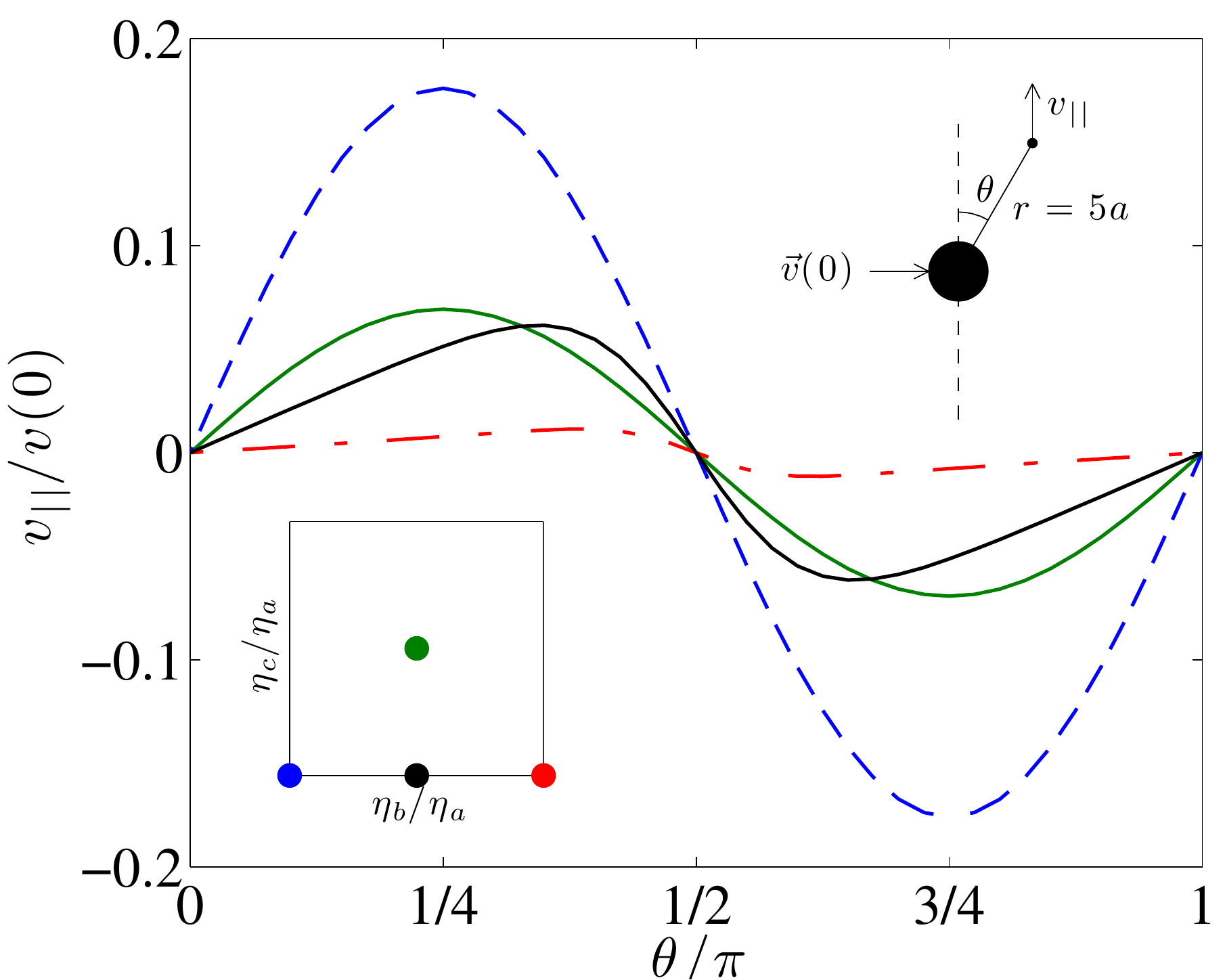}
\mylab{0.71\textwidth}{0.30\textwidth}{(\fff)}
\caption{
Flow velocity elicited by a sphere moving perpendicular to the nematic
director.
The velocities are calculated on the meridional plane $\phi=0$ at a distance
$r=5a$ away from the sphere centre, and represented as a function of the
inclination angle $\theta$:
(\aaa)-(\ccc), $v_{\perp,||}$; (\ddd)-(\fff), $v_{||}$.
(\aaa) and (\ddd), $\eta_a = \eta_c/10$; (\bbb) and (\eee), $\eta_a = \eta_c$;
(\fff) and (\ggg), $\eta_a = 10 \eta_c$.
Each line type corresponds to a different value of $\eta_b / \eta_c$, as
indicated in the inset maps:
\textcolor{blue}{\dashed}, $\eta_b = \eta_c/10$; \textcolor{black}{\solid},
$\eta_b = \eta_c$; \textcolor{red}{\chndot}, $\eta_b = 10 \eta_c$.  The
isotropic case is also plotted for reference (\textcolor{verde}{\solid}).
				 } 
\label{fig:veloperz}
%
%
\vspace{0.5cm}
%
\includegraphics*[width=0.329\columnwidth,keepaspectratio]{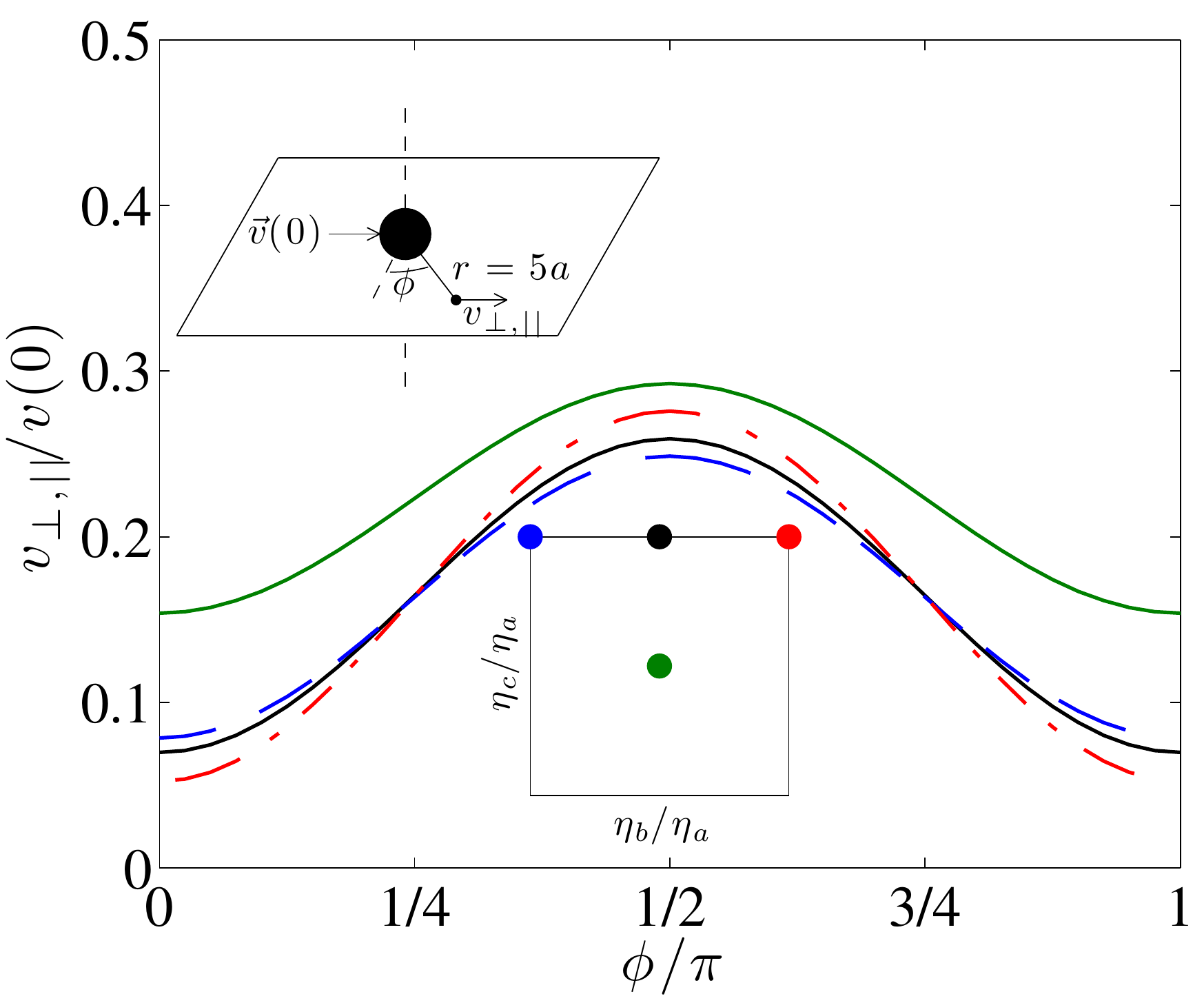}
\mylab{-0.3\textwidth}{0.275\textwidth}{(\aaa)}%
\includegraphics*[width=0.329\columnwidth,keepaspectratio]{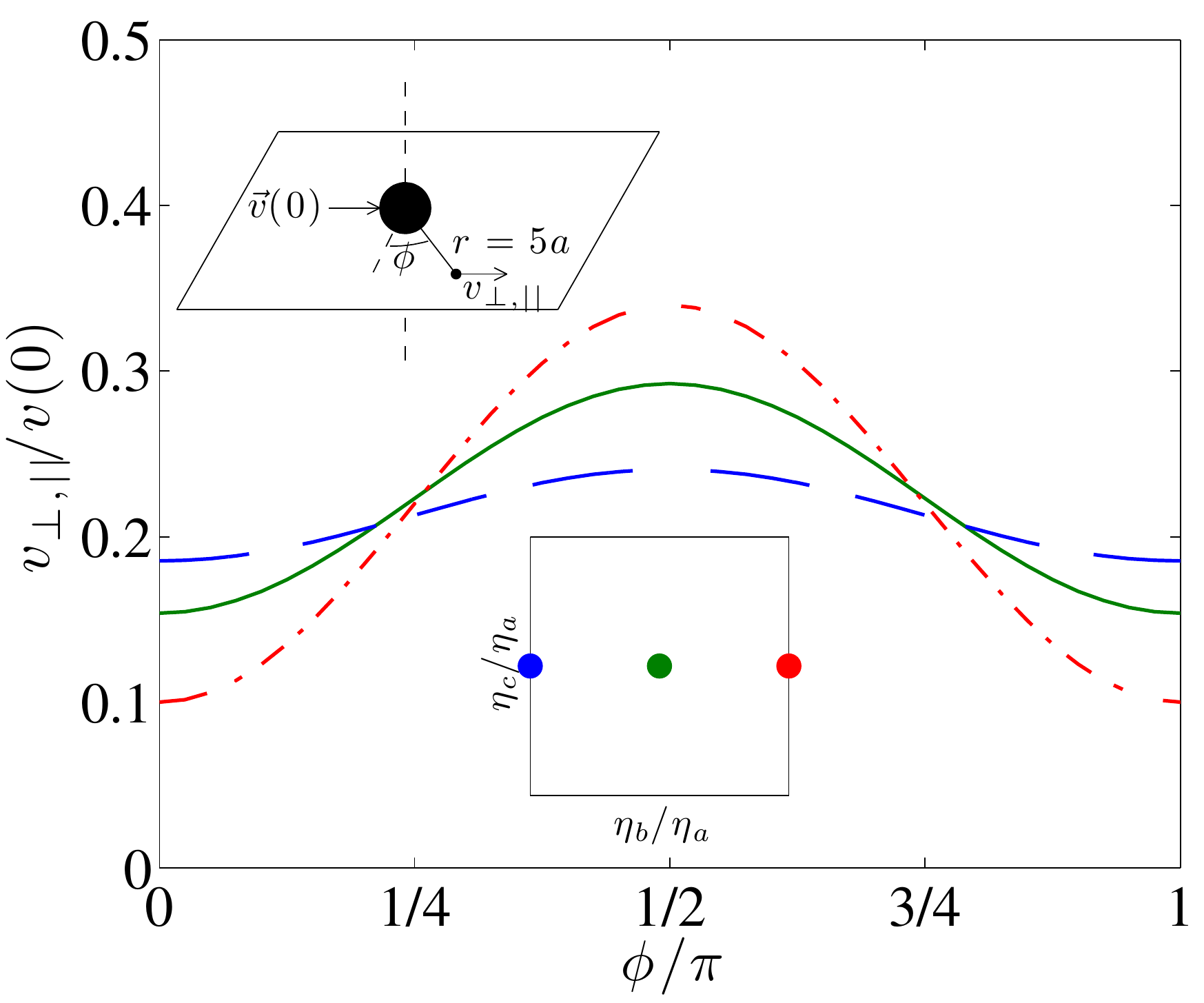}
\mylab{-0.3\textwidth}{0.275\textwidth}{(\bbb)}%
\includegraphics*[width=0.329\columnwidth,keepaspectratio]{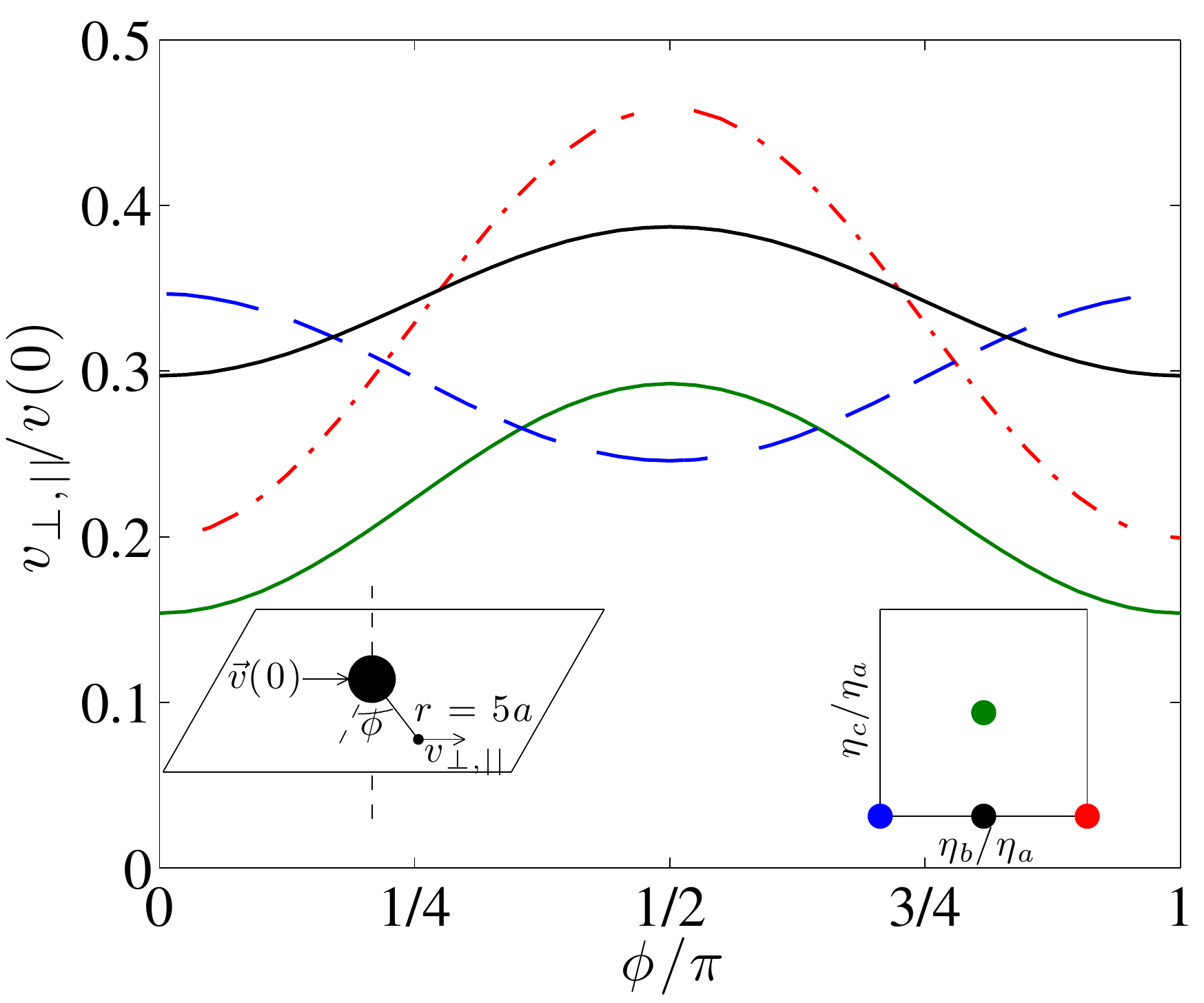}
\mylab{0.71\textwidth}{0.30\textwidth}{(\ccc)}%
\\
\includegraphics*[width=0.329\columnwidth,keepaspectratio]{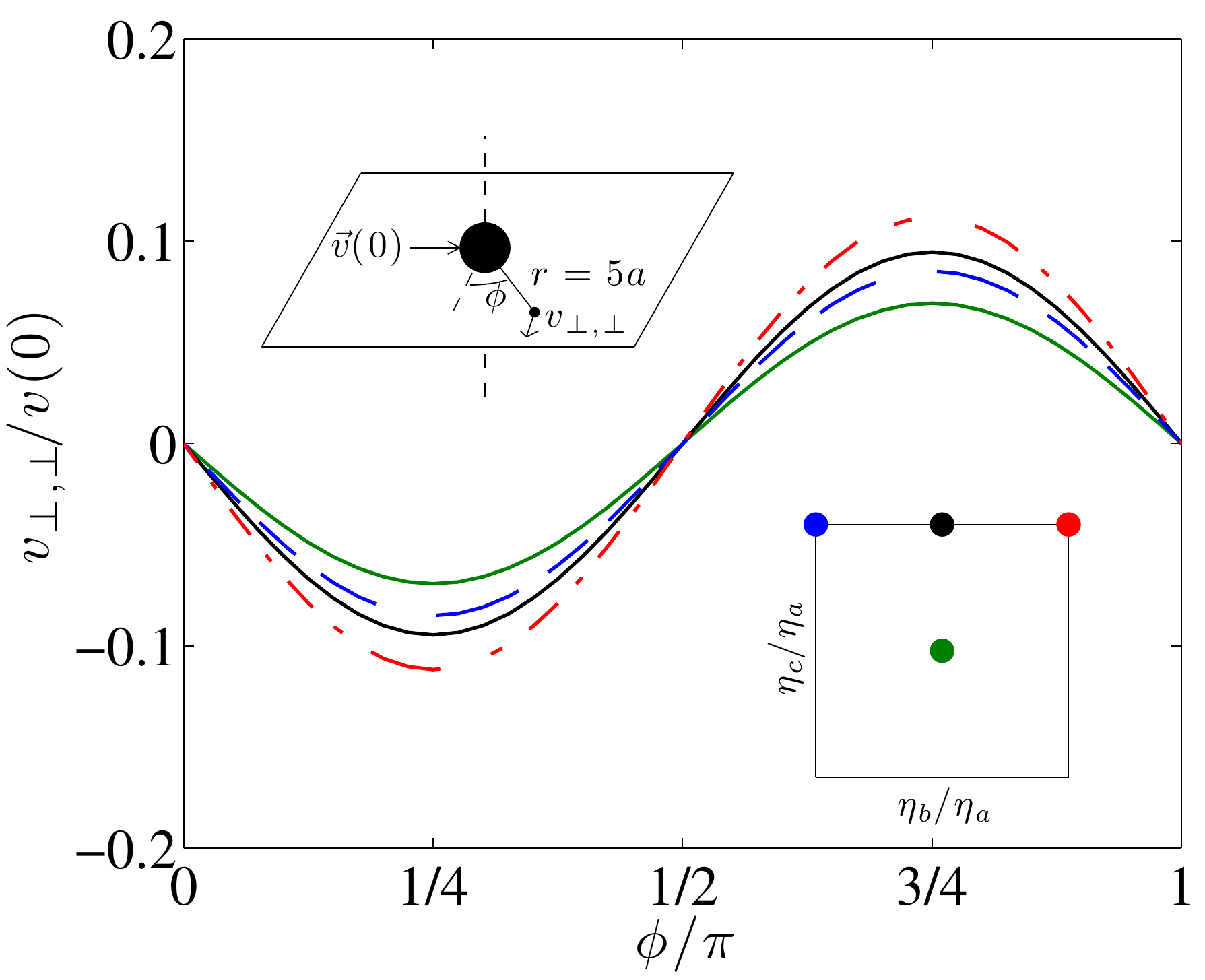}
\mylab{-0.3\textwidth}{0.275\textwidth}{(\ddd)}%
\includegraphics*[width=0.329\columnwidth,keepaspectratio]{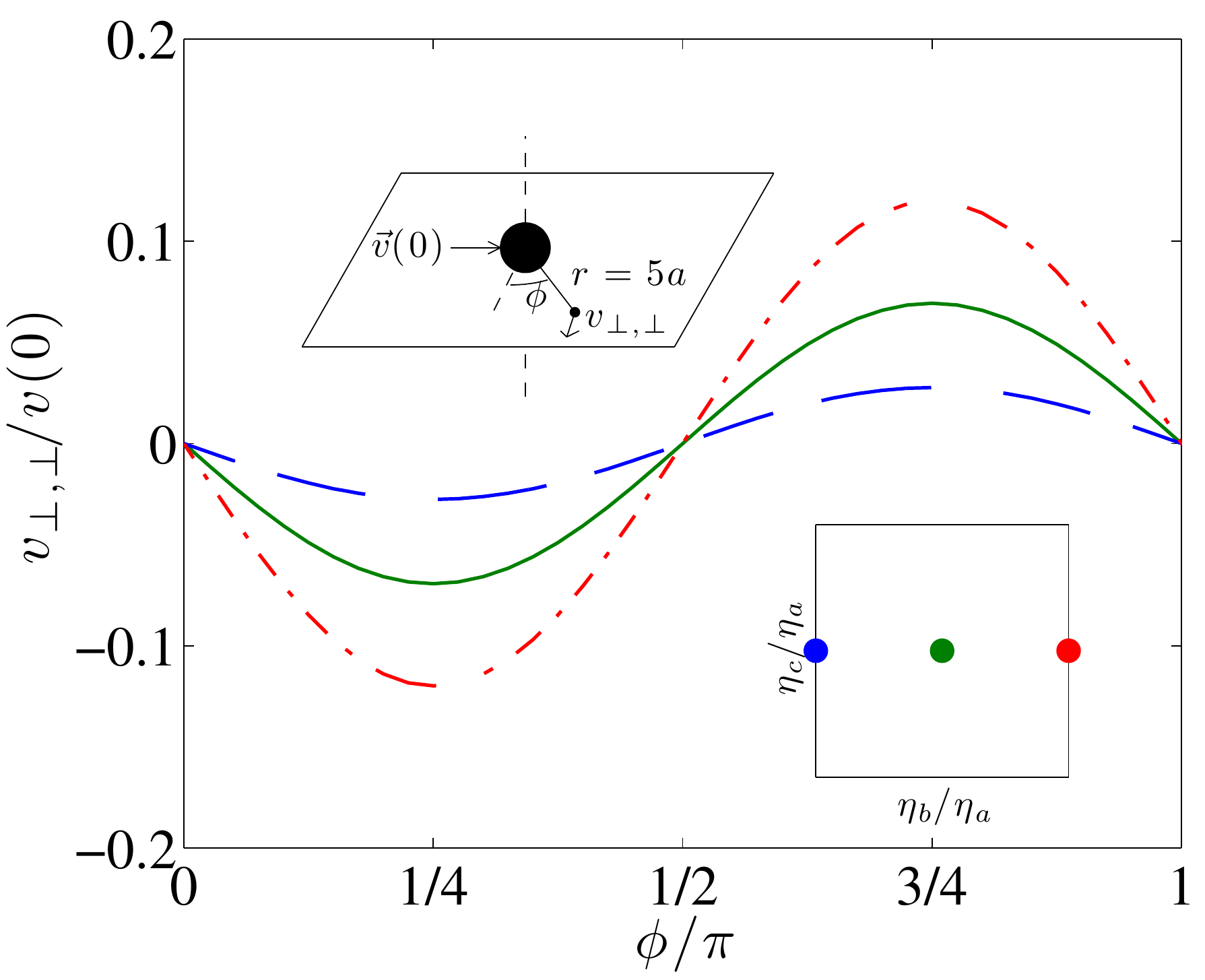}
\mylab{-0.3\textwidth}{0.275\textwidth}{(\eee)}%
\includegraphics*[width=0.329\columnwidth,keepaspectratio]{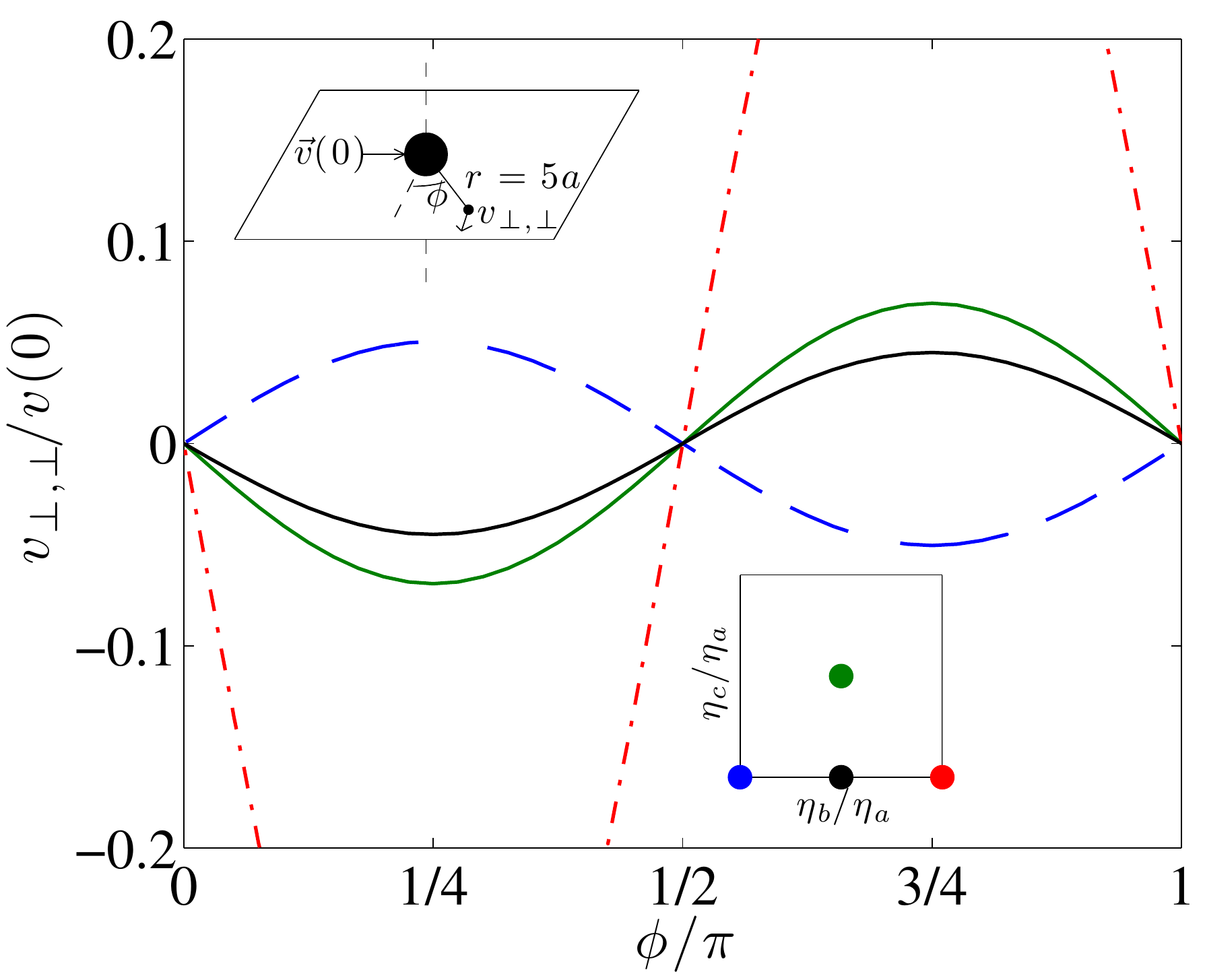}
\mylab{0.71\textwidth}{0.30\textwidth}{(\fff)}%
\caption{
Same as Fig. \ref{fig:veloperz} for the equatorial plane $\theta=0$.
Velocities are represented as a function of the azimuth angle $\phi$:
(\aaa)-(\ccc), $v_{\perp,||}$; (\ddd)-(\fff), $v_{\perp,\perp}$.
}
\label{fig:veloperx}
%
%
\end{figure*}

Figures \ref{fig:veloperz} and \ref{fig:veloperx} display angular profiles of
the velocity at a constant distance $r=5a$ from the origin, similar to figure
\ref{fig:velopar}. To characterise the spatial organization of the velocity
field in the present three-dimensional configuration, we plot data on two
planes. 
The first is the meridional plane $\phi=\pi/2$ (figure \ref{fig:veloperz})
and the second is the equatorial plane $\theta = \pi/2$ (figure 
\ref{fig:veloperx}).
These planes intersect along the west-east axis of translation of the sphere.
On them, $v_{||}$ is the velocity component parallel to the nematic director
(\ie northward), whereas $v_{\perp,||}$ is perpendicular to the director and
parallel to the sphere's trajectory (\ie eastward).  The remaining velocity
component, $v_{\perp,\perp}$, is perpendicular to both the director and the
sphere's trajectory.

Figure \ref{fig:lindecor_per} demonstrates the effect of anisotropy on the
streamline pattern. 
The streamlines in each panel are obtained by integrating the far velocity
field relative to the sphere from a common initial point in the far east,
upstream of the sphere. 
They are represented on the same meridional and equatorial planes used in
figure \ref{fig:veloperz} and \ref{fig:veloperx}. Apart from allowing direct
comparison between velocity profiles and streamlines, the choice of these two
planes has the additional advantage that the local cross-plane flow velocity is
zero, so that the streamlines remain in-plane.  The same is not true for other
initial points of integration given the three-dimensional nature of the flow
when $\vec v_0 \perp \vec n$.

\subsubsection{Case $\eta_a = \eta_c$}

A first look into figures \ref{fig:veloperz} and \ref{fig:veloperx} indicates
that, for $\eta_a = \eta_c$, the velocity profiles are qualitatively similar to
the isotropic condition (panels \bbb\, and \eee). 
The reason for this behavior is that the bending stresses in the equations of
fluid motion are zero for $\eta_a = \eta_c$, and the flow becomes rotationally
pseudo-isotropic with $\eta_b$ and $\eta_c$ acting respectively as $\eta_{||}$
and $\eta_\perp$ (see equation \ref{eq:rotiso}).
Comparing figures \ref{fig:velopar} and \ref{fig:lindecor_par} with panels
(\bbb),\,(\eee) of figures \ref{fig:veloperz} and \ref{fig:veloperx} and figure
\ref{fig:lindecor_per}(\bbb) suggests that anisotropy influences the flow
organization more deeply in the present rotationally pseudo-isotropic
configuration than in the $\vec v_0 \, ||\, \vec n$ case, consistent with the
fact that the bending stresses are not zero in that configuration.

The response function of a sphere in the case $\eta_a = \eta_c$ is
characterized in \S \ref{subsec:etaaeqetac} above.  Here we focus on the effect
of the ratio $\eta_b / \eta_a$ on the flow organization.  
We find that $v_{\perp,||}$ increases with $\eta_b / \eta_a$ on the axis of
translation of the sphere, $\theta = \pi/2$, whereas it decreases or varies
little on the perpendicular axes (figures \ref{fig:veloperz}{\bbb} and
\ref{fig:veloperx}\bbb ).  Thus, the profile of this velocity component becomes
sharper as $\eta_b / \eta_a$ increases.
More importantly, the magnitude of $v_{||}$ decreases with $\eta_b / \eta_a$ on
the meridional plane (figure \ref{fig:veloperz}\eee), while the magnitude of
$v_{\perp,\perp}$ increases on the equatorial plane (figure
\ref{fig:veloperx}\eee).
Overall, these velocity profiles suggest that the motion of the sphere leads to
an asymmetric deflection of the fluid particles that is larger in the plane of
lower viscosity and vice versa. 
This behavior can be easily appreciated by looking at the streamline pattern in
figure \ref{fig:lindecor_per}(\bbb).
Note also that these streamlines resemble the streamlines in the flow of
isotropic fluid around a prolate ellipsoid, consistent with the results in \S
\ref{subsec:etaaeqetac}.

\subsubsection{Case $\eta_c \gg \eta_a$}

\begin{figure*}
\vspace{0.5cm}
\hspace{-0.5cm}
\includegraphics*[width=0.36\columnwidth,keepaspectratio]{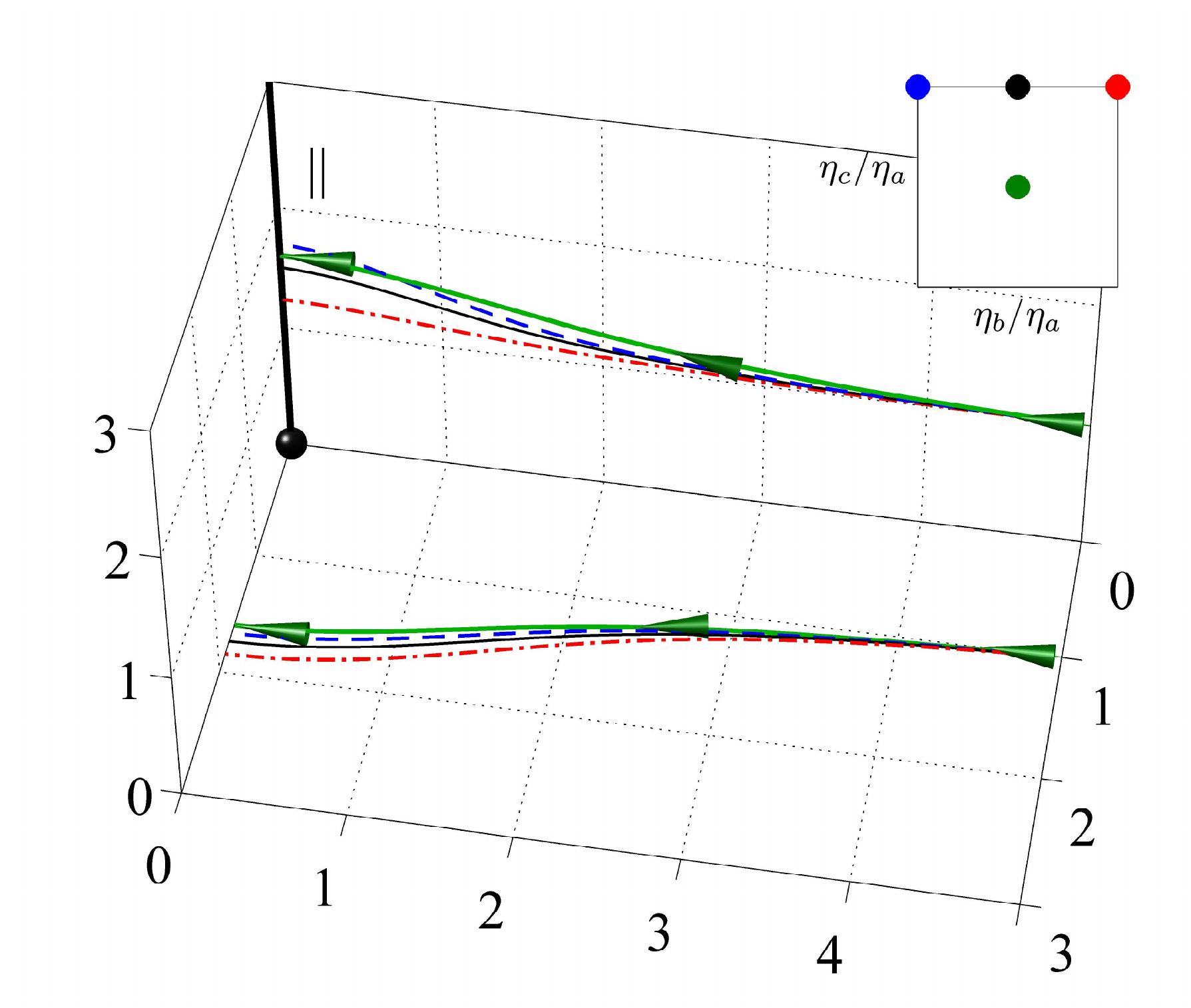}
\mylab{-0.3\textwidth}{0.3\textwidth}{(\aaa)}%
\hspace{-0.5cm}
\includegraphics*[width=0.36\columnwidth,keepaspectratio]{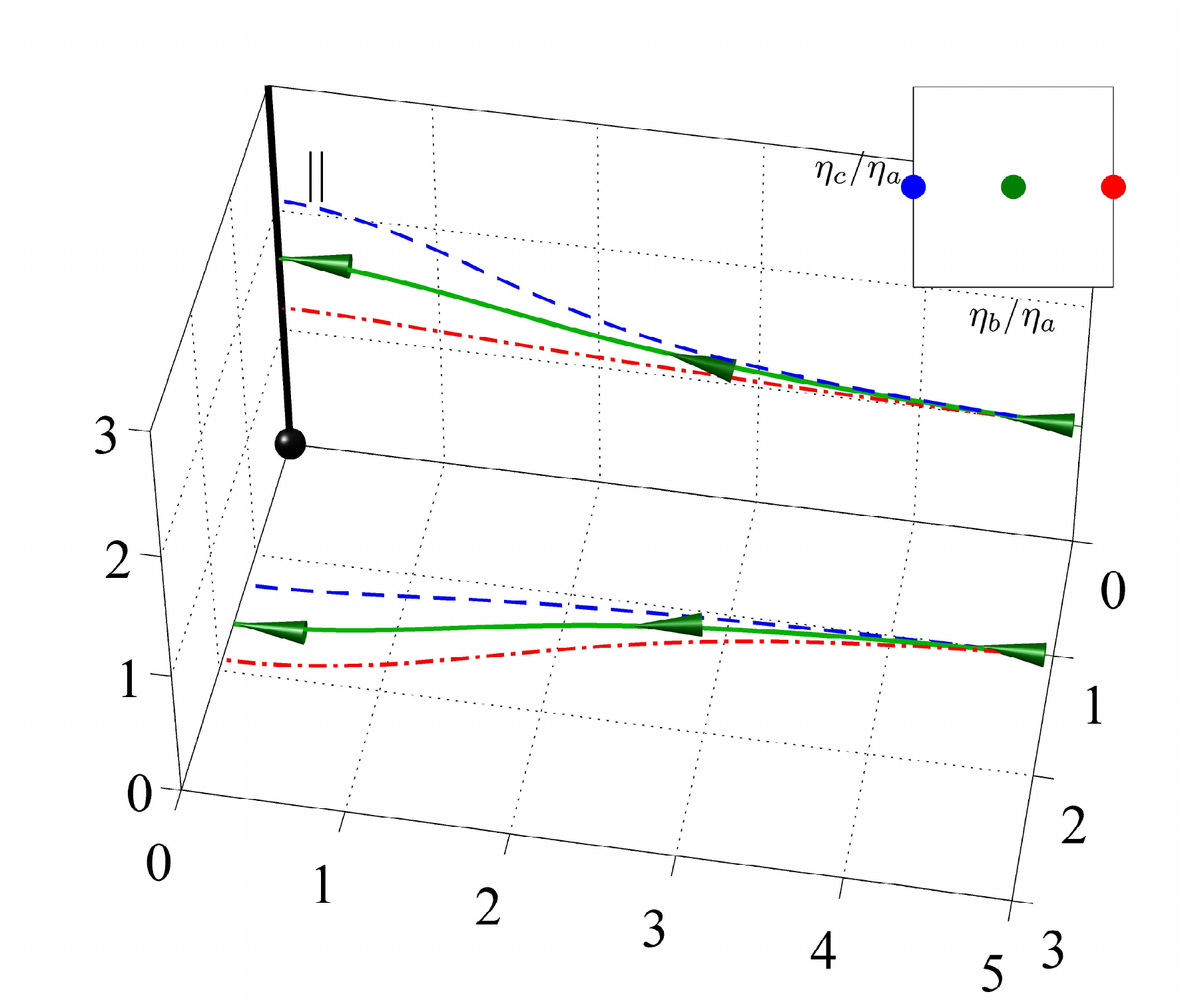}
\mylab{-0.3\textwidth}{0.3\textwidth}{(\bbb)}%
\hspace{-0.5cm}
\includegraphics*[width=0.36\columnwidth,keepaspectratio]{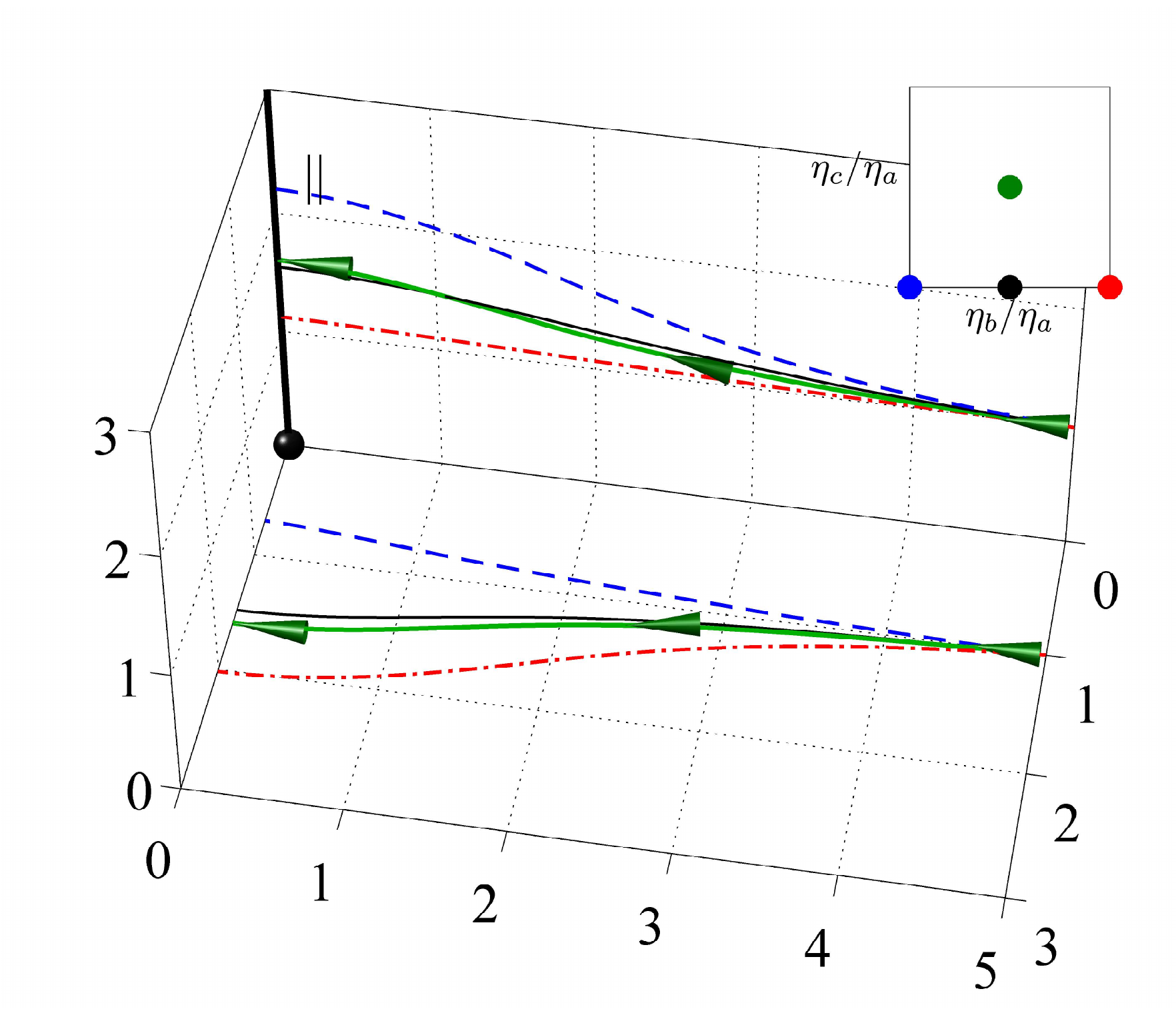}
\mylab{0.7\textwidth}{0.33\textwidth}{(\ccc)}%
\\
\caption{
Streamlines on the meridional and equatorial planes of a sphere moving
perpendicular to the nematic director (thick black vertical axis).  
The arrows indicate the flow direction in a reference frame moving with the
sphere.
The spatial coordinates are normalized with the sphere radius.
Each panel corresponds to one value of the first viscosity ratio:
(\aaa) $\eta_a = \eta_c/10$; (\bbb) $\eta_a = \eta_c$; (\ccc) $\eta_a = 10
\eta_c$.
Each line type corresponds to a different value of $\eta_b / \eta_c$, as
indicated in the inset map at the upper right corner of each figure:
\textcolor{blue}{\dashed}, $\eta_b = \eta_c/10$; \textcolor{black}{\solid},
$\eta_b = \eta_c$; \textcolor{red}{\chndot}, $\eta_b = 10 \eta_c$.  The
isotropic case is also plotted for reference (\textcolor{verde}{\solid}).
				 } \label{fig:lindecor_per}
%
\end{figure*}

For large values of $\eta_c / \eta_a$, the viscous bending stresses become
important in the equations of motion
\eqref{eq:fluid_equations_x}-\eqref{eq:fluid_equations_z}, causing changes in
the anisotropic velocity field that depend little on the other viscosity ratio,
$\eta_b / \eta_a$.   
This result is consistent with the fact that $\zeta_{\perp}/(a\eta_a)$ is
almost independent of $\eta_b / \eta_a$ in the limit $\eta_c \gg \eta_a$ (see
fig. \ref{fig:contour_zeta_22}\aaa). 
The largest variations in the velocity profiles occur on the meridional plane,
where the east-west velocity component $v_{\perp,||}$ increases substantially
near the polar axis (figure \ref{fig:veloperz}\aaa).  
This is also the region where the anisotropic stresses are maximum (data not
shown).
The northward velocity $v_{||}$ behaves qualitatively similar to the $\eta_c =
\eta_a$ case. It experiences a moderate increase in magnitude and shifts
towards the polar axis as $\eta_b/\eta_a$ decreases (figure
\ref{fig:veloperz}\ddd). 
As a consequence, the streamlines on the meridional plane are deflected more as
$\eta_b / \eta_a$ decreases, and this deflection occurs closer to the location
of the sphere (figure \ref{fig:lindecor_per}\aaa).

The flow in the equatorial plane is even less sensitive to the ratio $\eta_b /
\eta_a$, remaining very close to the isotropic case (figures
\ref{fig:veloperx}\aaa,\ddd).  In particular, the streamlines on this plane are
almost identical to the isotropic ones, with a slight increase in deflection
with $\eta_b / \eta_a$ (figure \ref{fig:lindecor_per}\aaa).

\subsubsection{Case $\eta_c \ll \eta_a$}

Similar to the rotationally pseudo-isotropic condition, the flows with $\eta_c
\ll \eta_a$ are rather sensitive to the ratio of momentum diffusivities in the
$||$ and $\perp$ directions, $\eta_b /\eta_a$.
For large values of this ratio, $v_{||}$ decreases to virtually zero on the
meridional plane (figure \ref{fig:veloperz}\fff), where the viscosity is higher
($\eta_b$). Conversely, $v_{\perp,\perp}$ increases significantly on the
equatorial plane (figure \ref{fig:veloperx}\fff), where the viscosity is lower
($\eta_a$).
Accordingly, the flow streamlines remain almost straight on the meridional
plane whereas they are highly deflected away from the sphere on the equatorial
plane (fig. \ref{fig:lindecor_per}\ccc).

The velocity field corresponding to $\eta_c \ll \eta_a = \eta_b$ is comparable
to the isotropic velocity field.  The only remarkable difference is that the
profile of $v_{\perp,||}$ is sharper than the isotropic one on the meridional
plane and flatter on the equatorial plane (figures
\ref{fig:veloperz}{\ccc} and \ref{fig:veloperx}\ccc).
Apart from that, anisotropy affects little the velocity components
perpendicular to axis of translation of the sphere in this case ($v_{\perp,||}$
and $v_{||}$, figures \ref{fig:veloperz}{\fff} and \ref{fig:veloperx}\fff), and 
the streamline pattern remains similar to the isotropic one (figure
\ref{fig:lindecor_per}\ccc).

Finally, we consider the flow elicited by a sphere moving perpendicular to the
nematic director in a fluid with $\eta_c \ll \eta_b \ll \eta_a$.  
As expected, $v_{||}$ significantly increases in the meridional plane because
of the $\eta_b \ll \eta_a$ condition (figure \ref{fig:veloperz}\fff), and the
flow streamlines are highly deflected away from the sphere on this plane
(figure \ref{fig:lindecor_per}\ccc).
However, this flow has a remarkable characteristic which separates it from all
other cases studied above, including the isotropic one; its streamlines
converge towards the sphere on the equatorial plane (figure
\ref{fig:lindecor_per}\ccc).
Consequently, the sign of $v_{\perp,\perp}$ on this plane is opposite to that
of all of the other flow conditions (figure \ref{fig:veloperx}\fff).
This somewhat counterintuitive behavior can be partially understood by
inspecting the balance of the different terms in the vorticity equation
\eqref{eq:vorti} at constant $\eta_b / \eta_a$ and varying $\eta_c / \eta_a$
(not shown).
It is found that the angular profiles of the torques coming from the anisotropy
of the stress-strain relation remain similar when varying $\eta_c / \eta_a$.
However, the torques coming from the bending of the fluid with respect to the
nematic change sign at $\eta_c = \eta_a$, attenuating the effect of the
stress-strain anisotropy for $\eta_c> \eta_a$ and magnifying it for $\eta_c <
\eta_a$. 

\subsubsection{A remark on $\eta_b / \eta_a$ dependence of the flow}
%
It should be noted that the above classification in terms of $\eta_c / \eta_a$
is exclusively intended to facilitate the presentation of the nine flows 
obtained by combining $\eta_c / \eta_a = 0.1,\, 1,\, 10$ and 
$\eta_b / \eta_a = 0.1,\, 1,\, 10$ in the $\vec v_0 \, \perp \, \vec n$ 
configuration.
Likewise, one may choose to classify these flows in relation to the order of
magnitude of $\eta_b / \eta_a$.  Such classification is indeed implicit in
figures \ref{fig:veloperz} to \ref{fig:lindecor_per}, where the strain
pseudo-isotropic condition is depicted with solid (black) lines, whereas the
$\eta_b / \eta_a \ll 1$ and $\eta_b / \eta_a \gg 1$ cases are depicted
respectively by dashed (blue) and chain-dotted (red) lines.
In broad strokes, although the general shape of the velocity profiles is mainly
dependent on $\eta_c / \eta_a$, the velocity magnitudes in the direction
perpendicular to the sphere's motion and, therefore, the asymmetry in
streamline deflection, are mostly dependent on $\eta_b / \eta_a$.

\begin{figure*}
\vspace{0.5cm}
\includegraphics*[width=0.95\textwidth,keepaspectratio]{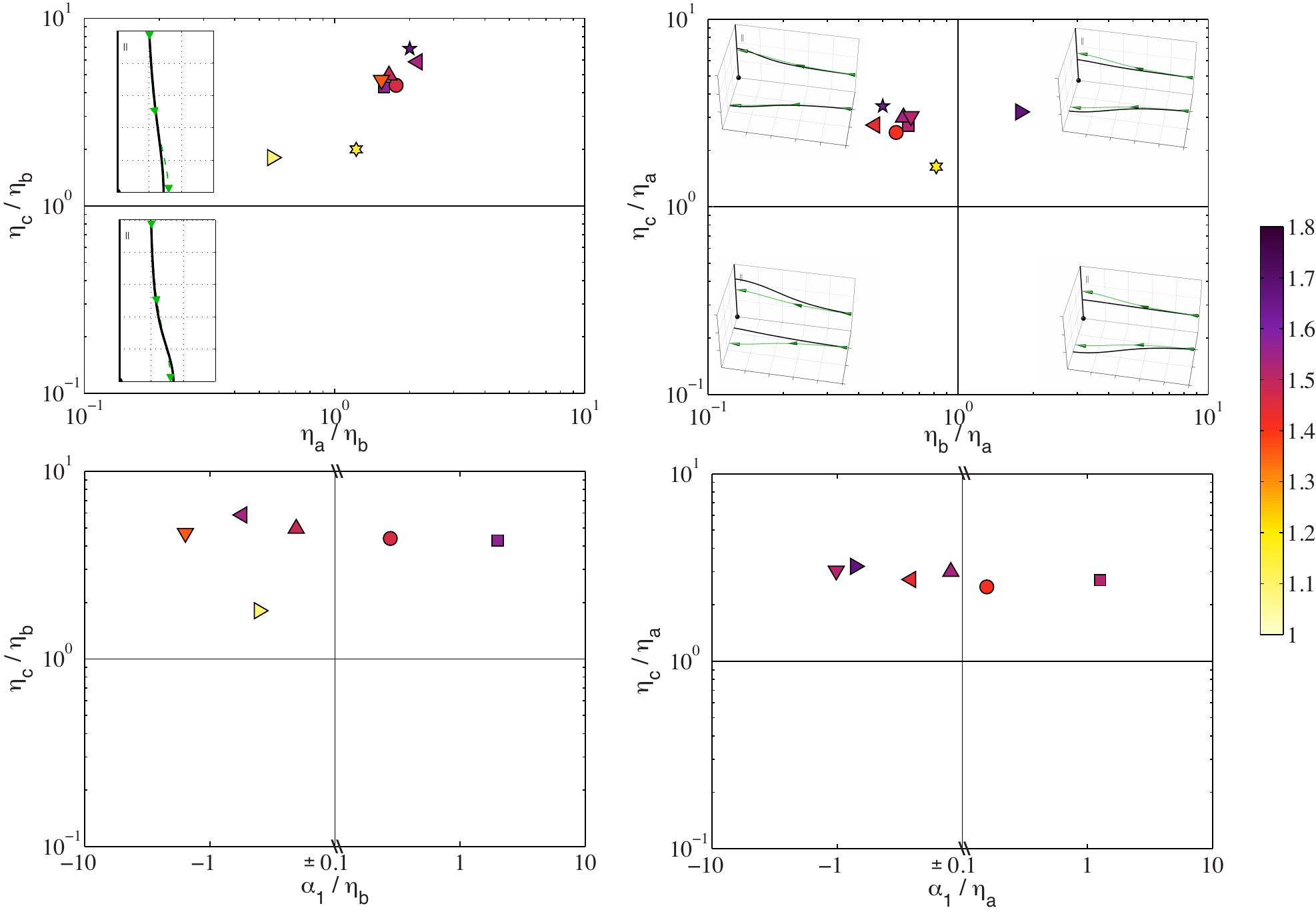}
\mylab{-0.88\textwidth}{0.67\textwidth}{(\aaa)}%
\mylab{-0.43\textwidth}{0.67\textwidth}{(\bbb)}%
\mylab{-0.88\textwidth}{0.333\textwidth}{(\ccc)}%
\mylab{-0.43\textwidth}{0.333\textwidth}{(\ddd)}%
	\caption{ 
Survey of Miesowicz viscosity coefficients reported for nematic liquid crystals
in the literature, plotted in the scaling of figures
\ref{fig:contour_zeta_11} and \ref{fig:contour_zeta_22}:
(\aaa, \ccc) $\zeta_{||}/6\pi a \eta_b$; (\bbb, \ddd) 
$\zeta_{\perp}/6\pi a \eta_a$.
Each symbol represents a different compound and has been coloured according to 
the corresponding value of the response function as predicted by equations
\eqref{eq:response_function_11_analytic}-\eqref{eq:response_function_22_analytic},
and according to the colourbar on the right hand side of the figure:
\squar, PPA \citep{Tseng.Silver.Finlayson.1972, Orsay.1971, Miesowicz.1946}; 
\circle, MBBA \citep{Gahwiller.1971, Langevin.1972};
\trian, 5CB \citep{Skarp.Lagerwall.Stebler.1980,
Herba.Szymanski.Drzymala.1985};
\dtrian, MBPP \citep{Herba.Szymanski.Drzymala.1985};
\leftrian, E7 \citep{Wang.Wu.Gauza.Wu.Wu.2006};
\rightrian, UCF-02 \citep{Wang.Wu.Gauza.Wu.Wu.2006};
$\bigstar$, N4 \citep{Beens.deJeu.1983};
$\ast$, 8OCB \citep{Janik.Moscicki.Czuprynski.Dabrowski.1998}. 
(\aaa, \bbb) also include the representative streamline pattern
corresponding to each region of the parameter space.  For reference, the
isotropic streamlines are also included in green.
				 } \label{fig:figure_summary}
%
\end{figure*}

\section{Conclusions}
%
\subsection{\textcolor{black}{Summary of findings}}
One of the greatest challenges that hinders the application of PTM to complex
anisotropic fluids is the lack of knowledge about the dynamics of the probing
particle. 
This paper studies the creeping flow generated by the motion of a point
particle in a nematic incompressible fluid defined by the Leslie-Ericksen
constitutive relation.
The equations governing this flow are generally applicable to a broad range of
nematic systems, ranging from liquid crystals to aligned biopolymer networks
that are strongly coupled to their solvent, and nematic visco-elastomers.

The response function $\overline{\overline{\zeta}}$ of a sphere in a nematic
fluid is a $3 \times 3$ tensor that provides the drag force $\vec F$
experienced by the particle as 
$$
\vec F = \overline{\overline{\zeta}} \cdot \vec v_0,
$$
where $\vec v_0$ is the velocity of the particle.  This tensor only has two
independent values, $\zeta_{||}$ and $\zeta_{\perp}$, which respectively
represent the resistance of the fluid to the motion of the particle in the
directions parallel and perpendicular to the nematic director, $\vec n$
\citep{De_Gennes.1993}.
In this work, we provide explicit analytical expressions for $\zeta_{||}$ and
$\zeta_{\perp}$, \ie the equivalent of Stokes' drag formula for a nematic
fluid, by performing a multipole expansion in Fourier space.  
These formulae depend on up to four viscosity coefficients: the three Miesowicz
viscosities $\eta_a$,  $\eta_b$ and  $\eta_c$,  and the first Leslie viscosity
$\alpha_1$.
Our solution compares well with the results from previous numerical simulations
\citep{Heuer.Kneppe.Schneider.1992} in the limited region of the parameter
space where the simulation data are available.
Similar to \cite{Kneppe.Schneider.Schwesinger.1991}, we observe that the flow
caused by a particle moving parallel to the nematic ($\vec v_0 || \vec n$) is
axially symmetric and $\zeta_{||}$ is independent of the first Miesowicz
viscosity.  
On the other hand, the flow originated by a particle moving perpendicular to
the nematic ($\vec v_0 \perp \vec n$) is three-dimensional and $\zeta_{\perp}$
depends on all four viscosity coefficients.
It is important to note, however, that the $\vec v_0 || \vec n$ flow is
independent of $\eta_a$ only if the velocities of the network and the solvent
are divergence-free.
This requirement is only satisfied when the viscous coupling between the
network and the solvent is strong and the network velocity matches that of the
incompressible solvent.

In a nematic two-fluid gel, anisotropy can present itself through the viscous
coupling between the network and the solvent, as well as through the network's
constitutive equations.
The anisotropy arising from the coupling has been illustrated by considering an
idealized network formed by a square array of circular cylinders, which opposes
twice as much resistance to the relative motion of the solvent in the $\perp$
direction than in the $||$ direction.  Further analysis of the effect of
anisotropic coupling will require relaxing the strong coupling assumption and
is beyond the scope of this study.
The anisotropy of the Leslie-Ericksen constitutive relation can, however, be
thoroughly studied in the present formulation.  We find that the response
function is affected mainly by the anisotropic diffusion of momentum along
different directions (\ie anisotropy in the stress-strain relationship), and by
the resistance of the fluid to bending.
Analysis of the equations of fluid motion indicates that momentum diffusivity
is anisotropic when $\eta_a \ne \eta_b$ and that the fluid opposes resistance
to bending when $\eta_a \ne \eta_c$.
We have studied these two mechanisms independently from each other by examining
the response function in a rotationally pseudo-isotropic fluid where the 
bending stresses are zero and in a strain pseudo-isotropic fluid where the 
stress-strain relation is isotropic.
The first of these fluids is governed by a pseudo-Stokes equation with
different viscosities $\eta_{||}=\eta_b$ and $\eta_\perp=\eta_a$ in the
directions parallel and perpendicular to the nematic director.
The second fluid is characterized by an isotropic momentum diffusivity
$\eta=\eta_a$ plus a resistance to bending $\Delta \eta=\eta_c-\eta_a$.
A comparative analysis of these two ``pure'' conditions indicates that
anisotropic momentum diffusivity leads to greater changes in $\zeta_{||}$,
whereas $\zeta_\perp$ is more influenced by the bending resistance of the
fluid.

It is also found that directional momentum diffusivity has a limited effect on
the response function in rotationally pseudo-isotropic fluids, so that
$\zeta_{||}/\eta_{||}$ and $\zeta_{\perp}/ \eta_{\perp}$ reach constant
asymptotic values for high levels of anisotropy with the only exception of
$\zeta_{||}/\eta_{||} (\eta_{||} \ll \eta_\perp)$.  
In fluids with zero resistance to bending,
this result provides some level of justification for the qualitative usage of
effective directional viscosities, which are defined by applying Stokes's drag
law separately along the $||$ and $\perp$ directions.
However, it is important to note that the thus estimated effective viscosities
differ from the actual viscosities of the fluid by over one order of magnitude
for moderate levels of anisotropy like those observed in reconstituted
biopolymer networks and the cytoskeleton of cultured animal cells.
Furthermore, the very notion of directional effective viscosities may be
misleading in fluids with an appreciable resistance to bending because the
effect of $\Delta \eta$ is felt in both components of the response function for
all levels of anisotropy.

The present study is finalized by a parametric description of the far velocity
field around a spherical particle under different anisotropy conditions.  For
simplicity and due to the observed small influence of the first Leslie
viscosity on the response function, the analysis is restricted to zero values
of this coefficient ($\alpha_1 =0$).
When the particle moves parallel to the nematic director, the axial symmetry of
the flow establishes a constraint between the bending stresses in the fluid and
the directional diffusion of momentum, which cannot vary independently of each
other in this flow configuration.
As a consequence, the flow pattern varies little with respect to the isotropic
case although the level of streamline deflection increases when $\eta_b =
\eta_{||} > \eta_\perp = \eta_c$ and vice versa.
When the sphere moves perpendicular to the nematic director, the flow is
three-dimensional and the directional diffusion of momentum can be varied
independently of the bending resistance of the fluid through the viscosity
ratios $\eta_b/\eta_a$ and $\eta_c/\eta_a$.  
When $\eta_c/ \eta_a \gg 1$, the fluid opposes high resistance to bending and
the flow differs little from the isotropic case, similar to the $\vec v_0 ||
\vec n$ flow.
However, when $\eta_c / \eta_a$ is of order unity, anisotropic momentum
diffusivity can lead to large changes in the flow structure which essentially
consist of asymmetric streamline deflection; for $\eta_b > \eta_a$, the
streamlines are deflected more in the equatorial plane than in the meridional
ones, and the opposite happens for $\eta_b < \eta_a$.
Intriguingly, when $\eta_c / \eta_a \ll 1$, the resistance to bending enhances
the flow asymmetry caused by anisotropic momentum diffusivity rather than
reducing it, so much as to cause the streamlines to converge towards the sphere
in the equatorial plane when $\eta_b \ll \eta_a$.

The parametric study of anisotropic viscosity effects on the flow of a nematic
fluid around a sphere would be incomplete without a survey of the experimental
values of the viscosity coefficients reported in the literature.
Such a survey is useful in determining which of the conditions explored in this
paper are more likely to be observed experimentally and, at the same time, it
can offer new insight about the dynamics of those fluids.
Unfortunately, there is little information about the anisotropic rheology of
nematic biopolymer gels, and most of this information comes from effective
viscosity approaches that suffer from the limitations described above.
Our analysis suggests that these biopolymers are relatively close to a
rotationally pseudo-isotropic fluid with $0.5 \lesssim \eta_{\perp}/\eta_{||}
\lesssim 100$ but these values should only be taken as rough estimations.
For nematic liquid crystals, the situation is better and we find values of the
Miesowicz viscosities for eight different compounds.  This information is
summarized in figure \ref{fig:figure_summary} using the same scaling as figures
\ref{fig:contour_zeta_11} and \ref{fig:contour_zeta_22}.
The vast majority of the compounds have $\eta_c / \eta_a \approx 3$ and 
$0.5 \lesssim \eta_b/\eta_a \lesssim 0.8$, which implies that their bending
resistance is relatively high and that, in contrast to biopolymer networks,
nematic liquid crystals can be considered to behave somewhat closer to a strain
pseudo-isotropic fluid than to a rotationally pseudo-isotropic fluid.
Interestingly, all of the liquid crystals considered are in the range where
$\alpha_1$ has a small effect on the flow.

Although the main motivation for this study is to provide fundamental
understanding of the dynamics of a microrheology probe in nematic fluids, many
of the results presented here have broader applications.
Modelling the motility of spermatozoa in the cervical mucus is just one example
of such applications.  This mucus varies through different phases along the
ovulatory cycle, including one in which the mucin network aligns parallel to
the direction of the reproductory tract thereby favouring the directed
migration of spermatozoa \citep{Chretien:2003}.
This anisotropic synergy could be studied in more detail by modeling each
swimmer as a superposition of regularized point forces using the anisotropic
Green's functions derived in this study. \\

\subsection{\textcolor{black}{Model limitations}}

\textcolor{black} {
The present formulation is linear, incompressible and assumes affine 
deformations. Thus, the results in this paper are valid only for small 
deformations and do not apply to regimes displaying strain-stiffening, 
negative normal stresses and other non-linear effects.}

A potential limitation of this study is that, for the sake of brevity, we
consider the response of the fluid to be purely viscous.  However, most
biopolymer gels have viscoelastic properties defined by a dependence of their
stress field on their deformation at previous instants of time.
A standard procedure to model these effects is to replace the constant
viscosity coefficients in equation \eqref{eq:Leslie_Eriksen} by time-dependent
memory functions, and the products by temporal convolutions.   
Owing to the linear nature of the problem, Fourier transforming in time renders
similar equations in which the constant, real-valued viscosity coefficients are
replaced by complex-valued functions of the frequency, $ \alpha_i \rightarrow
\widehat \alpha_i(s)$.  From there, the calculation of the response function in
the frequency domain follows the same steps as in the purely viscous case,
although its analysis is complicated by the fact that both the viscosity
coefficients and the response function are complex numbers.  Such analysis is
straightforward but is out of the scope of this paper.

One of the main potential limitations of this study is our assumption that the
particle does not distort the nematic director field.  
\textcolor{black}{ Molecular anchoring at the particle surface may distort the
director in liquid crystals, leading to dipole or Saturn ring configurations
next to the particle.
The drag force of a sphere in these configurations has been numerically
calculated by \cite{Ruhwandl.Terentjev.1996}, and \cite{Stark.Ventzki.2001},
showing considerable differences with respect to results obtained for a uniform
nematic.  
The influence of surface anchoring on the local director topology and particle
diffusion has been determined experimentally
\citep{Koenig.Ong.Cortes.Moreno-Razo.dePablo.Abbott.2009} and numerically
\citep{Moreno-Razo.Sambriski.Koenig.Diaz-Herrera.Abbott.dePablo.2011}.
These near-field effects disappear in two-point PTM
\citep{crocker:00,Levine.Lubensky.2001}, which is based on measurements of
correlated motion of distant particles.}

\textcolor{black}
{
Particle motion can also distort the director in the far field due to the
finite elastance of the nematic.
We have hypothesized that this effect is negligible in passive PTM of
biopolymer networks because the thermal energy that drives particle motion is
extremely low.  
However, this energy can be several orders of magnitude higher in active PTM.
In that case, the methodology presented here can be extended by prescribing a
suitable spatiotemporal nematic distribution \citep{Stark.Ventzki.2001},
including the Frank elasticity constants into the formulation
\citep{Frank.1958} or by calculating the dynamics of the nematic through the
Poisson-bracket approach \citep{Stark.Lubensky.2003}.
}

\textcolor{black}
{
Finally, assuming a uniform director is relatively reasonable for live cell PTM
experiments.  Adherent cells are known to align their cytoskeletal fibers in
response to directional mechanical stimuli \citep{galbraith:98,kaunas:05} and
subtratum stiffness \citep{Leeetal:2013}.  Even in the absence of global
alignment, the cytoskeleton will organize in smaller nematic domains in which
the director remains locally uniform.
} 

\section*{Acknowledgements}
%
This work was supported by NSF CAREER grant CBET-1055697. Manuel 
G\'omez-Gonz\'alez was partially funded by Fundaci\'on Ibercaja. The authors 
are indebted to Dr. Shu Chien, Dr. Juan C. Lasheras, Dr. Javier Jimenez and Dr. 
Javier Urzay for fruitful discussions on preliminary versions of this 
manuscript.

\appendix 
%
\section{Green's Functions}
\label{sec:green_ap}

\textcolor{black}{
In this appendix, we outline the procedures to obtain the Green's functions for
the flow velocity and pressure from the equations of motion
\eqref{eq:Motion_1}, \eqref{eq:Motion_2}, \eqref{eq:Motion_3} and
\eqref{eq:continuity}. In Fourier space, these equations of motion become
%
\begin{align}
	\label{eq:Motion_1_F} i k_1 \widehat{p} + \left( \frac{1}{2 \pi}
\right)^{3/2} \widehat{f_x} &= (\eta_c + \eta_b - \eta_a + \alpha_1) k_1^2
\widehat{u} + \eta_b (k_2^2 + k_3^2) \widehat u, \\
	\label{eq:Motion_2_F} i k_2 \widehat{p} + \left( \frac{1}{2 \pi}
\right)^{3/2} \widehat{f_y} &= ~~~~~~~~~~~~~~~~~~~~~~ \eta_c k_1^2 \widehat{v}
+ \eta_a (k_2^2 + k_3^2) \widehat{v}, \\
	\label{eq:Motion_3_F} i k_3 \widehat{p} + \left( \frac{1}{2 \pi}
\right)^{3/2} \widehat{f_z} &= ~~~~~~~~~~~~~~~~~~~~~~ \eta_c k_1^2 \widehat{w}
+ \eta_a ( k_2^2 + k_3^2) \widehat{w}, \\
	\label{eq:Motion_4_F} k_1 \widehat u  + k_2 \widehat v  + k_3 \widehat w 
&= 0.
\end{align}
}

\textcolor{black}{Assuming a linear solution of the form 
\eqref{eq:velocity_greens} - \eqref{eq:pressure_greens}, equations 
\eqref{eq:Motion_1_F} - \eqref{eq:Motion_4_F} simplify to
%
\begin{align}
	\label{eq:Motion_5_F} \left[ (\alpha_1 + \eta_c - \eta_a ) k_1^2 + 
\eta_b k^2 \right] \frac{\widehat{\mathcal{G}}_{1j}}{8 \pi}  &=  
i k_1 \frac{\widehat{\mathcal{P}}_j}{8 \pi} + \left( \frac{1}{2 \pi} 
\right)^{3/2} \delta_{1j}, \\
	\label{eq:Motion_6_F} \left[ (\eta_c - \eta_a ) k_1^2 + 
\eta_a k^2 \right] \frac{\widehat{\mathcal{G}}_{2j}}{8 \pi}  &=  
i k_2 \frac{\widehat{\mathcal{P}}_j}{8 \pi} + \left( \frac{1}{2 \pi} 
\right)^{3/2} \delta_{2j}, \\
	\label{eq:Motion_7_F} \left[ (\eta_c - \eta_a ) k_1^2 + 
\eta_a k^2 \right] \frac{\widehat{\mathcal{G}}_{3j}}{8 \pi}  &=  
i k_3 \frac{\widehat{\mathcal{P}}_j}{8 \pi} + \left( \frac{1}{2 \pi} 
\right)^{3/2} \delta_{3j}, \\
	\label{eq:Motion_8_F} k_i \widehat{\mathcal{G}}_{ij} &= 0 .
\end{align}
}

\textcolor{black}{After applying the divergence of equations 
\eqref{eq:Motion_5_F} - \eqref{eq:Motion_7_F} and using \eqref{eq:Motion_8_F}, 
we obtain
%
\begin{equation}
	\label{eq:Pressure_Green_prev} \frac{\widehat{\mathcal{P}}_j}{8 \pi} =
\left( \frac{1}{2 \pi} \right)^{3/2} \frac{i k_j}{k^2} - \left[ \alpha_1 k_1^2
+ (\eta_b - \eta_a) k^2 \right] \frac{i k_1}{k^2}
\frac{\widehat{\mathcal{G}_{1j}}}{8 \pi} .
\end{equation}
}

\textcolor{black}{Finally, plugging this result into equations \eqref{eq:Motion_5_F} - 
\eqref{eq:Motion_7_F} and \eqref{eq:Pressure_Green_prev} we obtain the Green's 
functions for the velocity and pressure,
%
\begin{align}
	\label{eq:Greens_vx} \frac{\widehat{\mathcal{G}}_{1j}}{\sqrt{8/\pi}} &=
\frac{ \delta_{1j}k^2 - k_1 k_j }{ \alpha_1 k_1^2 (k_2^2 + k_3^2) + \eta_b k^4
+ (\eta_c - \eta_b) k_1^2 k^2 }, \\
	\label{eq:Greens_vy} \frac{ \widehat{\mathcal{G}}_{2j} }{\sqrt{8/\pi}} &=
\frac{ \delta_{2j} }{ (\eta_c - \eta_a) k_1^2 + \eta_a k^2} - k_2 k_j \frac{ (1
- \delta_{1j}) \frac{ \alpha_1 k_1^2 + (\eta_b - \eta_a) k^2 } { (\eta_c -
  \eta_a) k_1^2 + \eta_a k^2 } + 1 }{ \alpha_1 k_1^2 (k_2^2 + k_3^2) + \eta_b
k^4 + (\eta_c - \eta_b) k_1^2 k^2 }, \\
	\label{eq:Greens_vz} \frac{\widehat{\mathcal{G}}_{3j}}{\sqrt{8/\pi}} &=
\frac{ \delta_{3j} }{ (\eta_c - \eta_a) k_1^2 + \eta_a k^2 } - k_3 k_j \frac{
(1 - \delta_{1j}) \frac{ \alpha_1 k_1^2 + (\eta_b - \eta_a) k^2 } { (\eta_c -
\eta_a) k_1^2 + \eta_a k^2 } + 1 }{ \alpha_1 k_1^2 (k_2^2 + k_3^2) + \eta_b k^4
+ (\eta_c - \eta_b) k_1^2 k^2 }, \\
	\label{eq:Greens_p} \frac{\widehat{\mathcal{P}}_{j}}{\sqrt{8/\pi}} &=
i k_j \left[ \frac{1 - \delta_{1j}}{k^2} + \frac{\frac{1 - \delta_{1j}}{k^2}
(\alpha_1 k_1^4 - \eta_a k^4) + \eta_a k^2 + (\eta_b - \eta_a) k_1^2 + 
(\eta_c - \eta_b) k_1^2 \delta_{1j}}{\alpha_1 k_1^2 (k_2^2 + k_3^2) + 
\eta_b k^4 + (\eta_c -\eta_b) k_1^2 k^2} \right].
\end{align}
}
%
\section{Expressions for the Functions Appearing in the Response Function}
\label{sec:functions_ap}
 
Equations \eqref{eq:response_function_11_analytic} and
\eqref{eq:response_function_22_analytic} for the response function of a
sphere in a nematic fluid refer to the non-dimensional functions
%
\begin{align*}
%
	A(\vec \eta) &= \frac{\alpha_1}{\eta_b} + \frac{\eta_c}{\eta_b} - 1, \\
	%
	B(\vec \eta) &= \sqrt{A(\vec \eta)^2 + 4\frac{\alpha_1}{\eta_b}},
\end{align*}
%
\begin{align*}
	C_{+}(\vec \eta) &= \sqrt{\frac{A(\vec \eta) + B(\vec \eta)}{2}}, \\
	%
	C_{-}(\vec \eta) &= \sqrt{\frac{A(\vec \eta) - B(\vec \eta)}{2}}, 
\end{align*}
%
\begin{align*}
	D_{+}(\vec \eta) &= A(\vec \eta) + 2 + B(\vec \eta), \\
	%
	D_{-}(\vec \eta) &= A(\vec \eta) + 2 - B(\vec \eta),
\end{align*}
%
\begin{align*}
	%
	E_{+}(\vec \eta) &= A(\vec \eta) D_{+}(\vec \eta) - 2 \left(
\frac{\eta_c}{\eta_b} - 1 \right), \\
	%
	E_{-}(\vec \eta) &= A(\vec \eta) D_{-}(\vec \eta) - 2 \left(
\frac{\eta_c}{\eta_b} - 1 \right),
\end{align*}
%
which arise from the integrals in equation \eqref{eq:gamma_integrals}. 

These equations degenerate in the limit $B(\vec \eta) \longrightarrow 0$, which
occurs whenever 
%
\begin{equation}
\frac{\eta_c} {\eta_b} \longrightarrow 1 - \frac {\alpha_1} {\eta_b} \pm 2
\sqrt{\frac{-\alpha_1}{\eta_b}}.
\label{eq:ap:singu}
\end{equation}
%
In this limit, $C_+ \rightarrow C_-$, $D_+ \rightarrow D_-$ and $E_+
\rightarrow E_-$, and the expressions that determine the two components of the
response function (equations
\ref{eq:response_function_11_analytic}-\ref{eq:response_function_22_analytic})
become undefined.  Solving the limit yields that $\zeta_{||}=\zeta_\perp=0$ 
along the branch associated with the $-$ sign in the square root of equation
(\ref{eq:ap:singu}), and the same happens for the $-$ branch when $\alpha_1 >
\eta_b$.  As a consequence, the region of parameter space corresponding to
%
\begin{equation}
\eta_c < \eta_b -\alpha_1- 2 \sqrt{-\alpha_1 \eta_b}, \;\;\; \alpha_1 < -\eta_b
\end{equation}
%
is associated with unphysical complex values of the response functions and
cannot be realized.  On the other hand, the signs of $\zeta_{||}$ and
$\zeta_\perp$ do not change along the branch associated with the $+$ sign in the
square root of equation \eqref{eq:ap:singu}.

\bibliographystyle{apa}
\bibliography{Bibliography}
%
\end{document}